\documentclass[titlepage]{article}
\newtheorem{lemma}{Lemma}
\usepackage{amsmath}
\usepackage[dvips]{graphicx}

\newcommand{\eqngraph}[2][1]{\mbox{
\raisebox{-0.45\totalheight}{\includegraphics[scale=#1]{#2.eps}}}}

\newcommand{\dif}{\mathrm{d}}

\renewcommand{\theequation}{\arabic{section}.\arabic{equation}}

\begin{document} 

\title{Angle and Volume Studies in Quantized Space}
\author{Michael Seifert \\ Swarthmore College}
\date{March 19, 2000}
\maketitle

\pagenumbering{roman}

\begin{abstract}
The search for a quantum theory of gravity is one of the major 
challenges facing theoretical physics today.  While no complete 
theory exists, a promising avenue of research is the \emph{loop quantum 
gravity} approach.  In this approach, quantum states are 
represented by \emph{spin networks}, essentially graphs with 
weighted edges.  Since general relativity 
predicts the structure of space, any quantum theory of gravity 
must do so as well;  thus, ``spatial observables'' such as area, volume, 
and angle are given by the eigenvalues of Hermitian operators on 
the spin network states.  We present results obtained in our 
investigations of the angle and volume operators, two operators 
which act on the vertices of spin networks.  We find that the 
minimum observable angle is inversely proportional to the square 
root of the total spin of the vertex, a fairly slow decrease to 
zero.  We also present numerical results indicating that the angle 
operator can reproduce the classical angle distribution.  The 
volume operator is significantly harder to investigate 
analytically;  however, we present analytical and numerical 
results indicating that the volume of a region scales as the 
$3/2$ power of its bounding surface, which corresponds to the 
classical model of space.
\end{abstract}

\tableofcontents \newpage
\pagenumbering{arabic}

\section{Introduction}
Most scientists would probably consider the two ``great theories'' of 
twentieth-century physics to be Einstein's theory of general relativity 
and the theory of quantum mechanics pioneered by Bohr, Heisenberg, and 
Schr\"{o}dinger.  General relativity describes the interaction of matter and 
energy with the fabric of spacetime (we experience this interaction as 
gravity), and accurately predicts the 
large-scale structure of the Universe.  Quantum mechanics, on the other 
hand, deals with small-scale interactions between matter and energy, 
successfully explaining the properties of electron, atoms, and other 
minutia.  

General relativity, in 
some sense, could be considered as the last of the classical theories of 
physics:  it considers space and time to be continuous, and, with 
sufficient information about the current state of a system, allows us to 
predict the results of any future experiments with certainty.  This is at 
odds with quantum mechanics, which often requires observable 
quantities to have discrete spectra;  which says (by the Uncertainty 
Principle) that 
one's knowledge of complementary variables will always have a certain 
degree of uncertainty --- we simply \emph{cannot} have complete 
information about the current state of a system; and which only allows 
us to predict the probabilities of experimental results.

One would hope, then, that there would be some way of reconciling these two 
world-views.  The program pursued by theoretical physicists since the 
establishment of quantum mechanics has been, given a classical theory, to 
attempt to construct a quantum theory which has the known classical theory 
as its large-scale limit;  this process is commonly known as 
\emph{quantization}.  In the case of electrodynamics, the other 
major classical field theory, this program has been successful:  the theory of quantum 
electrodynamics (also known as QED) was formulated in the 1950s by 
Feynman, Schwinger, and Tomonaga.  Out of QED grew the \emph{Standard 
Model}, which includes not only electromagnetism but also the
strong and weak nuclear forces (which have no corresponding 
classical theories).

However, there does not yet exist a consistent quantum theory of 
gravity.  This is partly 
due to a dearth of experimental data around which to build a theory.  To 
obtain data to help us in this quest, we would have to devise an 
experiment which would (directly or indirectly) measure the gravitational 
interaction on a small scale.  This pursuit is stymied, however, by the 
incredible weakness of the gravitational force compared to the other three 
forces of the Standard Model on these scales.
An illustration of the vast difference between the strength of the 
electrostatic force and the gravitational force is attributed to 
Feynman:  If you jumped off the Empire State Building, the gravitational 
force between you and the Earth would accelerate you to a speed of about 80 
m/s over the 350-metre height of the building.  However, when you hit the 
ground, the electrostatic repulsion between the molecules in your body and 
the molecules in the sidewalk would \emph{decelerate} you an equal amount 
over less than a centimetre --- a minuscule distance.

Moreover, the scales on which quantum gravitational effects might 
possibly be observed are minuscule.  The fundamental constants 
governing a quantum theory of gravity are $\hbar$ (Planck's 
constant), $c$ (the speed of light), and $G$ (the gravitational 
constant.)  If we express the units of each of these constants in 
terms of length, mass, and time, we obtain:
\begin{eqnarray*}
    \hbar & \to & \mathrm{ \frac{ (mass) (length)^{2}}{time}}  \\
    c & \to & \mathrm{\frac{length}{time}}  \\
    G & \to & \mathrm{\frac{(length)^{3}}{(mass) (time)^{2}}}
\end{eqnarray*}
Playing around with these constants, we discover that the only way to 
create a quantity with dimensions of length is
\begin{equation} \label{PlanckLength}
    l_{0} = \sqrt{\frac{\hbar G}{c^{3}}} \approx 1.62 \times 10^{-35} 
    \mbox{ m.}
\end{equation}
This length, known as the \emph{Planck length}, is the scale on which 
most researchers expect to observe quantum gravitational effects.  
This is an incredibly minuscule distance;  as a basis for comparison, 
the radius of a proton is on the order of $10^{-15}$ metres, a full 
\emph{twenty} orders of magnitude larger!  This fact adds to the 
difficulties in finding any experimental data to guide the search for 
a quantum theory of gravity.

Undeterred by this lack of physical data, many physicists have proposed 
theories which attempt to reconcile quantum mechanics with general 
relativity.  Some of these theories (which will not concern us here) 
attempt to expand or modify the Standard Model to include gravity;  
these approaches include the popular theories collectively known as 
\emph{string theory}.  Other 
attempts to reconcile quantum mechanics and general relativity are more 
conservative, merely attempting to construct a theory of quantum gravity 
without worrying about the forces included in the Standard Model (at least 
for the moment.)  The theory we will be dealing with in this work falls 
into the latter category:  it attempts only to create a quantum theory with 
general relativity as its large-scale (``classical'') limit.  This is the 
\emph{loop quantum gravity} approach, in which the fundamental 
variables of general relativity are encoded in loops (essentially 
closed paths through space);  the 
observables of this theory then depend only on these loops.  

Since general relativity makes predictions about the structure of space, we 
would expect that this theory would make predictions about the structure 
of space on a small scale.\footnote{
The reader may note here that general relativity does not just 
predict the structure of space, but the structure of spacetime.  
However, the theory of loop quantum gravity (as we will consider it) 
is a \emph{canonical} theory, i.e.~it
requires a separation of the kinematics and the dynamics of the 
system.  As a result, we will only consider the states of the system 
(corresponding to space only), 
and not their evolution.  We must also point out here that while some
theories of the dynamics of spin networks have been put forward, 
their time-evolution is still not fully understood.}
In particular, this theory makes predictions 
about the possible values one could observe for such ``spatial observables'' as 
areas, angles, lengths, and volumes.  In this work, we will describe recent 
research into to the spectra of operators corresponding to angle and 
volume.

Note that this work is intended to be understandable to the reader 
\emph{without} any significant background in general relativity.  As a 
result, a large portion of the initial theory (Section \ref{LoopTheory}) is 
couched in terms of a theory the average reader may be more familiar 
with, namely electrodynamics;  in Section \ref{HoloGR}, we make the connection to 
general relativity by (qualitatively) describing the major differences 
encountered in applying the described procedures to general relativity.  
In Section \ref{SpinNet}, we describe \emph{spin networks}, a complete and 
sufficient basis for the states of loop quantization theory, as well as 
the actions of operators on this basis.  

Finally, in Sections \ref{AngRes} and \ref{VolRes}, we describe the 
results obtained in our research on the angle and volume operators, 
respectively.  In particular, we find that the minimum observable angle 
for a given vertex is inversely proportional to the square root of its 
total spin --- a fairly slow convergence to zero.  We also calculate the 
``resolution'' of angles around a vertex given its total spin, and show 
that the distribution of angles around a vertex with sufficient total spin 
approximates the classical distribution.  The complexity of the volume 
operator hinders any serious analytical work;  however, we do use 
numerical methods to show that the volume of a region seems to scale as the 
3/2 power of the area of its bounding surface, which is also expected classically. 
\setcounter{equation}{0}

\section{Theory of Loop Quantization}
\label{LoopTheory}

\subsection{Canonical Electrodynamics}
\label{CanElec}

As stated in the introduction, we will begin by examining loop 
quantization theory in the context of classical electrodynamics.  The 
fundamental equations of electrodynamics are, of course, Maxwell's 
equations:
\begin{equation}
\begin{array}{c @{\qquad} c}
    \nabla \cdot \mathbf{E} = 0 & \nabla \times \mathbf{B} - 
    \frac{1}{c} \frac{\partial \mathbf{E}}{\partial t} = 0\\[6pt]
    \nabla \cdot \mathbf{B} = 0 & \nabla \times \mathbf{E} + 
    \frac{1}{c} \frac{\partial \mathbf{B}}{\partial t} = 0
\end{array}
\end{equation}
(note that here and throughout this section, we will be working with 
the source-free equations in Gaussian units.)  The symmetry of these 
equations is striking;  if we 
replace $\mathbf{E}$ with $\mathbf{B}$ and $\mathbf{B}$ with 
$-\mathbf{E}$, the equations are unchanged.  One might wonder, then, 
what the source of this symmetry might be.  

To answer this question, we need to return to one of the fundamental 
properties of $\mathbf{E}$ and $\mathbf{B}$:  their definition as 
vectors.  We 
can define any vector $\mathbf{x}$ in terms of its (Cartesian) coordinates 
$(x^{1}, x^{2}, x^{3})$.\footnote{
Note that these superscripts are merely indices, not powers of $x$.}  
In  this case,
\begin{equation}
    \mathbf{x} = x^{1} \hat{e}_{1} + x^{2} \hat{e}_{2} + x^{3} 
    \hat{e}_{3}
\end{equation}
To save effort, we will introduce the \emph{Einstein summation convention} 
here:  if an index is repeated in an expression, then we interpret 
this as summation over all possible values of that index.  Using this 
convention, the previous equation becomes
\begin{equation}
    \mathbf{x} = x^{i} \hat{e}_{i}
\end{equation}

Special relativity, however, tells us that we must also consider an 
event's position in time (not just in space.)  To take this into 
account, we modify the conventional idea of a vector to create a 
\emph{four-vector};  instead of having a component for 
each of the $x$-, $y$-, and $z$-directions, a four-vector also has a component 
for the $t$-direction.  We will denote these components as
$x^{0} = ct, x^{1} = x, x^{2} = y, \mbox{ and } x^{3} = z$.  An event 
with spacetime coordinates $(x^{0}, x^{1}, x^{2}, x^{3})$ can then be 
written as
\begin{equation}
    \mathbf{x} = x^{\mu} \hat{e}_{\mu}
\end{equation}
In general, we will use Greek indices such as $\mu, \nu, \ldots$ to 
denote components that must be summed over space and time, while Roman 
indices such as $i, j, k, \ldots$ will denote summation over space only.

Special relativity tells us that if we switch between two frames 
travelling at a relative velocity $\beta = v/c$ to each other in the 
$x$-direction, the components of a spacetime event $\mathbf{x}$ will 
transform as
\begin{equation} \label{LTrans1}
    \begin{split}
    \bar{x}^{0} & = \gamma (x^{0} - \beta x^{1})  \\
    \bar{x}^{1} & = \gamma (-\beta x^{0} + x^{1})  \\
    \bar{x}^{2} & = x^{2}  \\
    \bar{x}^{3} & = x^{3}
    \end{split}
\end{equation}
where the ``barred'' components are in the frame to which we are transforming. 
Using the summation convention outlined above, we can write 
(\ref{LTrans1}) as a single equation:
\begin{equation} \label{LTrans2}
    \bar{x}^{\mu} = \Lambda_{\nu}^{\mu} x^{\nu} \mbox{,}
\end{equation}
where $\Lambda$ is the Lorentz transformation matrix
\begin{equation}
    \Lambda = \left( \begin{array}{cccc}
    \gamma & -\gamma \beta & 0 & 0 \\
    -\gamma \beta & \gamma & 0 & 0 \\
    0 & 0 & 1 & 0 \\
    0 & 0 & 0 & 1
    \end{array} \right) \mbox{,}
\end{equation}
and $\Lambda_{\nu}^{\mu}$ is the component of the $\mu$th row and 
the $\nu$th column of this matrix.
 
If $\mathbf{E}$ and $\mathbf{B}$ are really vectors, then, they should 
transform according to equation (\ref{LTrans2}).  However, a Lorentz 
transformation on a given electromagnetic field gives the components 
of the transformed fields as
\begin{equation} \label{LTrans3}
    \begin{array}{c @{\qquad} c @{\qquad} c}
    \bar{E_{x}} = E_{x} & \bar{E_{y}} = \gamma \left( E_{y} - \beta 
    B_{z} \right) & \bar{E_{z}} = \gamma \left( E_{z} + \beta B_{y} 
    \right) \\[6pt]
    \bar{B_{x}} = B_{x} & \bar{B_{y}} = \gamma \left( B_{y} + \beta 
    E_{z} \right) & \bar{B_{z}} = \gamma \left( B_{z} - \beta E_{y} 
    \right) 
    \end{array}
\end{equation}
--- not at all a simple Lorentz transformation like those mentioned 
above.

The way that the components of $\mathbf{E}$ and $\mathbf{B}$ mix 
suggests that these two fields are really part of a larger entity.  
As it happens, they are components of a second-rank tensor.
Such a tensor has sixteen components, which we can denote in an array:\footnote{
Note that, technically, this tensor should not be considered a 
matrix, since it has two lowered indices;  if we were to multiply it 
by a four-vector $x^{\mu}$ as in (\ref{LTrans2}), we would not obtain another 
vector $y^{\nu}$.}
\begin{equation}
    F_{\mu \nu} \longrightarrow \left\{ \begin{array}{cccc}
    F_{00} & F_{01} & F_{02} & F_{03} \\
    F_{10} & F_{11} & F_{12} & F_{13} \\
    F_{20} & F_{21} & F_{22} & F_{23} \\
    F_{30} & F_{31} & F_{32} & F_{33}
    \end{array} \right\}
\end{equation}
Theoretically, each one of these components could be different;  
however, such an object has 16 degrees of freedom, whereas we 
only have six to account for from $\mathbf{E}$ and $\mathbf{B}$.  As 
it happens, if we require $F_{\mu \nu}$ to be antisymmetric (i.e. 
$F_{\mu \nu} = - F_{\nu \mu}$), such a tensor would only have six 
degrees of freedom:
\begin{equation}
    F_{\mu \nu} \longrightarrow \left\{ \begin{array}{cccc}
    0 & F_{01} & F_{02} & F_{03} \\
    -F_{01} & 0 & F_{12} & F_{13} \\
    -F_{02} & -F_{12} & 0 & F_{23} \\
    -F_{03} & -F_{13} & -F_{23} & 0
    \end{array} \right\}
\end{equation}

How would the components of such a tensor transform under a Lorentz 
transformation?  In analogy to the transformation of vectors in 
(\ref{LTrans2}), we have
\begin{equation} \label{LTrans4}
    \bar{F}_{\mu \nu} = \Lambda^{\kappa}_{\mu} \Lambda^{\lambda}_{\nu} 
    F_{\kappa \lambda}
\end{equation}
If we write out the transformations of each component of $F_{\mu \nu}$ 
explicitly, we obtain
\begin{equation}
    \begin{array}{c @{\qquad} c @{\qquad} c}
    \bar{F}_{01} = F_{01} & \bar{F}_{02} = \gamma \left( F_{02} - \beta 
    F_{12} \right) & \bar{F}_{03} = \gamma \left( F_{03} + \beta F_{31} 
    \right) \\
    \bar{F}_{23} = F_{23} & \bar{F}_{31} = \gamma \left( F_{31} - \beta 
    F_{03} \right) & \bar{F}_{12} = \gamma \left( F_{12} + \beta F_{02} 
    \right)
    \end{array}
\end{equation}
The reader will note that these are exactly the 
transformations stated in (\ref{LTrans3}), suggesting that we can 
write the electric and magnetic fields together in the \emph{field 
tensor} $F_{\mu \nu}$ as
\begin{equation}
    F_{\mu \nu} \longrightarrow \left\{ \begin{array}{cccc}
    0 & E_{x} & E_{y} & E_{z} \\
    -E_{x} & 0 & B_{z} & -B_{y} \\
    -E_{y} & -B_{z} & 0 & B_{x} \\
    -E_{z} & B_{y} & -B_{x} & 0
    \end{array} \right\} \mbox{.}
\end{equation}
We have thus succeeded in combining the electric and magnetic fields 
into a single tensor $F_{\mu \nu}$;  the peculiar transformations of $\mathbf{E}$ 
and $\mathbf{B}$ are merely consequences of the properties of this 
tensor.
However, the symmetry in Maxwell's equations suggests that 
this is not the only way to put the components of $\mathbf{E}$ and 
$\mathbf{B}$ into a second-rank antisymmetric tensor;  we could 
equally well have constructed the \emph{dual tensor} $G^{\mu \nu}$:
\begin{equation}
    G^{\mu \nu} \longrightarrow \left\{ \begin{array}{cccc}
    0 & B_{x} & B_{y} & B_{z} \\
    -B_{x} & 0 & -E_{z} & E_{y} \\
    -B_{y} & E_{z} & 0 & -E_{x} \\
    -B_{z} & -E_{y} & E_{x} & 0
    \end{array} \right\}
\end{equation}

In terms of these new tensors, Maxwell's equations become radically 
simpler.  They are now:
\begin{equation}
    \partial_{\nu} F^{\mu \nu} = 0 \qquad \mbox{and} \qquad \partial_{\nu} G^{\mu 
    \nu} = 0 \mbox{,}
\end{equation}
where we have introduced the operators $\partial_{\mu} = \partial / 
\partial x^{\mu}$.  That these equations are equivalent to Maxwell's 
equations is not entirely evident at first 
glance, so let us compute a couple of the equations (note that each 
of the equations above is really four separate equations, one for each 
value of $\mu$) to show this equivalence. 
Consider the first equation when $\mu = 0$;  this becomes
\begin{eqnarray}
    \partial_{\nu} F^{0 \nu} & = & \partial_{0} F^{00} + \partial_{1} 
    F^{01} + \partial_{2} F^{02} + \partial_{3} F^{03} \nonumber \\
    & = & \frac{\partial E_{x}}{\partial x} + \frac{\partial 
    E_{y}}{\partial y} + \frac{\partial E_{z}}{\partial z} \nonumber 
    \\
    & = & \nabla \cdot \mathbf{E} = 0 \mbox{,}
\end{eqnarray}
which is Gauss' Law.  In the case where $\mu = 1$, we have 
\begin{eqnarray}
    \partial_{\nu} F^{1 \nu} & = & \partial_{0} F^{10} + \partial_{1} 
    F^{11} + \partial_{2} F^{12} + \partial_{3} F^{13} \nonumber \\
    & = & - \frac{1}{c} \frac{ \partial E_{x}}{\partial t} + 
    \frac{\partial B_{z}}{\partial y} - \frac{\partial B_{y}}{\partial 
    z} \nonumber \\
    & = & \left( -\frac{1}{c} \frac{\partial \mathbf{E}}{\partial t} + \nabla 
    \times \mathbf{B} \right)_{x} = 0
\end{eqnarray}
which is the $x$-component of Ampere's law.  The $\mu = 2$ and $\mu = 
3$ cases of this equation give the $y$ and $z$ components of 
Ampere's law, and the tensor equation for $G^{\mu \nu}$ gives the two 
remaining Maxwell's equations.\footnote{
This method can be elegantly extended to include field sources (i.e. charges 
and currents), but this case will not concern us here.  The 
interested reader is referred to Griffiths \cite{Griff} (from which 
most of this derivation is taken) for further information.}

\subsection{The Vector Potential, Holonomies \& the Aharonov-Bohm effect}

To quantize a given theory, we need to decide on a particular set of 
configuration variables which describe the system.  In the 
simplest example, the quantum theory of a free particle uses the 
particle's position $x$ and momentum $p$ to describe the motion of the 
particle.  In general, we want to find some set of generalized 
``positions'' and ``momenta'' that can describe all the states of the 
system; however, doing so is not always easy, since a ``natural'' set 
of configuration variables may be related by some set of complicated 
constraints.  A general method for finding the constraints on an 
arbitrary set of configuration variables
 was pioneered by Bergmann and Dirac, and is presented in 
Appendix \ref{BDCA}, along with its application to electrodynamics.  
Here, however, we only need to note the final results:  the canonical 
configuration variable for electrodynamics is the vector 
potential $\mathbf{A}$, given by $\mathbf{B} = \nabla \times \mathbf{A}$.  
We will also make use of the four-vector potential, given by
\begin{equation}
    A_{\mu} = (V, A_{x}, A_{y}, A_{z})
\end{equation}
where $V$ is the electric potential.  
The field tensor $F_{\mu \nu}$ can be written quite elegantly in 
terms of this four-vector potential:
\begin{equation}
    F_{\mu \nu} = \partial_{\mu} A_{\nu} - \partial_{\nu} A_{\mu} 
    \mbox{.}
\end{equation}
To successfully quantize this theory, we must also have a conjugate 
momentum for the chosen configuration variable.  Using Bergmann-Dirac 
canonical analysis, we can show that the ``momentum'' of the vector 
potential $A_a$ is the electric field $E^a$.\footnote{
Note that in this equation, the components of $E$ have a 
superscript instead of a subscript;  this is due to the difference 
between covariant and contravariant tensors.  In this case, 
however, raising or lowering an index simply 
introduces a minus sign if and only if the index is 0 (i.e. $x^{0} = - 
x_{0}$, but $x^{1} = x_{1}$.)}  Our two configuration variables are 
the vector potential and the electric field.  

One might ask the question, however, whether the vector potential is 
necessarily the best choice for a configuration variable, especially 
since (in the classical theory) only $\mathbf{B} = \nabla \times 
\mathbf{A}$ is measurable.   
Furthermore, $A_a$ has the distinct disadvantage of possessing 
gauge freedom:  given a vector potential $A_a$, we can perform a 
transformation of the form
$$
A_a \to A_a + \frac{\partial \lambda}{\partial x^{a}}
$$
without affecting the physical state.  This freedom is worse than it 
appears:  suppose we have a physical system which we consider under 
two gauges $\lambda_{1}$ and 
$\lambda_{2}$, the first of which is a function of time and the 
second of which is not.  If we let the system evolve, the two gauges 
will generally predict different values of $A_{a}$ after a given 
amount of time;  hence, our theory does not uniquely predict the 
time-evolution of the configuration of the system.
If we insist on treating 
$A_a$ as a fundamental variable, then, we must abandon 
determinism --- a step that we are loath to take.  Further, this interpretation 
does not prohibit causal influences from travelling faster than light;  
in theory, a change in the field can cause an instantaneous change arbitrarily 
far away.  Thus, interpreting $A_a$ as a fundamental field forces us to 
eliminate determinism and causality, two of the most basic tenets of 
physics.  

In the face of this, most people would interpret $\mathbf{B} = \nabla 
\times \mathbf{A}$ (and $\mathbf{E}$) as physical fields;  this is the standard classical interpretation.  
This interpretation has the advantage of being deterministic (since gauge 
transformations do not change $\mathbf{B}$) and limiting causal influences to the 
speed of light.  Indeed, if we are defining our fields on a 
``well-behaved'' chunk of space, or only examining classical 
phenomena, then the choice of $\mathbf{B}$ as the fundamental variable is 
natural, and explains everything that needs explaining.\footnote{
The reader might wonder what we might possibly mean by 
``well-behaved'';  this will become evident shortly.}

\begin{figure}
    \begin{center}
    \includegraphics[width=4in]{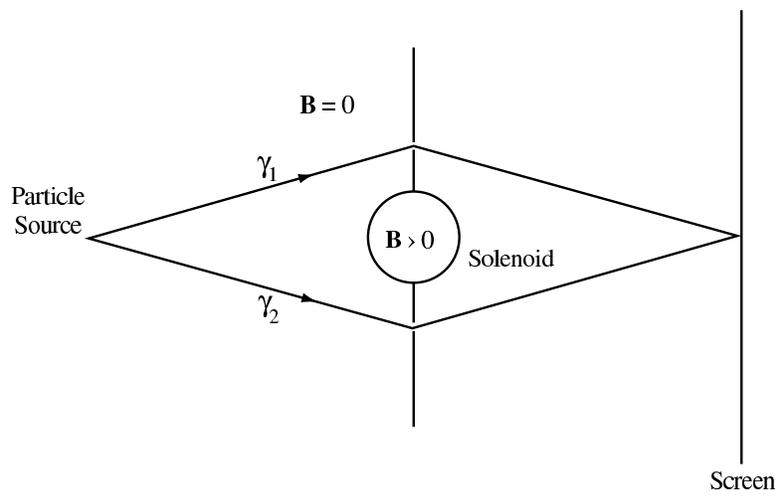}
    \end{center}
    \caption{\label{AharBohm}  The Aharonov-Bohm experiment.  Particles are 
    sent through a double-slit apparatus and create an interference 
    pattern on the screen.  The presence of the solenoid creates a phase 
    difference between the two paths $\gamma_1$ and $\gamma_2$, which 
    affects the interference pattern.  Note that although the magnetic field is 
    \emph{zero} in the region where the particle travels, the 
    solenoid still has an effect on the particles.}
\end{figure}

As the reader might have guessed from the above caveat, however, 
there are quantum effects that seem to contradict the primacy of 
$\mathbf{B}$.  The best-known of these is the \emph{Aharonov-Bohm effect}. In 
this experiment, a beam of charged particles with charge $e$ is sent 
through a double-slit interference experiment (see 
Figure \ref{AharBohm}.)  Between the slits we place an 
impenetrable solenoid (i.e. the particles' wave functions are zero inside 
the solenoid.)

Theory predicts that a particle travelling along a 
given path $\gamma$ will change its phase factor:
\begin{equation} \label{Phfactor}
| \phi (t) \rangle = \exp \left[ \frac{-i E t}{\hbar} + \frac{i e}{\hbar 
c} \oint_{\gamma} A_{i} \, \dif x^{i} \right] 
\end{equation}
where $E$ is the energy of the particles.
The first part of the phase factor corresponds to the normal time-evolution 
of the state;  the second part arises solely from the vector 
potential due to the solenoid (even though the particle is moving in 
a region of no magnetic field.)
This phase factor will have observable consequences in the interference 
pattern on the screen behind the slits.  For example, the phase difference 
between the two paths at the centre of the screen (i.e. equal time for 
both paths) will be given by
\begin{eqnarray}
\Delta \phi & = & \frac{e}{\hbar c} \int_{\gamma_1} A_{i} \, \dif x^{i} - 
\frac{e}{\hbar c} \int_{\gamma_2} A_{i} \, \dif x^{i} \nonumber \\
& = & \frac{e}{\hbar c} \oint_{\gamma}  A_{i} \, \dif x^{i} \label{Hol0}
\end{eqnarray}
where $\gamma$ is the closed loop created by travelling along 
$\gamma_{1}$, then along $\gamma_{2}$ in reverse.  The integrand of 
this equation can be expressed more familiarly as $\mathbf{A} \cdot 
\dif \mathbf{x}$; we can thus apply Stokes' Law to this to give 
\begin{eqnarray}
\Delta \phi & = & \frac{e}{\hbar c} \int_{S} \nabla \times \mathbf A 
\cdot \dif \mathbf{S} \nonumber \\
& = & \frac{e \Phi}{\hbar c}
\end{eqnarray}
where  $\gamma$ is the border of the new surface $S$, and 
$\Phi = \int \mathbf{B} \cdot \dif \mathbf{S}$ is the magnetic flux through 
this surface (which is equal to the flux through the solenoid, since 
$\mathbf{B} = 0$ outside it.)  

This experiment has been performed (the experimenters measured shifts in the interference 
pattern as $\Phi$ was increased);  the results confirm the theoretical 
prediction, that the interference pattern does indeed depend on $\int 
\mathbf{A} \cdot \dif \mathbf{x}$.
This result seems to catch us in a contradiction.  On the one hand, 
choosing the vector potential $A^{\mu}$ as a fundamental 
configuration variable forces us to abandon 
determinism and causality;  on the other hand, the Aharonov-Bohm 
effect suggests that the vector potential holds some physical 
importance.  Without a third option, we are forced to choose the 
former --- while a non-deterministic theory is not very nice, one 
that doesn't correspond to the physical world is worthless.

Fortunately, a third option does exist: \emph{holonomies}.  Recall that 
the observable result of the Aharonov-Bohm effect is not dependent on 
the vector potential \emph{per se}, but instead on its path integral 
as defined in (\ref{Hol0}).  Thus, given a path in space 
$\gamma$, we define the holonomy of $\gamma$ to be
\begin{eqnarray}
U_{\gamma} & = & \exp \left[ i g \int_0^1 A_{i} (\gamma(t)) 
{\dot{\gamma}}^{i}(t) \, \dif t \right] \nonumber \\
\label{hol1} & = & \exp \left[ i g \int_{\gamma} A_{i} \, dx^{i} \right]
\end{eqnarray}
where $g$ is a coupling constant to make the argument of the 
exponential dimensionless.  The reader will note that if we 
set $g = e/\hbar c$, this factor reduces to that given in the second 
term of (\ref{Phfactor}).

We can thus escape our earlier dilemma by regarding the holonomies as 
the fundamental configuration variables.  The consequences of doing 
this are discussed 
at length in an excellent article by Belot \cite{Belot}.  He notes 
that the 
holonomies, like $\mathbf{B}$, are gauge-invariant;  however, they 
possess another important, related quality which the magnetic field 
does not.  In a chunk of 
``well-behaved'' space, two fields correspond to the same $\mathbf{B}$ 
field if and only if they differ by one of the usual gauge 
transformations (i.e. a term of the form $\partial \lambda/\partial 
x^{a}$ for some $\lambda$.)   This is because normal Euclidean 
space is \emph{simply connected}: any loop in space (i.e. 
a path through space that begins and ends at the same point) can be 
contracted continuously to a point.  However, if a space is 
not simply connected, then there are non-gauge transformations 
that are interpreted as identical $\mathbf{B}$ fields.  The 
holonomies, on the other hand, do not have this problem;  this is 
thus a point in their favour.\footnote{The reader who might be wondering if 
non-simply-connected spaces actually correspond to any physical 
situations is reminded that the Aharonov-Bohm effect takes place in 
such a space --- a loop around the solenoid cannot be contracted ``through'' 
the solenoid.}

However, using the holonomies as the fundamental variables of the 
theory violates locality in an even more fundamental 
way than our first interpretation.  Instead of considering the field 
$A_a$ at a collection of points, we must now consider the values 
of the field over a region with actual spatial extent;  this means 
that if we don't know the field \emph{everywhere}, we cannot determine 
all of the possible holonomies.  The philosophical issues this fact 
raises are fascinating, but are unfortunately not the primary concern 
of this paper;  the interested reader is encouraged to consult Belot's 
paper for more information on this subject.

\subsection{Holonomies and Reconstruction Theorems} \label{HaCE}

We have shown that the construction of a quantum theory from the canonical 
variables $A_a$ and $E^a$ depends in some way on holonomies;  
we will now make this notion somewhat more formal.  We 
define a path $\gamma$ on a space $\Sigma$ to be a continuous map from the closed 
unit interval $[0,1]$ to $\Sigma$.  In the case of Euclidean space we can 
think of $\Sigma$ as equivalent to $\mathbf{R}^3$;  however, $\Sigma$ may 
equally well be a ``curved'' manifold, as in general relativity.  If the 
property $\gamma(0) = \gamma(1)$ holds, then we say that $\gamma$ is a 
\emph{closed loop}.  

Given a (not necessarily closed) path $\gamma$ and a vector potential $A_a$, we can then define 
a holonomy $U_{\gamma}$ of the path $\gamma$:
\begin{eqnarray}
U_{\gamma} & = & \exp \left[ i g \int_0^1 A_a (\gamma(t)) 
{\dot{\gamma}}^{a}(t) \, dt \right] \nonumber \\
& = & \exp \left[ i g \int_{\gamma} A_a \, dx^{a} \right]
\end{eqnarray}
where $g$ is merely a coupling constant introduced to make the argument of 
the exponential dimensionless.  In other words, the holonomy of a given 
path is merely (in this simple case) the exponential of $i$ times the path 
integral of the vector potential along the path in question.

An important feature of closed-loop holonomies is that it is generally possible 
to reconstruct the vector potential from them;  in other words, the 
representation of the vector potential in terms of its holonomies and in 
terms of the field itself are completely equivalent.  The conditions under 
which this holds are the subject of mathematical \emph{reconstruction theorems}.  
We will not go into detail on how the existence of a vector potential 
corresponding to a given set of loops is proven, nor the actual process of 
the reconstruction itself;  for more details, the interested reader is 
referred to Barrett \cite{Barrett} or Giles \cite{Giles}.  

\subsection{Canonical variables \& holonomies in general relativity}
\label{HoloGR}

Before we examine the nature of the spin networks mentioned in Section 
\ref{HaCE}, we must discuss the difficulties in applying this technique to 
Einstein's theory of general relativity.  As originally conceived by 
Einstein, general relativity consisted of a spacetime manifold $\Sigma$, at each 
point of which a metric $g_{\mu \nu}$ is defined.  Early attempts to quantize 
general relativity used the metric as one of the parameters of the 
configuration space.  Unfortunately, this approach failed, mainly because 
the constraints generated by this approach (as in Appendix \ref{BDCA}) 
were intractable.

There the matter rested until the 1986, when 
Ashtekar \cite{Ash1, Ash2} and Sen \cite{Sen}
reformulated general relativity in terms of the so-called ``new 
variables'' . In this formulation, the metric was no longer the fundamental 
variable of the theory.  Instead, the two variables used in this new 
formulation are the \emph{connection} $A_a^i(x)$ (akin to the vector 
potential in electrodynamics) and the \emph{inverse 
densitized triad} $E^a_i(x)$ (akin to the electric field).
As our notation would suggest, the 
connection is viewed as the configuration variable (akin to the vector 
potential), while the inverse densitized triad is its canonically 
conjugate momentum.  These quantities have ``clean'' Poisson brackets, as 
we require:
\begin{eqnarray}
\left\{E^a_i(x), E^b_j(y) \right\} & = & 0 \\
\left\{A^i_a(x), A^j_b(y) \right\} & = & 0 \\
\left\{A^i_a(x), E^b_j(y) \right\} & = & \beta G \delta_i^j \delta_a^b 
\delta^3(x-y)
\end{eqnarray}

For those readers familiar with general relativity, we can also express 
these variables in terms of more familiar quantities.  $E^a_i$ is related 
to the metric of constant-time surfaces $q_{ab}$ by
\begin{equation}
|q| q^{ab} = E^a_i E^b_i \mbox{,}
\end{equation}
where $q_{ab}$ is the spatial part of the metric and $|q|$ is the determinant 
of $q_{ab}$.  We can also express these in terms of the triad from the 
ADM formalism:
\begin{equation}
E^a_i = \sqrt{|q|} e^a_i
\end{equation}
The connection $A_a^i(x)$ can be expressed as
\begin{equation}
A_a^i = \Gamma_a^i + \beta K^i_a \mbox{,}
\end{equation}
where $\Gamma_a^i$ is the connection associated with the (undensitized) 
triad (defined by $\partial_{[a} e^i_{b]} = \Gamma_{[a}^i e_{b]j}$), $K^i_a$ is 
the extrinsic curvature of the 3-surface, and $\beta$ is an as-yet 
undetermined parameter known as the \emph{Immirzi parameter}, first 
put forward in \cite{Imm}.  

While there is a resemblance between electrodynamics and general 
relativity in this regard, there are also differences (of course) between 
these two theories.  The reader will no doubt have noticed by now 
that our relativistic variables bear an extra index (i.e. $i, j, 
\ldots$).  This index corresponds to the \emph{internal space} of the 
configuration variables.  Each point on $\Sigma$ has an internal space 
associated with it; a vector based at a given point can be thought of as 
existing in this space.  Since two vectors at two different points 
exist in different spaces, they cannot immediately be compared.  We 
need to define some self-consistent way of comparing two objects at 
two different points;  this, as it turns out, is done by the 
connection $A^i_a$.  

Suppose, then, we take a vector at some point $x_0$ on the manifold, and 
attempt to move it around the loop (using the connection $A^i_a$ to 
translate from the neighbourhood of each point to the manifold.)  This 
movement will be performed in such a way that as the vector passes through 
a small neighbourhood of each point, it will point in the same direction.  
This process is called \emph{parallel transport}, since the vector stays 
parallel to itself along each infinitesimal segment.

\begin{figure}
    \begin{center}
    \includegraphics[width=4in]{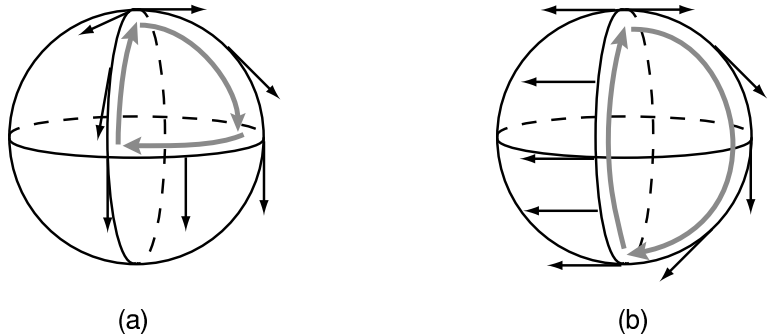}
    \end{center}
    \caption{\label{Sphere}  Parallel transport of vectors on a non-flat 
    surface.  In (a), the vector is transported from the north pole to the 
    equator, 90 degrees along the equator, and back to the north pole;  the 
    resulting vector is rotated by 90 degrees.  In (b), the vector is 
    transported from the north pole to the south pole along a line of 
    longitude, and back to the north pole along another line of longitude 
    90 degrees from the first;  the resulting vector is rotated by 180 
    degrees.  Note that the vector has been transported ``parallel to 
    itself'' at each point;  it is the curvature of the surface itself 
    that causes these rotations. }
\end{figure}

Transporting a vector around a loop in this way does not always 
yield a vector that is parallel to the original vector --- even if we keep 
the vector parallel to itself along each ``step of the way.''  In the 
special case of flat space, the parallel-transported vector is 
equivalent to the original vector.  However, if space has some amount of 
curvature, we will not necessarily get the same vector back after we 
transport it around a loop.  (For an example of this, see Figure 
\ref{Sphere}, in which a vector is parallel-transported around two 
different loops on a sphere.)  With this in mind, we associate a 
three-dimensional rotation with each path in $\Sigma$;  these 
rotations will be the holonomies of general relativity.
In three dimensions, we can represent these rotations by 
members of the group $SO(3)$, the group of all orthogonal $3 \times 3$ 
matrices with determinant 1.  However, this group has the same structure as 
(i.e. is isomorphic to) $SU(2)$, the group of all $2 \times 2$ unitary matrices 
with determinant 1;  since these matrices are simpler, 
general relativity is more commonly though of in terms of $SU(2)$ than in 
terms of $SO(3)$.  

This representation of the holonomies in terms of matrices has two 
important consequences that complicate the simple derivation we used for 
canonical electrodynamics.  The first complication arises from the 
structures of the groups used in each theory.
In the case of canonical electrodynamics, the holonomies were complex 
numbers of unit modulus;  these can be thought of, in the interest of 
drawing parallels, as members of $U(1)$ (the set of all unitary $1 \times 
1$ matrices.)  $U(1)$ happens to be a simple group, in that any two elements $a, 
b \in U(1)$ commute (i.e. $ab = ba$.)  However, $SU(2)$ is not so simple;  
in general, elements of $SU(2)$ do not commute (i.e. $ab \neq 
ba$.)\footnote{
The reader with a modicum of background in group theory will recognize 
that $U(1)$ is \emph{Abelian}, while $SU(2)$ is \emph{non-Abelian.}}
Because of this, it turns out that we can no longer use the simple equation 
in (\ref{hol1}) to find the holonomy.  Instead of simply taking the 
exponential of a path integral, we must use the \emph{path-ordered 
exponential} defined as
\begin{eqnarray}
\lefteqn{U_{\gamma}[A] = \mathcal{P} \exp \left[ \int_0^1 A_{a} (\gamma(t)) 
{\dot{\gamma}}^{a}(t) \, dt \right]} \\
& & = 1 + \sum_{n=1}^{\infty} \int_0^1 \! \! \! dt_1 
\int_{t_1}^1 \! \! \! dt_2 
\cdots \int_{t_{n-1}}^1 \! \! \! \! \! dt_n  \nonumber \\ 
& & \qquad \qquad \qquad \qquad \times A_{a_n} (\gamma(t_n)) \ldots A_{a_1} 
(\gamma(t_1)) {\dot{\gamma}}^{a_n}(t_n) \ldots {\dot{\gamma}}^{a_1}(t_1) 
\end{eqnarray}
Readers familiar with time-dependent perturbation theory may recognize 
this expression as being similar to the time-ordered exponential (also known as 
the Dyson series.)

The second important complication introduced by the matrix representation 
arises when we attempt to account for gauge transformations.  Under gauge 
transformations, the holonomies transform as
\begin{equation}
U_\gamma \mapsto U'_{\gamma} = g^{-1}(\gamma(1)) \cdot U_{\gamma} \cdot
g(\gamma(0))
\end{equation}
where $g$ is a smooth $SU(2)$-valued function on $\Sigma$.  This kind of 
thing just won't do;  we want a gauge-invariant quantity.  To turn this 
quantity into such a quantity, we restrict ourselves to closed loops, and 
take the trace of the holonomy, also known as the \emph{Wilson loop}:
\begin{equation}
T_{\gamma} = \mbox{Tr} (U_\gamma)
\end{equation}
This quantity is invariant under gauge transformations:
\begin{eqnarray}
T'_{\gamma} & = & \mbox{Tr} (g^{-1}(\gamma(1)) \cdot U_{\gamma} \cdot
g(\gamma(0))) \nonumber \\
& = & \mbox{Tr} ( g(\gamma(0)) \cdot g^{-1}(\gamma(1)) \cdot U_{\gamma}) 
\nonumber \\
& = & \mbox{Tr} (U_\gamma) = T_{\gamma} \mbox{,}
\end{eqnarray}
where we have used the cyclic properties of the trace and the fact that 
$\gamma(0) = \gamma(1)$.

\setcounter{equation}{0}

\section{Spin Networks}
\label{SpinNet}

The construction of a quantum theory from closed-loop holonomies is 
known as the \emph{loop quantization} technique.  However, there is a 
problem with loop quantization: the loop basis is overcomplete.  In other 
words, certain linear combinations of closed-loop holonomies are always equal to 
each other.  This means that if we construct a theory out 
of these loops, we will end up with linearly dependent basis states (since 
we will not have taken these equalities into account.) 

In attempting to resolve this difficulty, Rovelli and Smolin 
\cite{RovSmoNet} 
discovered that it could be circumvented entirely by the use of \emph{spin 
networks}.  To construct such a network, we must first make another 
definition:  we define a \emph{graph} $\Gamma$ as a finite collection $\{ 
\gamma_1, \gamma_2, \ldots, \gamma_n \}$ of smooth curves in $\Sigma$ that 
intersect only at their ends (if at all.)  With each curve, we associate its 
holonomy, i.e.
\begin{equation}
U_{\gamma_{i}}(A) = \mathcal{P} \exp \left[\int_{\gamma_i} A_a \, \dif x^{a} 
\right]
\end{equation}
Our quantum states will then be functions of these holonomies:
\begin{equation}
\Psi(A) = f (U_1, U_2, \ldots, U_n)
\end{equation}

Since these networks play a central role in our theory, we will take 
some time to examine them now.  Much of the following exposition is 
taken from Major \cite{Maj1}.

\subsection{Definition \& Properties}
\label{DefProp}

Spin networks were first invented by Penrose, \cite{Penrose}
and essentially consist of graphs with weighted edges.  Their fundamental 
units are strands, which are defined to be equivalent to $2 \times 2$ 
matrices:
\begin{equation}
    \delta_{A}^{B} = \eqngraph[0.8]{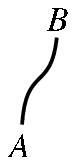} \qquad 
    \tilde{\epsilon} = i \epsilon^{AB} = \eqngraph[0.8]{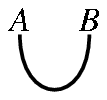} \qquad 
    \tilde{\epsilon} = i \epsilon_{AB} = \eqngraph[0.8]{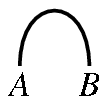}
\end{equation}
Indeed, any $2\times 2$ matrix can be ``placed'' into a spin network, in 
the form of a ``tagged'' line:
\begin{equation}
\psi_A^B = \eqngraph[0.8]{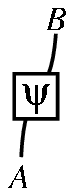}
\end{equation}
To join two strands together, we multiply two matrices with a common index:
\begin{equation}
\delta_A^B \cdot \tilde{\epsilon}^{BC} \cdot \tilde{\epsilon}_{CD} \cdot \delta_D^E = 
\eqngraph[0.8]{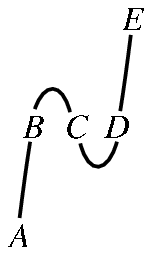} = \eqngraph[0.8]{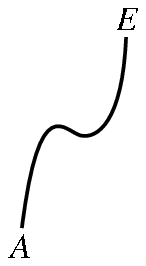} = \delta_A^E
\end{equation}

Using these rules, we can show that these strands behave as would thin 
strings in a plane, with one exception:  the equation $\delta_A^D 
\delta_B^C \tilde{\epsilon}_{CD} = - \tilde{\epsilon}_{AB}$ leads us to 
the odd strand equation
\begin{equation}
\eqngraph[0.8]{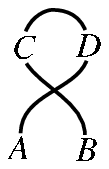} = \delta_A^D \tilde{\epsilon}_{CD} \delta_B^C = 
\tilde{\epsilon}_{BA} = - \tilde{\epsilon}_{AB} = - \eqngraph[0.8]{EpsStrand2}
\end{equation}
To resolve this, we associate a negative sign with each strand crossing.  
With this modification, the strands of a spin network behave exactly as 
would thin (i.e. untwistable) threads in a plane.

The mathematical definition of the strands in terms of matrices gives rise 
to an interesting identity, known as the ``binor identity'' or the ``skein 
relation.''  We recall the matrix identity
\begin{equation}
\epsilon_{AC} \epsilon^{BD} = \delta_A^B \delta_C^D - \delta_A^D 
\delta_C^B
\end{equation}
When translated into strands, this identity becomes
\begin{equation}
\eqngraph{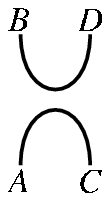} + \eqngraph{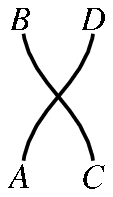} + \eqngraph{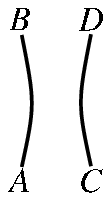} = 0
\end{equation}

The fact that a given set of strands may satisfy one or more linear 
relations (as in the binor identity above) leads us to construct a basis 
in which such linear relations do not exist.  This is accomplished by the 
use of antisymmetric combinations of multiple strands.  The simplest 
example of such a combination is the 2-strand:
\begin{equation}
\eqngraph[0.8]{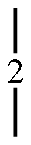} = \frac{1}{2} \left[ 
\eqngraph[0.8]{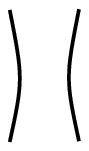} - \eqngraph[0.8]{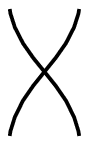} 
\right]
\end{equation}
In general, we construct an $n$-strand by summing over all permutations of 
the lines, changing the sign for each strand crossing:
\begin{equation} \label{nstrand}
\eqngraph[0.8]{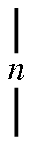} = \frac{1}{n!} \sum_{\sigma \in S_n} (-1)^{|\sigma|} 
\eqngraph[0.8]{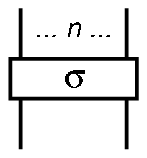}
\end{equation}
where $\sigma$ stands for a permutation of the lines.\footnote{
The exact correspondence between elements of $S_n$ and ``permutations of 
the lines'' can be established as follows:  represent an element of $S_n$ 
by its permutation on $n$ letters.  Write the letters $1 2 \ldots n$ in a 
row, and write the permuted letters $\sigma(1) \sigma(2) \ldots \sigma(n)$ 
just below them.  The lines connecting the same elements in the top and 
bottom rows correspond to the weaving of the strands.}

These networks would be boring if they only consisted of simple 
$n$-strands.  However, we can define a trivalent vertex of these edges by 
splitting three $n$-strands and reassembling them as follows:
$$
\eqngraph[0.7]{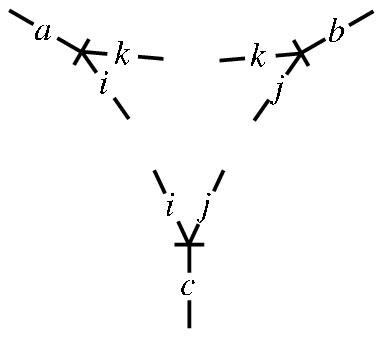} \longrightarrow \eqngraph[0.7]{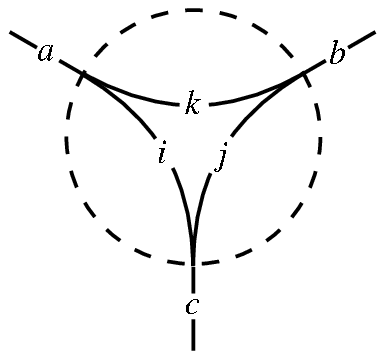} \longrightarrow 
\eqngraph[0.7]{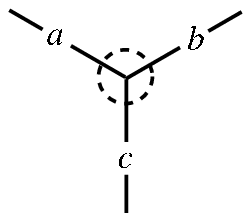}
$$
where the labels $i$, $j$, and $k$ are ``internal'' labels, and are related 
to the ``external'' labels $a$, $b$, and $c$ by the relations
\begin{eqnarray*}
i & = & \frac{1}{2}(a+c-b) \\
j & = & \frac{1}{2}(b+c-a) \\
k & = & \frac{1}{2}(a+b-c)
\end{eqnarray*}
The dotted lines in the above diagrams indicate the boundaries between 
the internal structure of the vertices and the external structure (i.e.\ 
the strands themselves.)
For these internal edges (and hence the vertex itself) to exist, $a$, $b$, 
and $c$ must satisfy two conditions.  First, the sum $a+b+c$ must be 
even;  second, $a$, $b$, and $c$ must satisfy the triangle inequalities:
$$
a+b \geq c \quad b+c \geq a \quad c+a \geq b
$$

\subsection{Intertwiner Bases \& Recoupling Theory}
\label{IntTwin}

Even with the existence of trivalent vertices, life is rather boring;  
we ask ourselves, then, whether it is possible to construct vertices 
with arbitrary valence.  We could try using the same technique we used 
in constructing the trivalent vertices, by ``splitting'' the incoming 
strands and reassembling them in all combinations:
$$
\eqngraph[0.7]{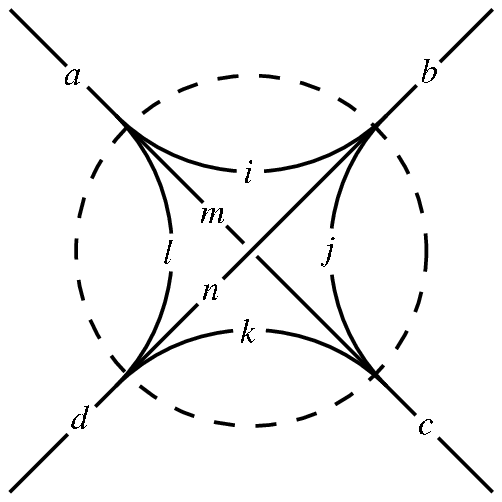} \longrightarrow \eqngraph[0.7]{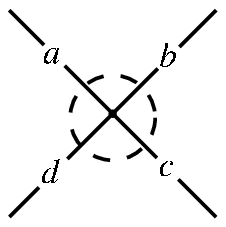}
$$
However, we quickly notice that the internal edges are 
underdetermined:  there are six internal edges $i,j,k,l,m,$ and $n$, and 
only four external edges to determine them.  Indeed, for a general 
$n$-valent vertex, there will be $n(n-1)/2$ internal edges to be 
determined by only $n$ parameters.

While we could still proceed with our analysis using all of the extra 
internal edges, it will be to our advantage to introduce a different 
way of denoting the internal edges of a vertex.  Since internal edges 
of a trivalent vertex are completely determined, we will attempt to 
build a four-valent vertex out of trivalent vertices instead:
\begin{center}
    \includegraphics[scale=0.8]{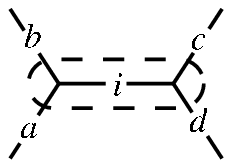}
\end{center}
Instead of having to deal with a mishmash of internal edges, we 
will now only only have to deal with the one internal edge 
$i$.  Indeed, from here on the phrase ``internal edges'' will 
be used exclusively to refer to edges such as $i$, and not to the 
``split strands'' used to construct the trivalent vertices.

However, this construction is useless if we cannot actually find an 
edge $i$ that allows the internal trivalent vertices to be 
constructed.  The first requirement to deal with is the ``evenness'' 
requirement.  For the trivalent vertices to exist, the restrictions 
on the triples $(a,b,i)$ and $(c,d,i)$ are:
\begin{equation}
    a + b \equiv i \pmod 2  \qquad \mbox{and} \qquad c + d \equiv i \pmod 
    2 \mbox{,}
\end{equation}
which together imply that the sum $a+b+c+d$ must be even.  Similarly, 
from the triangle inequalities $a+b \geq i$ and $i+c \geq d$ together imply 
that $a + b + c \geq d$;  we can similarly construct three other 
inequalities, creating a total of four ``generalized'' triangle 
inequalities:
\begin{equation}
    \begin{split}
    a+b+c & \geq d \\
    a+b+d & \geq c \\
    a+c+d & \geq b \\
    b+c+d & \geq a
    \end{split}
\end{equation}
Alternately, these four inequalities can be summarized in one 
statement:
\begin{equation} \label{GenIneq1}
    a + b + c + d \geq 2x \mbox{ for } x \in \{a, b, c, d\}
\end{equation}

There is still an ambiguity in the internal edge, however;  a bit of 
thought reveals that the value of $i$ can lie in the range
\begin{equation}
    \max \{ |a-b|, |c-d| \} \leq i \leq \min \{ a+b, c+d \}
\end{equation}
For example, in the case where $a=b=c=d$, $i$ can take on any (even) 
value between 0 and $2a$.  We can think of these sets of vertices as 
state vectors, which span the space of all possible four-valent 
vertices with ``external'' edges $a$, $b$, $c$, and $d$.  

We can define 
the inner product between two vertex states by joining together 
corresponding external edges.  For example, if we take the inner 
product between a state labelled by $i$ and a state labelled by $i'$, 
we find that it is given by the value of the following network:
\begin{center}
    \includegraphics[scale=0.8]{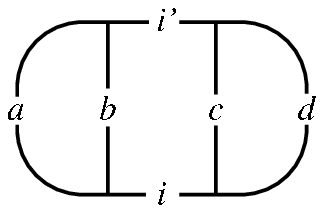}
\end{center}
If we apply the ``bubble'' identity (\ref{bubble}) to right-hand bubble 
in this diagram, we find that the value of this diagram is 0 unless $i = i'$:
\begin{eqnarray}
    \eqngraph[0.8]{InnerProduct} & = & \delta_{i i'} 
    \frac{\theta(i,c,d)}{\Delta_{i}} \eqngraph[0.8]{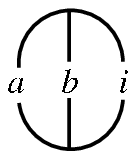} \nonumber \\
    & = & \delta_{i i'} \frac{\theta(i,a,b) \theta(i,c,d)}{\Delta_{i}}
\end{eqnarray}
Thus, these basis states 
are orthogonal.\footnote{They are certainly not normalized;  however, 
this fact will have no effect on the properties we examine.}

The reader may have wondered, back when we were defining the 
four-valent vertex in terms of internal trivalent vertices, why we 
chose to do so in this manner.  After all, we could have equally well 
joined together the vertices $a$ and $d$ in one internal vertex and 
the vertices $b$ and $c$ in the other:
\begin{center}
    \includegraphics[scale=0.7]{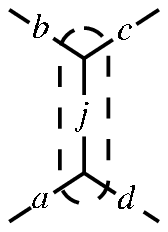}
\end{center}
If we had defined the internal edge in this way, we would have 
obtained a different set of state vectors;  however, it is not 
immediately evident that these vectors will span the same space as our 
original set.  That these new vectors actually \emph{do} span the same 
space is a result of the \emph{recoupling theorem} (see Appendix 
\ref{zoo}):
\begin{equation}
\eqngraph[0.7]{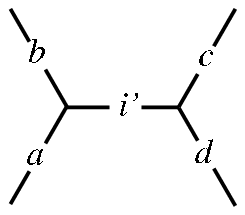} = \sum_{i} \left\{ \begin{array}{ccc}
a & b & i \\
c & d & i'
\end{array} \right\} \eqngraph[0.7]{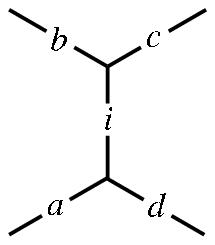}
\end{equation}
Thus, given a vertex with an internal decomposition in either form, 
we can express it as a linear combination of vertices with an internal 
decomposition of the other form.  These internal decompositions are 
called \emph{intertwiners};  what we have just shown is that the set 
of all intertwiners of either form is a basis for the space of all 
intertwiners, and that the recoupling theorem is a transformation 
between these bases.  Moreover, this transformation happens to have a 
nice property.  In \cite{DPRVol1}, De Pietri and Rovelli show that the 
recoupling theorem is a unitary transformation, i.e.\ vector norms are 
preserved under it.  We can thus switch back and forth between bases 
fairly simply;  this ability will become invaluable to us when we 
examine the actions of the angle and volume operators in Sections 
\ref{AngleOp} and \ref{VolOp}, respectively.

We have now defined what we mean by an 4-valent vertex;  however, our 
goal was to create a vertex of arbitrary valence.  It is fairly 
obvious how this can be done:  for each new edge we wish to 
add, we add a trivalent vertex to the intertwiner.  Thus, a 
5-valent vertex contains three trivalent vertices in its 
intertwiner, a 6-valent vertex contains four trivalent vertices, 
and (in general) an $n$-valent vertex contains $n-2$ trivalent 
vertices in its intertwiner.  It can also be shown that given a set of 
incoming edges with labels $\{ a_{1}, a_{2}, \ldots, a_{n} \}$, it is 
possible to create an $n$-valent vertex out of them if and only if the 
following conditions are satisfied:
\begin{equation}
    \sum_{i=1}^{n} a_{i} \mbox{ is even}
    \label{GenEven}
\end{equation}
\begin{equation}
    \sum_{i=1}^{n} a_{i} \geq 2 a_{j} \mbox{ for } 1 \leq j \leq n
    \label{GenTri}
\end{equation}
The first of these is the usual ``evenness'' requirement, 
while the second is a generalized triangle inequality similar to 
(\ref{GenIneq1}).

\begin{figure}
    \begin{eqnarray*}
    \lefteqn{ \eqngraph[0.7]{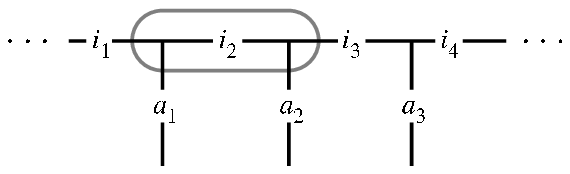} = \sum_{j_{2}} \left\{ 
    \begin{array}{ccc} a_{1} & i_{1} & j_{2} \\ i_{3} & a_{2} & i_{2} \end{array} 
    \right\} \eqngraph[0.7]{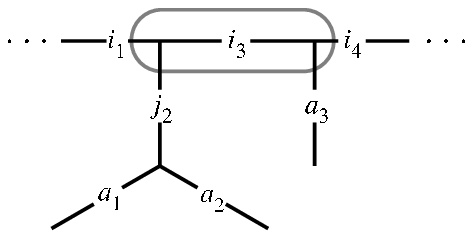} } \\
    & = & \sum_{j_{2}} \sum_{j_{3}} \left\{ \begin{array}{ccc} j_{2} & i_{1} & 
    j_{3} \\ i_{4} & a_{3} & i_{3} \end{array} \right\} \left\{ 
    \begin{array}{ccc} a_{1} & i_{1} & j_{2} \\ i_{3} & a_{2} & i_{2} \end{array} 
    \right\} \eqngraph[0.7]{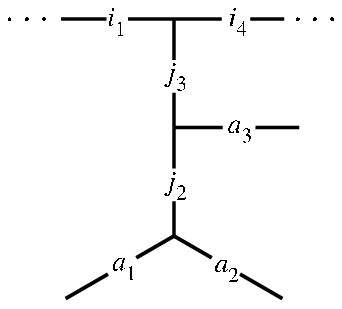}
    \end{eqnarray*}
    \caption{\label{BranchTrans} The ``branch transformation'', in 
    which an internal branch is added to the intertwiner.  Internal 
    edges are marked with $i$s and $j$s, while external edges are 
    marked with $a$s.  The gray outlines indicate the internal edge 
    to which the recoupling theorem is applied in the next step.}
\end{figure}

\begin{figure}
    \raisebox{-0.5\height}{
    \begin{minipage}[b]{2.25in}
    \begin{center}
    \includegraphics[scale=0.8]{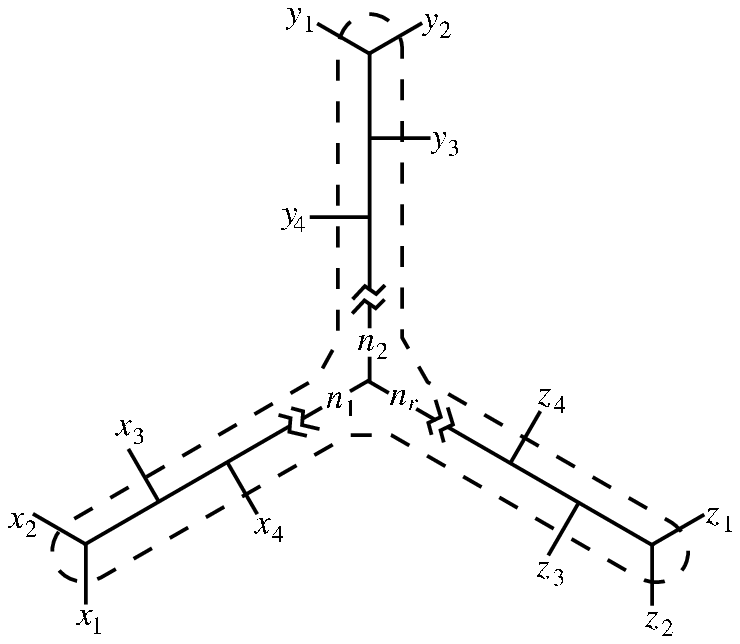}
    
        (a)
    \end{center}
    \end{minipage}  
    }
    \hfill 
    \raisebox{-0.5\height}{
    \begin{minipage}[b]{2.25in}
    \begin{center}
    \includegraphics[scale=0.8]{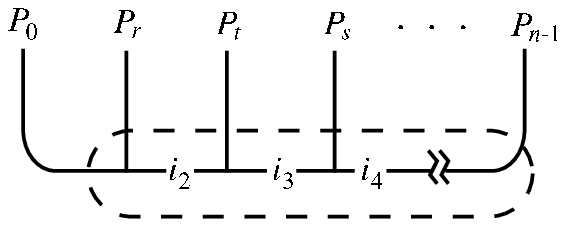}
    
        (b)
    \end{center}
    \end{minipage}
    }
    \caption{\label{SnowComb}  Two useful shapes for intertwiner 
    bases.  The first, in (a), is the ``snowflake'' basis.  Its 
    essential feature is that the edges of a given $n$-valent vertex 
    are partitioned into three groups (denoted $\{ x_{i} \}$, 
    $\{y_{i}\}$, and $\{z_{i}\}$ in the diagram);  each of the groups is then 
    ``collected'' into a single edge (labelled $n_{1}$, $n_{2}$, and 
    $n_{r}$);  and finally, these three collecting edges form a 
    trivalent vertex at the core of the vertex.  The second basis, 
    in (b), is the ``comb'' basis;  in it, all the edges are lined up 
    along a central ``spine'' of intertwiner edges (labelled $i_{2}$, 
    $i_{3}$, $i_{4} \ldots$ in the diagram.)}
\end{figure}

As might be expected, these higher-valence vertices also have 
intertwiner bases (although these bases are of much higher dimension.)  
However, we can again use the recoupling theorem to shift between 
these bases.  There are two general types of these transformations 
which will be very useful to us later on.  The first of these is 
illustrated in Figure \ref{BranchTrans}, and allows us to transform 
between bases of different ``shapes.''  
By creating (or, in reverse, 
removing) branches from an arbitrary internal decomposition, we can 
transform an intertwiner of arbitrary shape into any other shape we 
wish.  Two such useful shapes are the ``snowflake'' and ``comb'' 
bases, depicted in Figure \ref{SnowComb}.

\begin{figure}
    \begin{eqnarray*}
    \lefteqn{ \eqngraph[0.7]{Branch1} = \sum_{j_{2}} \left\{ 
    \begin{array}{ccc} a_{1} & i_{1} & j_{2} \\ i_{3} & a_{2} & i_{2} \end{array} 
    \right\} \eqngraph[0.7]{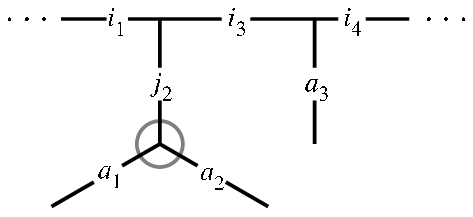} } \\
    \qquad & = & \sum_{j_{2}} \left\{ 
    \begin{array}{ccc} a_{1} & i_{1} & j_{2} \\ i_{3} & a_{2} & i_{2} \end{array} 
    \right\} \lambda^{a_{1}a_{2}}_{j_{2}} \eqngraph[0.7]{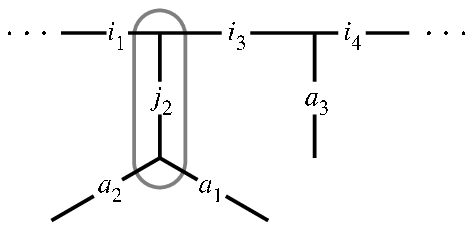} \\
    \qquad & = & \sum_{j_{2}} \sum_{k_{2}} \left\{ 
    \begin{array}{ccc} a_{1} & i_{1} & j_{2} \\ i_{3} & a_{2} & i_{2} \end{array} 
    \right\} \lambda^{a_{1}a_{2}}_{j_{2}} \left\{ \begin{array}{ccc} 
    a_{1} & a_{2} & k_{2} \\ i_{1} & i_{3} & j_{2} \end{array} \right\} 
    \eqngraph[0.7]{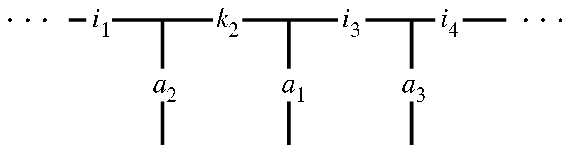}
    \end{eqnarray*}
    \caption{\label{SwapTrans} The ``swap transformation'', in 
    which two edges along an internal spine are switched.  Note that in the third step
    we have applied the lambda move (\ref{Lambda}) to the vertex indicated in the 
    previous step.}
\end{figure}

The second such transformation is illustrated in Figure 
\ref{SwapTrans}.  This transformation allows us to switch the order 
of two adjacent edges along a branch of the intertwiner, while leaving 
the other branches in their original order.  Thus, any order of edges 
is possible along a given intertwiner shape;  this fact will become 
vitally important to us in our discussion of the volume operator in 
Section \ref{VolOp}.

\subsection{Spin Networks \& Angular Momentum Coupling}
\label{AngCo}

The trivalent vertex rules introduced in Section \ref{DefProp}, and their generalized 
versions in (\ref{GenEven}) and (\ref{GenTri}), are not entirely transparent.  However, 
a simple analogy exists which may make these rules more intuitive to the reader;  there
is a striking parallel between the coupling of angular momenta in quantum mechanics
and spin networks.\footnote{Indeed, this similarity is the very source of the name 
``spin'' networks.}

To draw this parallel, let us relabel all the edges of a given trivalent vertex with half
their original values.  In other words, an edge with colour 1 becomes an edge with colour
$\frac{1}{2}$;  an edge with colour 2 becomes an edge with colour 1;  and in general, an edge
with colour $i$ is reassigned a new colour $i' = i/2$. In 
terms of these half-integer edge colours, the constraints that the edges must satisfy become:
\begin{equation}
    a'+b' \geq c' \quad b'+c' \geq a' \quad c'+a' \geq b' \quad \mbox{(as before)}
    \label{HalfTri}
\end{equation}
\begin{equation}
    a' + b' + c' \mbox{ is an integer}
    \label{HalfEven}
\end{equation}

A it happens, these rules are exactly the same as those for the coupling of particles with
angular momentum.  If we wish to couple together a spin-$a'$ and a spin-$b'$ particle to form 
a spin-$c'$ particle, the three spins $a'$, $b'$, and $c'$ must satisfy the triangle 
inequalities, as in (\ref{HalfTri});  as well, the sum of the three spins must be an 
integer, as in (\ref{HalfEven}).

As a concrete example of this parallel, consider a trivalent vertex with an edge with 
colour 1, an edge with colour 2, and the third edge colour $m$ as yet undetermined.
According to the vertex rules, we must have $m = 1$ or $m = 3$;  any value less than 1 or 
greater than 3 would violate the triangle inequalities, and a value of 2 would violate the
evenness requirement.  The angular momentum coupling scheme which correspond to this
vertex would be the coupling of a spin-$\frac{1}{2}$ particle to a spin-1 particle;  
according to the rules of such couplings, the resulting spin must be either $\frac{1}{2}$
or $\frac{3}{2}$.

We can draw a similar parallel between 4-valent vertices and three-particle coupling
schemes.  Consider again the following vertex:
\begin{center}
    \includegraphics[scale=0.8]{FourValent3.eps}
\end{center}
Suppose $a = 2$, $b = 3$, and $c = 3$.  According to the vertex rules, this means that $d$
must have a value of 0, 2, 4, 6, or 8.  Similarly, if we couple together one spin-1
particle and two spin-$\frac{3}{2}$ particles, the resulting total spin can be 0, 1, 2, 3, or 4.
In fact, this rule holds for a general vertex with valence $n$, with incoming spins $a_1,
a_2, \ldots, a_n$;  such a vertex can be translated into a coupling scheme of $n-1$ spins
$a'_1, a'_2, \ldots, a'_{n-1}$ to yield a total spin of $a'_n$, where $a'_i = a_i/2$ for
$i = 1, 2, \ldots, n$.

The reader may be asking at this point whether any analogy for the internal edges of a
higher-valence vertex exists in angular momentum coupling schemes.  In fact, such an
analogy does exist.  In the above diagram, we can view the total coupling as happening
in two stages:  first particle $a$ and particle $b$ couple to form a particle $i$, then 
particle $i$ and particle $c$ couple to form particle $d$.  Of course, the coupling does not
necessarily have to take place in this order.  For instance, we could equally well have 
coupled particles $b$ and $c$ together first, to form a particle $j$, and then couple
particle $j$ to particle $a$ to form particle $d$;  this is an equally valid way to couple the
three particles together.  Similarly, we can represent a vertex with edges $a$, $b$, $c$, 
and $d$ in a second form:
\begin{center}
    \includegraphics[scale=0.7]{OtherBasis.eps}
\end{center}
The parallel between the intertwiner of this vertex and this second angular momentum coupling
scheme is evident.

Finally, we note the following:  In angular momentum coupling, we translate between 
these two coupling schemes (i.e.\ particle $i$ vs.\ particle $j$) using the \emph{Wigner 6-$j$ 
symbols};  in spin networks, we use the recoupling theorem and the \emph{Kauffman-Lins 6-$j$
symbols} to translate between intertwiners.  Given the above parallels, it should not surprise
us to learn that the Wigner and the Kauffman-Lins 6-$j$ symbols are related;  for more details,
see Appendix \ref{KLvsW}.

\subsection{Spin Networks \& Holonomies}
\newcommand{\Tr}{\mathrm{Tr}}

\begin{figure}
    \begin{center}
    \begin{tabular}{c @{\extracolsep{1cm}} c @{\extracolsep{1cm}} c}
        \multicolumn{3}{c}{\includegraphics{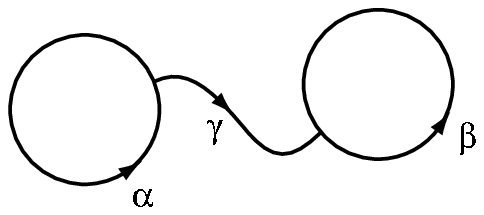}} \\[12pt]
        \includegraphics[scale=0.7]{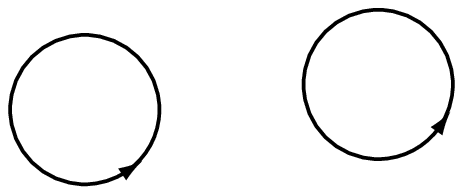} & 
        \includegraphics[scale=0.7]{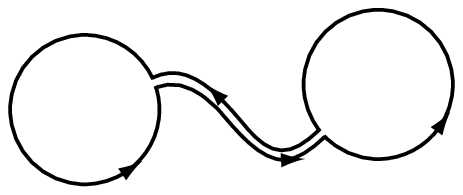} &
        \includegraphics[scale=0.7]{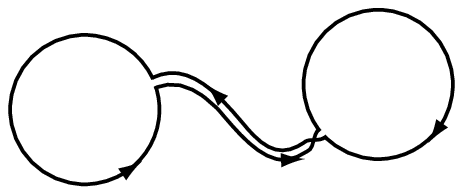} \\
        $\alpha \cup \beta$ & $\alpha \circ \gamma \circ \beta \circ 
        \gamma^{-1}$ & $\alpha \circ \gamma \circ \beta^{-1} \circ 
        \gamma^{-1}$
     \end{tabular}
    \end{center}
    \caption{\label{Mandelstam}  A simple network in space, and three 
    paths (or sets of paths) within it.}
\end{figure}

The reader may be wondering at this point what, exactly, the 
connection between the spin networks we have just described and the 
holonomies described in Section \ref{LoopTheory} is.  To answer this 
question, consider a simple network in space, shown in Figure 
\ref{Mandelstam}, along with three paths (or sets of paths) in this network.
To label these paths, we must introduce a few concepts.  We define
the composition of two paths to be 
the path obtained by traversing the two paths in order;  for example, 
the path $\alpha \circ \gamma$ is obtained by going around $\alpha$, 
then going along $\gamma$.  We also introduce the concept of 
inverses of a path, which we define (intuitively) to be the same path 
traversed in the opposite direction;  for example, the path $\alpha 
\circ \gamma \circ \gamma^{-1}$ is obtained by traversing $\alpha$ 
forwards, then $\gamma$ forwards, then $\gamma$ backwards 
(i.e.~retracing our steps along $\gamma$).   Using this notation, the 
three paths can be labelled $\alpha \cup \beta$, $\alpha \circ 
\gamma \circ \beta \circ \gamma^{-1}$, and $\alpha \circ \gamma 
\circ \beta^{-1} \circ \gamma^{-1}$.  

One might think that the state vectors corresponding to these three 
loops would be independent;  unfortunately, this is not the case. 
Recall that each 
path in the manifold $\Sigma$ has a matrix from the group $SU(2)$ 
associated with it.  Given any two matrices $A, B \in SU(2)$, their 
traces satisfy the following identity:
\begin{equation} \label{TraceId}
    \Tr (A) \Tr (B) - \Tr (AB) - \Tr (A B^{-1}) = 0
\end{equation}
Since this linear combination is equal to zero, the 
quantum states corresponding to each one of these terms satisfy an 
analogous identity:
\begin{equation} \label{StateId}
    \left\langle \alpha \cup \beta \right| - \left\langle \alpha \circ 
    \gamma \circ \beta \circ \gamma^{-1} \right| - \left\langle \alpha 
    \circ \gamma \circ \beta^{-1} \circ \gamma^{-1} \right| = 0.
\end{equation}
Hence, these three states are linearly dependent.  

For a fairly simple network (such as the one just described), it might 
be possible to keep track of all the identities satisfied by loops in 
the network.  However, for a more complicated network the number of 
these identities skyrockets;  it would be nice, then, if we could 
find a way to avoid these identities altogether.

As the reader may have guessed by now, spin networks allow us to do 
this.  Consider the states
\begin{equation}
    \begin{split}
    & \left\langle \alpha \cup \beta \right| = \left\langle \alpha \circ 
    \gamma \circ \beta \circ \gamma^{-1} \right| + \left\langle \alpha 
    \circ \gamma \circ \beta^{-1} \circ \gamma^{-1} \right| \quad 
    \mbox{and} \\
    & \left\langle \alpha \circ 
    \gamma \circ \beta \circ \gamma^{-1} \right| - \left\langle \alpha 
    \circ \gamma \circ \beta^{-1} \circ \gamma^{-1} \right| .
    \end{split}
\end{equation}
These two states are independent of each other, and (just as 
importantly) span the entire space of loops on this network --- any 
loop or set of loops on this network can be written as a linear 
combination of these states (using the identity in \ref{StateId} if 
necessary.)  However, the second one of these states can be rewritten 
in a more familiar form:
\begin{equation}
    \begin{split}
    &\left\langle \alpha \circ \gamma \circ \beta \circ \gamma^{-1} \right| - \left\langle \alpha 
    \circ \gamma \circ \beta^{-1} \circ \gamma^{-1} \right| \\
    & \quad = 
    \left\langle \eqngraph[0.66]{Loop2} \, \right| - \left\langle 
    \eqngraph[0.66]{Loop3} \, \right| \nonumber \\
    & \quad = \left\langle \eqngraph[0.66]{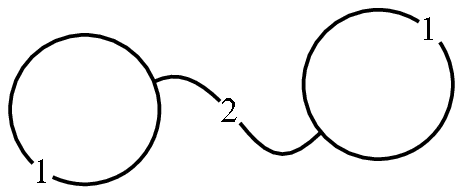} \, \right|
    \end{split}
\end{equation}
Hence, given a network in $\Sigma$, the introduction of spins on the 
edges eliminates the identities that might otherwise be present.  In 
this way, spin networks represent a basis for the states of loop 
quantum gravity. 
\subsection{Operators on Spin Networks}
We have now defined our quantum states in terms of spin networks.  
A successful theory of quantum gravity should be able to make 
predictions about ``spatial observables'' such as area, volume, and 
angle --- geometric quantities that we would normally be able to 
measure on a classical manifold.  As in any quantum 
theory, this is done by Hermitian operators which act on the spin networks 
and return eigenvalues corresponding to observable quantities.  
However, it is not at all clear what these operators should be.

We will describe the actions of four operators --- angular momentum, area, 
angle, and volume --- in varying degrees of detail.  In general, however, 
the construction of quantum operators from first principles is a complex 
process;  hence, we will not try to do so in this work.  The reader 
wanting further details as to the construction of these operators is 
referred to the derivations in \cite{Maj1} (for angular momentum), 
\cite{FLR, RSAreaVol} 
(for area), \cite{Maj2} (for angle), and \cite{DPRVol1} (for volume).

\subsubsection{Angular Momentum Operators}

One of our primary concerns in this thesis will be to describe the 
angles associated with a given vertex;  to define angles, however, we 
need to be able to associate a direction (i.e.~a vector) with an edge 
or a group of edges.
In Penrose's original formulation of spin networks, the labels on the edges 
actually refer to angular momenta ``carried'' by the edges (see Section 
\ref{AngCo});  hence, one of the most natural directions 
to associate with an edge is its angular momentum vector. 

These angular momenta can be measured by the 
angular momentum operators, which act upon the edges.  The angular 
momentum operators are expressed in terms of the Pauli spin matrices $\sigma_i$:
\begin{equation}
\sigma_1 = \left( \begin{array}{cc} 0 & 1 \\ 1 & 0 \end{array} \right) 
\mbox{,} \quad 
\sigma_2 = \left( \begin{array}{cc} 0 & -i \\ i & 0 \end{array} \right) 
\mbox{,} \quad
\sigma_3 = \left( \begin{array}{cc} 1 & 0 \\ 0 & -1 \end{array} \right)
\end{equation}
The operators themselves are then defined in terms of these matrices 
``grasping'' the edges:
\begin{equation}
\hat{\sigma_i} = \eqngraph[0.7]{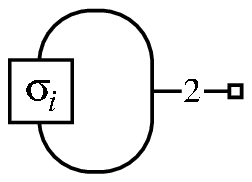}
\end{equation}
where the small square at the end indicates that this operator will 
``grasp'' the edge.  If the strand in question happens to be a multiple 
strand, then this grasping will apply to each strand in it; hence, we can 
show that a 2-edge grasping an $m$-edge is equivalent to $m$ times a 
trivalent vertex with edges $m$, $m$, and 2:
\begin{equation} 
\eqngraph[0.7]{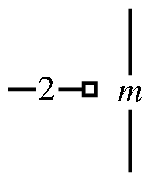} = m \eqngraph[0.7]{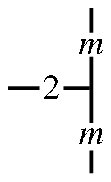}
\end{equation}
Using this, we can define angular momentum operators $\hat{J_i} = 
\frac{\hbar}{2} \hat{\sigma_i}$ which act on the edges;  these operators 
have the usual properties of angular momentum operators.\footnote{
The actual eigenvalues of the operators $\hat{J_i}$ will not concern us 
here;  however, it is worth noting that doing so requires (loosely 
speaking) giving orientations to the edges.  Since this is not required in 
the spin-network formulation of quantum gravity, we will not go into it 
here.} 
As we might expect, we can also construct the $\hat{J^2}$ operator.  This is 
equal to the sum of the squares of the three $\hat{J_i}$ operators, i.e. 
$\hat{J^2} = \hat{J_1^2} + \hat{J_2^2} + \hat{J_3^2}$.  We can use 
diagrammatics and graspings to find the eigenvalues of this operator; 
however, it is much simpler to note that the Pauli matrices have the 
following identity, which is easily expressible in terms of spin networks:
\begin{equation}
\frac{1}{2} \sum_{i=1}^{3} \sigma_{i} {}_A^B \sigma_{i} {}_C^D = 
\frac{1}{2} \left( \epsilon_{AC} \epsilon^{BD} - \delta_A^D \delta_C^B 
\right) = \frac{1}{2} \left(\eqngraph[0.7]{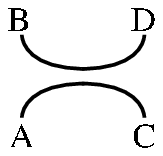} - 
\eqngraph[0.7]{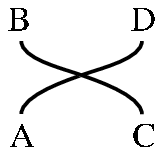} \right) = 
\eqngraph[0.7]{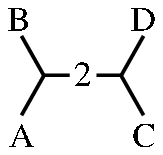}
\end{equation}
Hence, instead of expressing $\hat{J^2}$ in terms of the sum of three 
operators, we can express it very simply in diagrammatic form:
\begin{equation}
\hat{J^2} = \frac{\hbar^2}{2} \eqngraph[0.7]{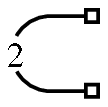}
\end{equation}
where we now grasp with two ``hands'' on the edge in question.  Using the 
``bubble'' identity (\ref{bubble}), this becomes
\begin{eqnarray*}
\hat{J^2} \eqngraph[0.7]{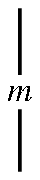} & = & \frac{\hbar^2}{2} m^2 
\eqngraph[0.7]{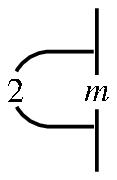} \\
& = & \frac{\hbar^2}{2} m^2 \frac{(-1)^m \theta(m,m,2)}{m+1} 
\eqngraph[0.7]{mStrand}
\end{eqnarray*}
Note that the factor of $m^2$ in the first step comes from the two 
graspings of the 2-edge.  If we apply the identity (\ref{mm2id}) to 
this equation, we get
\begin{eqnarray}
\hat{J}^{2} \eqngraph[0.7]{mStrand} & = & \frac{\hbar^2}{2} m^2 \frac{(-1)^m}{m+1} \frac{(-1)^{m+1} 
(m+2)(m+1)}{2m} \eqngraph[0.7]{mStrand} \nonumber \\
& = & - \hbar^2 \frac{m(m+2)}{4} \eqngraph[0.7]{mStrand} \label{JOpRes}
\end{eqnarray}
If we make the identification 
$j = m/2$, then the eigenvalues simplify to the easily recognizable form 
$\hbar^2 j (j+1)$.\footnote{
The reader will notice that the sign of this expression is 
incorrect.  This stems from a difference between the properties of the 
``angular momentum'' spin networks originally studied by Penrose and 
those used for quantum gravity.}
This operator (and its eigenvalues) will become important 
in our discussion of the area and angle operators.

\subsubsection{The Area Operator}
\label{AreaOp}

\begin{figure}
    \begin{center}
    \includegraphics{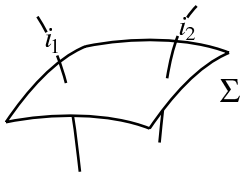}
    \end{center}
    \caption{\label{AreaFig}  A surface $\Sigma$ whose area we wish to 
    find.  The contribution to the area from the portion of the 
    surface shown would be the sum of the contribution from each of 
    the two edges $i_{1}$ and $i_{2}$ intersecting this portion.}
\end{figure}
In the spin-network version of quantum gravity, the area operator is the 
simplest to define.  If we have a closed surface $\Sigma$ which intersects our spin 
network in $N$ points, its area is can be decomposed into a contribution 
from each of these intersections:
\begin{equation}
A[\Sigma] = \sum_{i=1}^N A_i
\end{equation}

As it turns out, it is much easier to derive the action of the 
area-squared operator $A_i^2$ than to directly find the action of the area 
operator;  this action turns out to be the double grasping of a two-edge 
times the square of the Planck length $l_0$, defined in 
(\ref{PlanckLength}):
\begin{equation}
\hat{A_i^2} = l_0^2  \eqngraph[0.7]{Jop}
\end{equation}
The reader will notice that we have already done the work to 
find the eigenvalues of this operator in the previous section;  the 
result, from (\ref{JOpRes}), is then
\begin{equation}
\hat{A_i^2} \eqngraph[0.7]{mStrand} = l_0^2 \frac{m(m+2)}{4} \eqngraph[0.7]{mStrand}
\end{equation}
The total area of $\Sigma$ is then given by
\begin{equation}
A[\Sigma] = l_0 \sum_i \sqrt{\frac{m_i(m_i+2)}{4}}
\end{equation}
where the summation of $i$ is over all edges that intersect $\Sigma$, and 
$m_i$ is the edge label of the edge $i$.

\subsubsection{The Angle Operator}
\label{AngleOp}

\begin{figure}
    \begin{center}
    \includegraphics[width=2in]{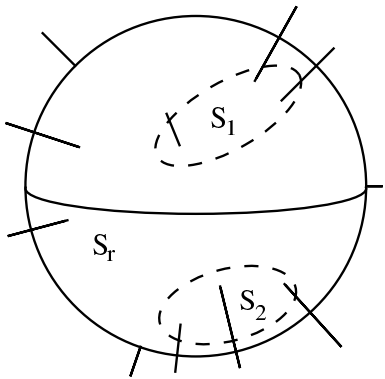}
    \end{center}
    \caption{\label{Partition}  Partition of the edges surrounding a 
    vertex into three groups $S_1$, $S_2$, and $S_3$.  (The vertex itself 
    is inside the sphere.)  The angle operator will return the angle 
    between the patches $S_1$ and $S_2$.}
\end{figure}

\begin{figure}
    \begin{center}
    \includegraphics[width=3.5in]{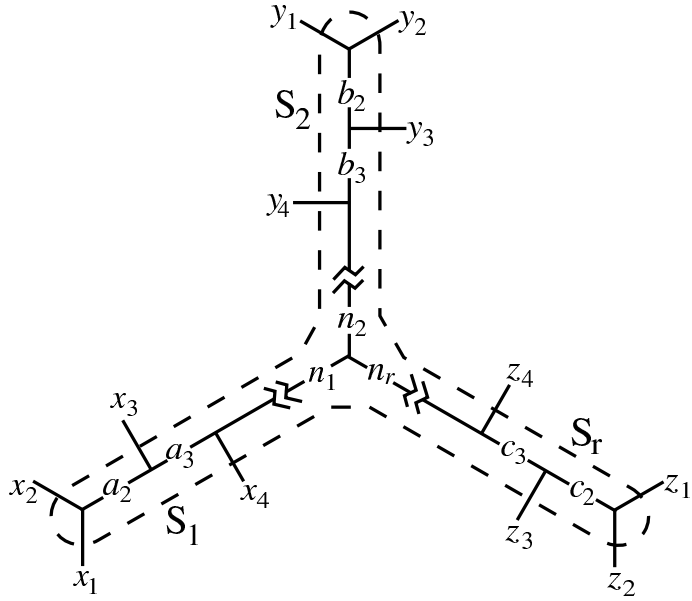}
    \end{center}
    \caption{\label{SnowFlake}  Internal decomposition of a vertex 
    operated on by the angle operator;  the dotted line denotes the 
    boundary between external and internal edges.  All of the edges $x_1, 
    x_2, \ldots$ which intersect the surface $S_1$ are ``collected'' into the 
    internal edge labelled with $n_1$;  similarly, all the edges which 
    intersect $S_2$ are collected into the edge labelled $n_2$, and the 
    edges which intersect $S_r$ are collected into the edge labelled 
    $n_r$.  In the text, we denote $s_1 = \sum x_i$, $s_2 = \sum y_i$, and 
    $s_r = \sum z_i$.}
\end{figure}

The angle operator acts on a vertex as follows:  we can imagine a 
surface encompassing a vertex of arbitrary valence.  We select two 
``patches'' on the surface $S_{1}$ and $S_{2}$, and call the 
remaining area of the surface $S_{r}$ (as in Figure~\ref{Partition}).  
This defines a partition of the edges into three groups corresponding 
to the three patches $S_1$, 
$S_2$, and $S_r$.  The intertwiner basis for this vertex is decomposed in 
such a way that all edges belonging to one group are ``collected'' into 
one internal edge;  the three internal edges created in this way then join 
at a trivalent ``core;''  see Figure \ref{SnowFlake}.\footnote
{Note that even if the internal structure of the intertwiner is not of this 
form for the partition we choose, it can be viewed as a superposition of 
such intertwiners (using the techniques stated in section \ref{IntTwin}.)}
For future use, we will also define the quantities $s_1$, $s_2$, and $s_r$ 
as the sum of the labels of all the edges traversing $S_1$, $S_2$, and 
$S_r$, respectively;  similarly, we will define $n_1$, $n_2$, and $n_r$ as 
the internal labels of their respective collecting edges.

Although a rigourous derivation of the provenance of the angle operator is 
rather complicated, it is possible to gain an intuitive understanding as 
follows.  The angle operator will measure the angle between the ``bunches'' of 
edges that traverse $S_1$ and those that traverse $S_2$.  The most 
``natural'' direction that is associated with the edges traversing 
$S_{1}$, $S_{2}$, and $S_{r}$ is the 
angular momentum vectors associated with their internal edges (which 
we define as $\vec{J_1}$, $\vec{J_2}$, and $\vec{J_r}$ respectively).  
Classically, we would expect the angle 
between $\vec{J_1}$ and $\vec{J_2}$ to be given by
\begin{equation} \label{ClassAng}
\cos \theta = \frac{\vec{J_1} \cdot \vec{J_2}}{\left|\vec{J_1}\right| 
\left|\vec{J_2}\right|}
\end{equation}
However, since we don't want a net angular momentum to be associated with 
a point in empty space, we require that
\begin{equation} 
\vec{J_1} + \vec{J_2} + \vec{J_r} = 0 \mbox{.}
\end{equation}
Using this, we can put the quantity in (\ref{ClassAng}) in terms of $J^2$ 
operators:
\begin{eqnarray}
(\vec{J_r})^2 & = & (\vec{J_1} + \vec{J_2})^2 \nonumber \\
J_r^2 & = & J_1^2 + J_2^2 + 2 \vec{J_1} \cdot \vec{J_2} \nonumber \\
\vec{J_1} \cdot \vec{J_2} & = & \frac{1}{2} \left(J_r^2 - J_1^2 - J_2^2\right)
\end{eqnarray}
The angle operator is then defined as
\begin{equation}
\hat{\theta} = \arccos \left( \frac{J_r^2 - J_1^2 - J_2^2}{2 \sqrt{J_1^2} 
\sqrt{J_2^2}} \right)
\end{equation}
Since we know the action of the $J^2$ operators already, we can readily state the 
eigenvalues of the angle operator:
\begin{equation}
\theta = \arccos \left(\frac{n_r(n_r+2) - n_1 (n_1+2) - n_2(n_2 + 2)}{2 
\sqrt{n_1 (n_1+2)}\sqrt{n_2(n_2+2)}} \right)
\end{equation}

We note briefly that there are important philosophical considerations 
which arise from the discrete spectrum of this operator as it relates to 
diffeomorphism invariance.  These considerations are discussed 
in Appendix \ref{Diffeo};  we will not go into them here.

\subsubsection{The Volume Operator}
\label{VolOp}

\begin{figure}
    \begin{center}
    \includegraphics[width=2.5in]{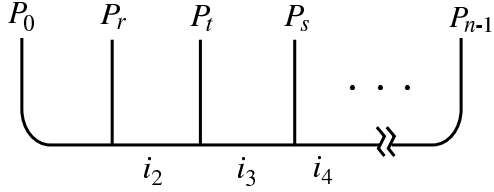}
    \end{center}
    \caption{\label{Comb}  Internal decomposition of a vertex to be 
    operated on by the volume operator.  The external edges $P_0, P_1, 
    \ldots, P_{n-1}$ are ``lined up'' along an internal ``spine.''  The 
    edges $i_2, i_3, \ldots, i_{n-2}$ are internal.}
\end{figure}

The volume operator also acts on a vertex.  In this case, however, it 
is more convenient to use the basis pictured in Figure \ref{Comb}, 
where all the edges are lined up sequentially.  
As we did for the 
area operator, we will also find it easier to define the 
volume-squared operator (known as $W$) first;  the volume eigenvalues 
are then the square roots of the absolute values of the 
eigenvalues of $W$.

The action of $W$ itself is rather complicated.  For a vertex with 
$n$ edges labelled $\{P_{0}, P_{1}, \ldots, P_{n-1}\}$, we define an operator 
$W_{[rst]}$ for each triad of edges such that $0\leq r<s<t\leq n-1$.  The action 
of $W_{[rst]}$ consists of three 2-edges, joined at a trivalent intersection, 
which grasp edges $r$, $s$, and $t$ as follows:
\begin{equation}
\hat{W}_{[rst]} = \eqngraph[0.6]{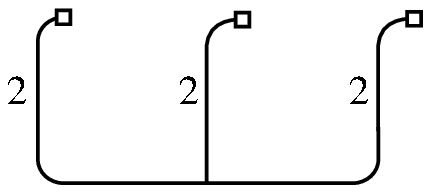}
\end{equation}
\begin{equation}
\hat{W}_{[rst]} \eqngraph[0.6]{Combi}  = P_{r} P_{s} P_{t} \eqngraph[0.6]{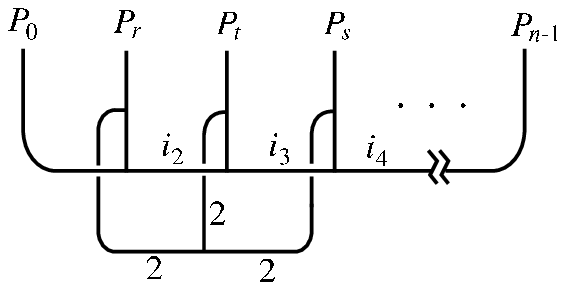}
\end{equation}
Since the ``comb'' basis spans the space of all intertwiners, we can 
write the action of the $W$ operator as sending the original vertex to a 
superposition of other vertices in the same basis:
\begin{equation}
\hat{W}_{[rst]} \eqngraph[0.6]{Combi} = \sum_{k_{2}, \ldots, k_{n-2}} W_{[rst]} {}_{i_{2}, 
\ldots, i_{n-2}}^{k_{2}, \ldots, k_{n-2}} \eqngraph[0.6]{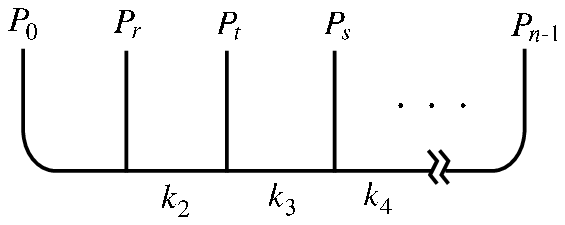}
\end{equation}
This is essentially a matrix equation;  the entries of the matrix are 
now indexed by the internal intertwiner edges.  We can find these 
entries through extensive use of the identities in Section 
\ref{SpinNet}.  This result was first found in \cite{DPRVol1}; the 
matrix entries turn out to be
\begin{eqnarray}
\lefteqn{W^{(n)}_{[rst]} {}^{k_2 \cdots k_{n-2}}_{i_2 \cdots 
i_{n-2}} = {}} \nonumber \\
& & \quad - P_r P_s P_t 
\left\{  \begin{array}{ccc}
k_2 & P_t & k_3 \\
i_2 & P_t & i_3 \\
2 & 2 & 2
\end{array} \right\} \lambda_{k_2}^{i_2 2} \delta_{i_4}^{k_4} 
\cdots \delta_{i_{n-2}}^{k_{n-2}}\nonumber \\
& & \quad \, \times \frac{\mbox{Tet} \left[ \begin{array}{ccc}
P_r & P_r & P_0 \\
k_2 & i_2 & 2 
\end{array} \right] \mbox{Tet} \left[ \begin{array}{ccc}
P_s & P_s & k_4 \\
k_3 & i_3 & 2 
\end{array} \right] \Delta_{k_2} \Delta_{k_3}}
{\theta(k_2, i_2, 2) \theta(k_3, i_3, 2) \theta(P_0, P_r, k_2) 
\theta(k_2, k_3, P_t) \theta(k_3, k_4, P_s)}\label{Weq}
\end{eqnarray}
This expression simplifies somewhat if we express it in terms of the 
Kauffman-Lins 6-$j$ symbols (see Appendix \ref{KLvsW}):
\begin{eqnarray}
W^{(n)}_{[rst]} {}^{k_2 \cdots k_{n-2}}_{i_2 \cdots i_{n-2}} & = & - P_r P_s P_t 
\left\{  \begin{array}{ccc}
k_2 & P_t & k_3 \\
i_2 & P_t & i_3 \\
2 & 2 & 2
\end{array} \right\}  \lambda_{k_2}^{i_2 2} \delta_{i_4}^{k_4} 
\cdots \delta_{i_{n-2}}^{k_{n-2}} \nonumber \\
& & \, \times \frac{ \left\{ \begin{array}{ccc}
P_r & P_r & P_0 \\
k_2 & i_2 & 2 
\end{array} \right\} \left\{ \begin{array}{ccc}
P_s & P_s & k_4 \\
k_3 & i_3 & 2 
\end{array} \right\} }{\theta(k_2, k_3, P_t)} \label{Weq2}
\end{eqnarray}
In both these formulas we have used the 9-$j$ symbol, which is equivalent to the 
following spin diagram:
\begin{equation}
\left\{  \begin{array}{ccc}
k_2 & P_t & k_3 \\
i_2 & P_t & i_3 \\
2 & 2 & 2
\end{array} \right\} = \eqngraph[0.7]{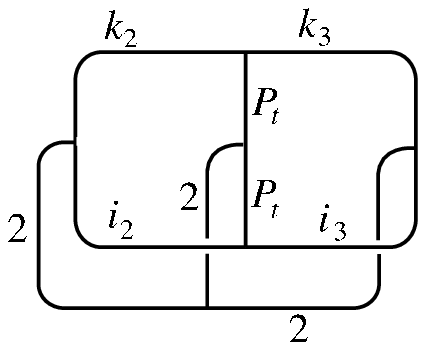} 
\label{9jdef}
\end{equation}

The eigenvalues of the volume operator are then proportional to the sum of the 
absolute values of the $W$-eigenvalues:
\begin{equation}
\hat{V} = \sqrt{\sum_{0\leq r < s < t \leq n-1} \left| \frac{i}{16} 
\hat{W}_{[rst]} \right|}
\end{equation}

The reader may have noticed that this definition for the $W$ 
matrices will not work in the case where the valence of the vertex is 
three or four.  Of course, the grasping action of the $W$ operator will be the 
same in these cases;  the only differences will be 
in the way we calculate the matrix entries.

In the case of the trivalent vertex, there is only one possible 
grasping:
\begin{equation}
\hat{W}_{[rst]} \eqngraph[0.7]{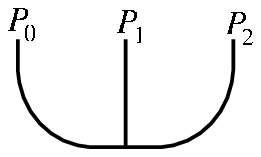} = P_{0} P_{1} P_{2} \eqngraph[0.7]{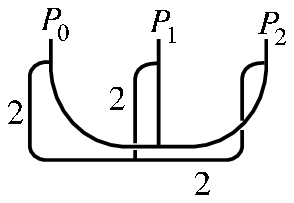}
\end{equation}
Since there is only one possible intertwiner for a trivalent vertex, 
the space in which we are operating is one-dimensional;  hence, we 
will end up with the volume-squared eigenvalue times the original 
vertex:
\begin{equation}
\eqngraph[0.7]{TriGrasp} = W^{(3)} \eqngraph[0.7]{TriVert}.
\end{equation}
This matrix entry evaluates to be
\begin{equation}
W^{(3)} = \frac{P_{0} P_{1} P_{2} \left\{ \begin{array}{ccc} P_{0} & 
P_{1} & P_{2} \\ P_{0} & P_{1} & P_{2} \\ 2 & 2 & 2 \end{array} 
\right\}}{\theta(P_{0}, P_{1}, P_{2})}
\end{equation}
However, the 9-$j$ symbol is antisymmetric under exchange 
of rows;  hence, if two rows are the same (as in this case), the 
9-$j$ symbol evaluates to 0.  Thus, the volume associated with any 
trivalent vertex is zero (a result first found by Loll \cite{Loll}).

In the case of a 4-valent vertex, things are a little more 
complicated.  As in the $n$-valent vertex (for $n \geq 5$), we will 
end up with a superposition of basis states:
\begin{equation}
    \begin{split}   
    \hat{W}_{[012]}^{(4)} \eqngraph[0.7]{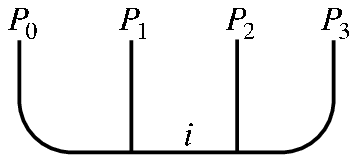} & = P_{0} P_{1} P_{2} 
    \eqngraph[0.7]{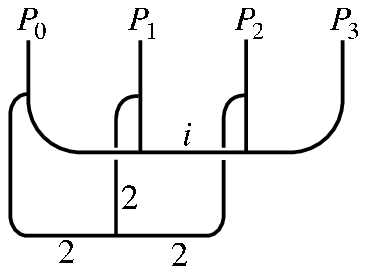} \\
    & = \sum_{k} W_{[012]} {}_{i}^{k} \eqngraph[0.7]{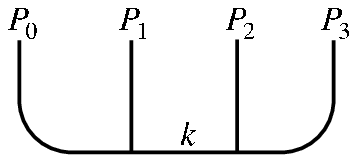}
    \end{split}
\end{equation}
The coefficients $W_{[012]} {}_{i}^{k}$ can be evaluated using similar 
methods to those used to obtain (\ref{Weq});  the result is
\begin{equation}
W_{[012]} {}_{i}^{k} = \frac{P_{0} P_{1} P_{2} \left\{ \begin{array}{ccc} P_{0} & 
P_{1} & k \\ P_{0} & P_{1} & i \\ 2 & 2 & 2 \end{array} 
\right\} \Delta_{j}}{\theta(P_{0}, P_{1}, j) \theta(P_{2}, P_{3}, j) 
\theta(i,j,2)} .
\end{equation}
De Pietri \cite{DPVol2} also gives an explicit polynomial formula for 
the matrix entries for an arbitrary 4-valent vertex with incoming 
edges $a$, $b$, $c$, and $d$:
\begin{eqnarray}
W^{(4)}_{[012]} {}^{t+\epsilon}_{t-\epsilon} & = & \frac{-\epsilon 
(-1)^{(a+b+c+d)/2}}{32\sqrt{t(t+2)}} \nonumber \\
& & \quad \left[ (a+b+t+3)(c+d+t+3)(1+a+b-t) \right. \nonumber \\
& & \qquad (1+c+d-t)(1+a+t-b)(1+b+t-a) \nonumber \\
& & \qquad \left. (1+c+t-d)(1+d+t-c) \right]^{\frac{1}{2}} \mbox{.} 
\label{4val}
\end{eqnarray}
where $t+\epsilon$ and $t-\epsilon$ correspond to the internal edge of the 
vertex, and $\epsilon = \pm 1$ (in other words, the matrix entries are non-zero
if and only if the two states differ by 2.)  
Since this expression is polynomial in form (i.e. doesn't 
contain any combinatoric expressions), it is much easier to work 
with.  We will return to this formula in Section \ref{4vbounds}. 
\setcounter{equation}{0}

\section{The Angle Operator}
\label{AngRes}

Although the form of the angle and volume operators are known, many 
of their properties have not yet been examined.  In this section and 
Section \ref{VolRes}, we describe the results of 
original research done into the properties of these operators.

\subsection{Spectrum of the Angle Operator}
\label{AngSpectrum}

Recall from Section \ref{AngleOp} that the eigenvalues of the angle operator are 
given by the formula
\begin{equation}
\theta = \arccos \left( \frac{j_r (j_r + 1) - j_1 (j_1 + 1) - j_2 (j_2 + 1)}
{2 \sqrt{j_1(j_1 + 1)j_2(j_2+1)}}\right)\mbox{.}
\end{equation}
We will find it more convenient to use the core edge colours $n_1$, $n_2$, and 
$n_r$ instead of their associated spins;  in terms of these labels, the 
angle becomes
\begin{equation} \label{angformn}
\theta = \arccos \left( \frac{n_r (n_r + 2) - n_1 (n_1 + 2) - n_2 (n_2 + 2)}
{2 \sqrt{n_1(n_1 + 2)n_2(n_2+2)}} \right) \mbox{,}
\end{equation}
where
$$
j_x = \frac{n_x}{2}\mbox{.}
$$
Note that the quantities $n_1$, $n_2$, and $n_r$ are the intertwiner edges 
that ``collect'' the spins from each of the three surface patches;  they 
are related to the total edge flux intersecting each patch by the relations
\begin{equation}
n_1 \leq s_1\mbox{, }n_2 \leq s_2\mbox{, and }n_r \leq s_r
\end{equation}
where $s_1$, $s_2$, and $s_r$ are the respective surface fluxes.

\begin{figure}
    \begin{center}
    \includegraphics[width=4in]{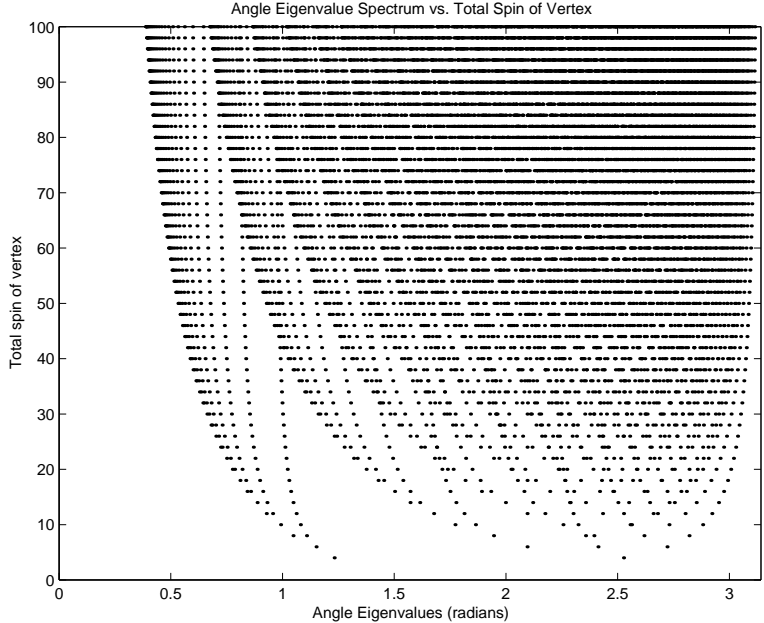}
    \caption{\label{AngSpect}  Angle operator spectrum for increasing 
    total vertex spin.}
    \end{center}
\end{figure}
\begin{figure}
    \begin{center}
    \includegraphics[width=4in]{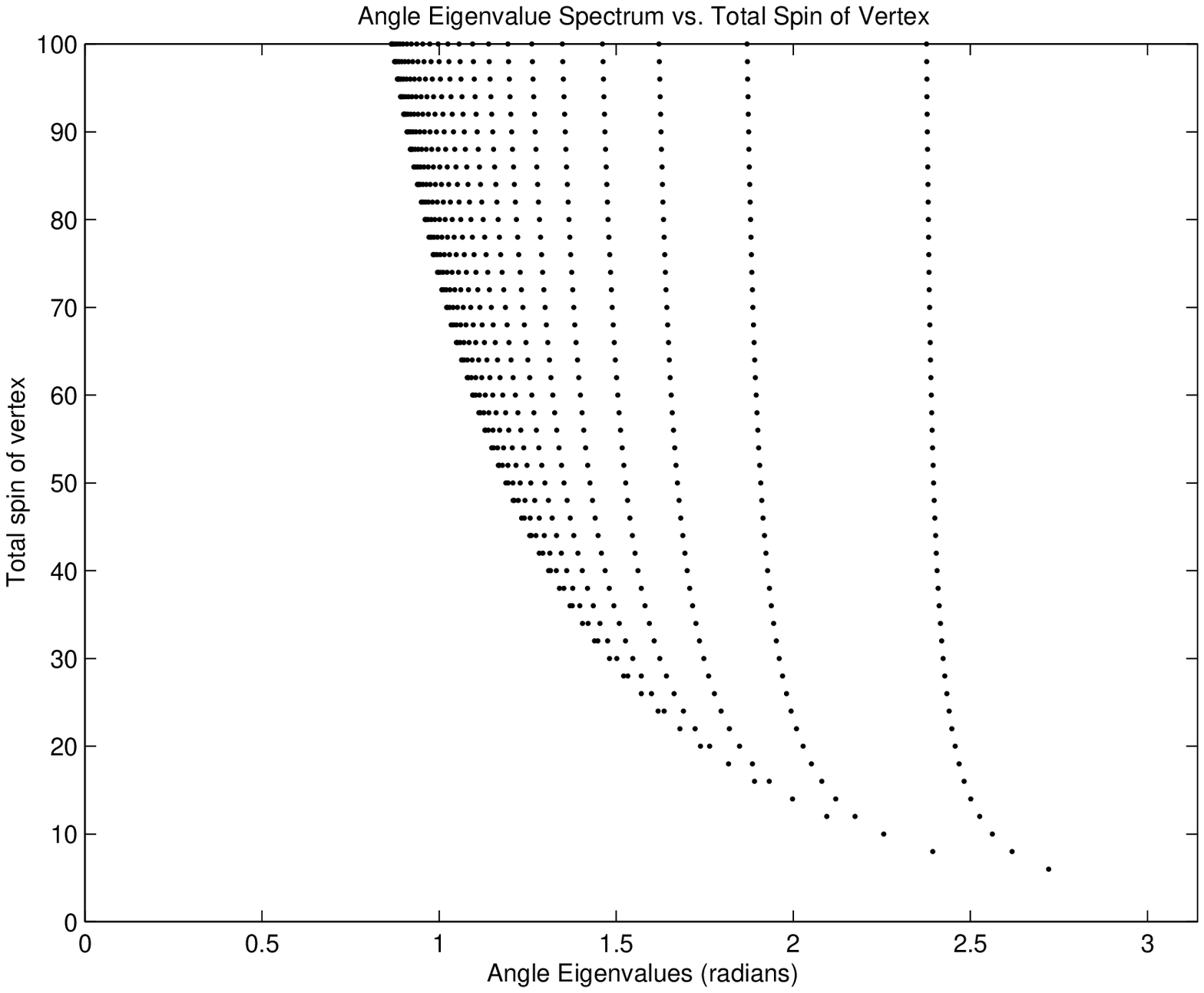}
    \caption{\label{Fan}  One of the fan-like structures that develop 
    as $m$ increases.  This particular fan consists of sequences for 
    which $n_1+n_2-n_r = 4$.}
    \end{center}
\end{figure}

The easiest way to visualize the angle spectrum is to plot its possible 
eigenvalues versus the total core spin (defined as $n_T = n_r + n_1 + 
n_2$).  Such a 
plot is shown in Figure \ref{AngSpect}.  This plot shows two interesting features.  
First, we note the existence of fan-like structures, consisting of several 
sequences splitting off from a sort of ``spine''.  (Such a structure has 
been isolated in Figure \ref{Fan}.  Second, and perhaps more importantly, there 
seem to be very few eigenvalues corresponding to small angles.  We will 
deal with the latter of these phenomena first.

To obtain a small angle, we want the argument of the inverse cosine 
in (\ref{angformn}) to be as close to 1 as possible:
\begin{equation}
Q \equiv \frac{n_r (n_r + 2) - n_1 (n_1 + 2) - n_2 (n_2 + 2)}
{2 \sqrt{n_1(n_1 + 2)n_2(n_2+2)}}
\end{equation}
However, the internal edge spins are subject to the triangle inequalities:
$$
2n_1, 2n_2, 2n_r \leq n_1 + n_2 + n_r
$$
To maximize $Q$, then, we wish to maximize its numerator and minimize its 
denominator while simultaneously respecting the triangle inequalities.  
This occurs when
\begin{equation}
n_r = \frac{n_T}{2} \mbox{ and } n_1 = n_2 \approx \frac{n_T}{4}
\end{equation}
where $n_T$ is as defined above.\footnote{
Note that while the value for $n_r$ is exact, the values of $n_1$ and $n_2$ 
are only approximate;  this is because $n_T$ must be even, but may not 
necessarily be divisible by 4, and because $n_1$, $n_2$, and $n_r$ must be 
integers.  However, for large values of $n_T$, the values of $n_1$ and $n_2$ 
that maximize $Q$ will approach $n_T/4$.}  
At these values of $n_1$, $n_2$, and $n_r$, the argument of the inverse 
cosine then reduces to 
\begin{equation}
Q = \frac{n_T}{n_T + 8}.
\end{equation}
For any angle $\epsilon$, then, we have the relation
$$
\epsilon \geq \arccos \left( \frac{n_T}{n_T + 8} \right)
$$
For small $\epsilon$, $\cos x$ is a uniformly decreasing function;  hence, 
for $\epsilon \ll 1$, we have 
\begin{eqnarray}
\cos \epsilon & \leq & \frac{n_T}{n_T + 8} \nonumber \\
1 - \frac{\epsilon^2}{2} & \leq & 1 - \frac{8}{n_T + 8} \nonumber \\
\epsilon & \geq & \frac{4}{\sqrt{n_T + 8}} \label{SmallAng}
\end{eqnarray}

Hence, for large values of $n_T$, the minimum possible angle between two 
patches is roughly proportional to $n_T^{-1/2}$.  Moreover, we note that 
if $s_R > s_1 + s_2$, then $n_T \leq 2(s_1 + s_2)$;  this implies that for a 
``macroscopic'' vertex, where $s_r \gg s_1, s_2$, the minimum observable 
angle depends only on the edge flux through the patches whose angular 
separation we are measuring.

The first observation (i.e.~the fan-like structures) is somewhat harder to
quantify.  A bit of investigation shows that the ``feathers'' of the fans
in Figure \ref{Fan} are sequences of the form
$$
n_r = m \mbox{, }n_1 = m - r \mbox{, and } n_2 = q \mbox{,}
$$
where $m$ varies and $r$ and $q$ are constant.  From the triangle 
inequalities, we find that 
$$
|r| \leq q \mbox{ and } r + q \mbox{ is even.}
$$
The angles associated with such a vertex are given by
\begin{equation} \label{arbseq}
\cos \theta = \frac{2mr - r^2 - q^2 - 2q}{2\sqrt{q(q+2)(m^2 - 2m(r-1) + 
r(r-2))}} \mbox{,}
\end{equation}
and for $r,q \ll m$, the sequences will approach the value 
\begin{equation}
\cos \theta = \frac{r}{\sqrt{q(q+2)}}
\end{equation}
It is natural to ask whether any real number in the interval $[-1, 1]$ 
is arbitrarily close to a number of this form;  if this were true, then it 
would follow that any angle could be approximated by a vertex with 
sufficiently large internal spin.  It turns out that this is true:

\begin{lemma}  For any real number $x \in [-1, 1]$ and any $\epsilon > 0$, 
there exist integers $q$ and $r$ such that 
$$
\left| x - \frac{r}{\sqrt{q(q+2)}} \right| < \epsilon \mbox{.}
$$
\end{lemma}
\textbf{Proof:}  Choose integers $b$ and $c$ such that 
$$
\left| x - \frac{b}{c} \right| < \frac{\epsilon}{2} \mbox{;}
$$
this can be done because the rational numbers are dense in the reals.  
Further, choose another integer $n$ such that
$$
\left| \frac{b}{c} \left(1 - \frac{1}{\sqrt{1+2/nc}} \right) \right| < 
\frac{\epsilon}{2} \mbox{;}
$$
this can be done since the quantity on the left-hand side of the equation 
can be viewed as a function which approaches 0 as $n \to \infty$.  So, if we 
set $r = nb$ and $q = nc$, we have
\begin{eqnarray*}
\left| x - \frac{r}{\sqrt{q(q+2)}} \right| & = & \left| x - \frac{nb}
{\sqrt{nc(nc+2)}} \right| \\
& = & \left| x - \frac{b}{c} \left( \frac{1}{\sqrt{1 + 2/nc}} \right) 
\right| \\
& \leq & \left| x - \frac{b}{c} \right| + \left| \frac{b}{c} \left(1 - 
\frac{1}{\sqrt{1+2/nc}} \right) \right| \\
& < & \epsilon \mbox{.}
\end{eqnarray*}
Hence, we can approximate any angle $\theta$ to arbitrary accuracy as long 
as we have a vertex of sufficiently high spin.  We note that this is 
merely a confirmation of Moussouris' spin geometry theorem \cite{Mouss}.

It is important to note, however, that while this proof shows the existence 
of vertices which yield any angle to arbitrary accuracy, 
it does nothing to associate a minimum total spin to this angle and 
accuracy.  As it turns out, this task would be fairly difficult to 
accomplish by using these sequences.  If we expand the expression 
in (\ref{arbseq}) to first order in $m^{-1}$, we end up with
\begin{eqnarray}
\cos \theta & = & \frac{1}{2\sqrt{q(q+2)}}(2mr - r^2 - q^2 - 2q)(m^2 - 2m(r-1) + 
r(r-2))^{-1/2} \nonumber \\
& = & \frac{1}{2m\sqrt{q(q+2)}}(2mr - r^2 - q^2 - 2q) \left(1 + 
\frac{r-1}{m} \ldots \right) \nonumber \\
& \approx & \frac{r}{\sqrt{q(q+2)}}\left(1 + \frac{r-1}{m}\right) - \frac{r^2 + 
q^2 + 2q}{2m\sqrt{q(q+2)}} \nonumber \\
\label{seqapp} & = & \frac{r}{\sqrt{q(q+2)}} + \frac{1}{m} \frac{r(r-2) + q (q+2)}{2 \sqrt{q(q+2)}}
\end{eqnarray}
We see that in general the rapidity with which this sequence approaches its 
limit is very difficult to obtain:  it depends non-trivially on the choice 
of $r$ and $q$, which in turn depend non-trivially on the angle we wish 
to approximate and the accuracy to which we wish to approximate it.  

In the special case where $x$ can actually be written in the form 
$r/\sqrt{q(q+2)}$, however, it is fairly simple to determine the minimum 
spin required:  simply plug $r$ and $q$ into the second term of 
(\ref{seqapp}) and find the minimum $m$ that reduces this term below the 
desired accuracy.  Given this fact, the natural question to ask is whether 
any real number $x$ can be written in this form.  
Unfortunately, this is not the case:  any number of the form $r/\sqrt{q(q+2)}$ 
is algebraic, so if $x$ is transcendental it cannot be written in this form.  
Let us examine a case that is even more specialized:  suppose $x$ is of the 
form $\sqrt{a/b}$, where $a$ and $b$ are integers.  We wish to find integer 
solutions to the equation
$$
\frac{a}{b} = \frac{r^2}{q^2 + 2q}\mbox{,}
$$
which is equivalent to the second-order Diophantine equation (in $q$ and $r$)
\begin{equation} \label{Dioph}
a q^2 + 2aq - br^2 = 0
\end{equation}
If we make the substitution $p=q+1$, this equation takes on the form
\begin{equation} \label{Dioph2}
a p^2 - b r^2 - a = 0
\end{equation}
which has the (trivial) solution $p = 1, r = 0$.  As it happens, there are 
recursion relations for Diophantine equations of this sort 
\cite{Edwards};  if we 
know one solution to a given equation, we can find other solutions using 
these relations.  In this case, if there exist integer solutions $u$ 
and $v$ to the equation 
\begin{equation} \label{Pell}
u^2 - abv^2 = 1\mbox{,}
\end{equation}
then $p = u, r = av$ is a solution to (\ref{Dioph2}):
\begin{equation}
a u^2 - b a^2 v^2 - a = a(u^2 - abv^2 - 1) = 0\mbox{.}
\end{equation}

Equation (\ref{Pell}) is known as a \emph{Pell equation};  it has 
solutions when $ab$ is not a perfect square.\footnote{
Note that when $ab$ is a perfect square, $\sqrt{a/b} = a/\sqrt{ab}$ is 
rational.
} However, the solutions are 
not always small, even for fairly small values of $a$ and $b$.  For example, if $a=1$ and 
$b = 61$, the smallest non-trivial solution to (\ref{Dioph}) is:
$$
r = 226153980\mbox{, }q = 1766319050\mbox{.}
$$
Also, solutions to (\ref{Pell}) are found by the expansion of $\sqrt{ab}$ 
in a continued fraction, i.e. an expression of the form
$$
\sqrt{ab} = a_0 + \frac{1}{a_1+\frac{1}{a_2 + \cdots}}
$$
Since the coefficients $\{a_0, a_1, a_2, \ldots\}$ vary widely depending on 
the choice of $ab$, so do the solutions.\footnote
{How these continued fractions 
relate to the solutions of the Pell equation is a fascinating subject 
which is, unfortunately, outside the scope of this piece.  The interested 
reader is referred to Chrystal \cite{Chrystal}.}  Hence, we are forced to 
conclude that finding any angle $\theta$ to an arbitrary accuracy $\epsilon$ using 
this ``fan'' method is possible, but the solutions are difficult to find and have 
an extremely complicated dependence on $\theta$ and $\epsilon$.  

\subsection{Mean Angular Resolution}
\label{AngResolution}

Despite the difficulties in approximating a specific angle to arbitrary 
accuracy, it is possible to get an idea of the average separation between 
angles over the interval $[0, \pi]$.  There are two ways to look at the 
mean angular separation.  The simplest way of finding the ``angular 
resolution'' for a given vertex is to note that an $n$-valent vertex has 
$n-2$ trivalent vertices in its intertwiner.  Since each one of these 
internal vertices can be looked at as a ``core'' vertex, we can conclude 
that such a vertex has at most $3(n-2)$ possible eigenvalues for the angle 
operator (the factor of three comes from regarding each of the three 
edges at each vertex as the $n_{r}$ edge.)  We define the mean angular 
resolution $\delta$ to be the average 
separation between each possible angle and the next greatest possible angle;  
in other words, $\delta$ is the average width of a ``gap'' between angles in 
the interval $[0, \pi]$.  Since the $3n-6$ eigenvalues associated with the 
vertex divide the interval into $3n-5$ gaps, we conclude that the mean 
angular separation between the eigenvalues of the angle operator for an 
$n$-valent vertex is
\begin{equation} \label{eigres}
\delta = \frac{\mbox{total of all gaps}}{\mbox{number of gaps}} = 
\frac{\pi}{3n-5}
\end{equation}

However, this method does not take into account the fact that internal 
vertices may yield the same angles.  For example, the internal vertices of 
a 4-valent vertex with all edges the same colour are identical;  hence, when 
each of these internal vertices is treated as the core vertex 
by their respective angle operators, there will only be 2 possible angles 
instead of the expected 6.  Since this degeneracy depends on the internal 
structure of the vertex, it is impossible to say much about it in general;  
the best we can do is to modify (\ref{eigres}) to read
\begin{equation}
\delta \geq \frac{\pi}{3n-5}
\end{equation}

This method, while legitimate, is somewhat limited:  it only examines the 
separation between the eigenvalues of the angle operators that have a given 
vertex as their eigenstate.  However, other angle measurements are 
possible for a given vertex.  For example, if we measure the angle between 
two edges that do not immediately intersect in the vertex's intertwiner, 
this measurement will not always yield a definite eigenvalue;  instead, 
there will be an expectation value 
associated with this measurement.  The question then becomes:  for a given 
vertex, can we find the total number of possible expectation values 
for all possible measurements?

We note that given any vertex, we can partition its edges into three 
groups (one for $S_{1}$, one for $S_{2}$, and one for $S_{r}$.)  If 
we switch basis to the ``snowflake'' basis, we will generally obtain a 
superposition of edges of this form;  however, for a given 
superposition (corresponding to our original vertex) and a given 
partition, we will find a single expectation value.  Hence, for a 
given vertex, the number of expectation values for the angle operator 
is equal to the number of partitions of its edges into three distinct 
groups.

Finding this number for a vertex with an arbitrary number of edges with 
arbitrary colours, however, is another question entirely.  We note that 
the possible range of colours for the collecting edge is not dependent on 
the order of the edges along the branch;  hence, the number of distinct 
partitions is given by
\begin{equation}
\mathcal{N} = \prod_{i=1}^{\infty} B(q_i) \mbox{,}
\end{equation}
where $q_i$ is the number of edges with colour $i$, and $B(n)$ is the 
number of distinct ways of partitioning $n$ into three distinct bins (one bin 
for each branch.)\footnote{
One might still wonder whether the expectation value of the operator could 
somehow depend on the order of the valent edges along the branches.  To show 
that it does not, let us consider two states 
$$
|\psi_1 \rangle = \sum_{i} a_i |n_1, n_2, n_r, \alpha_i \rangle
\quad \mbox{and} \quad
|\psi_2 \rangle = \sum_{i} b_i |n_1, n_2, n_r, \beta_i \rangle
$$
where $n_1, n_2, n_r$ are the collecting edges, and $\alpha_i$ and 
$\beta_i$ are all other indices that can vary within the $n_1, n_2, n_r$ 
subspace.  (Note that the $\alpha$ indices and the $\beta$ indices are not 
necessarily in the same basis, but both bases span the $n_1, n_2, n_r$ 
subspace.)  The expectation values for each of these states are then
$$
\langle \psi_1 | \theta | \psi_1 \rangle = f(n_1, n_2, n_r) \sum_i |a_i|^2
\quad \mbox{and} \quad
\langle \psi_2 | \theta | \psi_2 \rangle = f(n_1, n_2, n_r) \sum_i |b_i|^2
$$
where $f(n_1, n_2, n_r)$ is the function of $n_1, n_2, n_r$ given in 
(\ref{angformn}).  Since application of the recoupling theorem is a 
unitary basis transformation, $\sum_i |a_i|^2 = \sum_i |b_i|^2$;  hence, 
the expectation values of both states are the same.
}
Elementary combinatorics tells us that the number of ways to do this is
\begin{equation}
B(n) = {(n + 3) - 1 \choose n} = \frac{(n+1)(n+2)}{2}
\end{equation}
However, we must divide the overall product by two, since exchange of the 
sets corresponding to the $n_1$ branch and the $n_2$ branch yields the same 
angle.  Thus, the maximum number of possible expectation values for a 
given vertex is
\begin{equation}
\mathcal{N} = \frac{1}{2} \prod_{i=1}^{\infty} \frac{(q_i+1)(q_i+2)}{2}
\mbox{.}
\end{equation}
In the case where all the edges have the same colour, this formula reduces 
to\footnote
{In general, we will call a vertex whose valent edge colours are all the 
same a \emph{monochromatic} vertex;  we will also refer to a monochromatic 
vertex whose edges are a specific color as a monochromatic vertex with edge 
colour $m$, or sometimes as an $m$-chromatic vertex.}
\begin{equation}
\mathcal{N} = \frac{(n+1)(n+2)}{4}
\end{equation}
and the mean angular resolution (between expectation values) is given by
\begin{equation} \label{AngResol}
\delta \geq \frac{4 \pi}{n^2 + 3n + 6}
\end{equation}
where we have again used an inequality because of possible angle 
degeneracies.

It is important not to confuse the ``mean 
angular resolution'' as we have defined it with the average distance from a
point in the interval $[0, \pi]$ to the nearest angle (averaged over 
this interval).  The quantity $\delta$ that we
have found allows us to get an idea of how tightly, on average, the 
eigenvalues of the angle operator for a given vertex are spaced.  To find 
this other value (which, if found, would help us to estimate the level to 
which an arbitrary angle can be approximated) would
\begin{equation}
\langle \Delta \rangle = \sum_{\mathrm{all \: gaps}} {\mbox{prob.~of being} 
\choose \mbox{located in the gap}} {\mbox{mean separation from} \choose 
\mbox{either end of gap}}
\end{equation}
This is equivalent to the expression
\begin{equation}
\langle \Delta \rangle = \sum_{a} \frac{w_a^2}{4\pi}
\end{equation}
where $w_a$ is the width of the gap $a$.  This expression depends heavily 
on the precise location of the eigenvalues in the interval $[0, \pi]$, and 
is therefore difficult to compute analytically.  

\subsection{Angle Distribution}

In the classical continuum model of space, the distribution of solid 
angles is proportional to $\sin \theta$:
\begin{equation}
\mathcal{P}(\theta) \, \mathrm{d} \theta = \sin \theta \, \mathrm{d} \theta
\end{equation}
If our quantized angle operator is to be of any physical worth, it must 
reproduce this distribution in some ``classical limit.''  To find out 
whether this is so, we must examine not only the location of possible 
eigenvalues in the interval $[0, \pi]$ (as we have in the past two 
sections), but also the likelihood with which these angles occur.  
Due to the complexity of this problem, we will 
consider only the simplest possible case which could still 
conceivably have a classical limit: that of an $n$-valent monochromatic 
vertex with edge colour 1.

We note that every vertex can be transformed into the snowflake basis by
repeated application of the recoupling 
theorem.  If we characterize these states by the internal spins $a$, 
$b$, and $c$, this fact can be written algebraically as
\begin{equation}
| \psi \rangle = \sum_{a, b, c} C_{abc} | a_1, a_2, \ldots, b_1, b_2, 
\ldots, c_1, c_2, \ldots \rangle
\end{equation}
where $a$, $b$, and $c$ (without index) denote the sets $\{a_1, a_2, 
\ldots, a_i \}$, $\{b_1, b_2, \ldots, b_j\}$, and $\{c_1, c_2, \ldots, 
c_k\}$ respectively.  The most difficult part of this formula to analyze 
for arbitrary $n$ is the coefficients $C_{abc}$;  finding these for a 
specific internal intertwiner structure could be done, although such a 
calculation would be mind-bogglingly tedious.  For a random vertex, however,
we do not know the exact internal structure \emph{a priori};  hence, there
is no reason to assume that any given intertwiner is preferred.  We will
therefore assume that all intertwiners are equally likely. With this simplification in mind, 
we can conclude that the probability of measuring the angle associated with 
an intertwiner core $n_1, n_2, n_r$ is given by
\begin{equation}
\mathcal{P} (\theta(n_1, n_2, n_r)) \propto {\mbox{number of intertwiners} 
\choose \mbox{with core } n_1, n_2, n_r}
\end{equation}
where $\theta(n_1, n_2, n_r)$ is given by (\ref{angformn}).

\begin{figure}
    \begin{center}
    \includegraphics[scale=0.7]{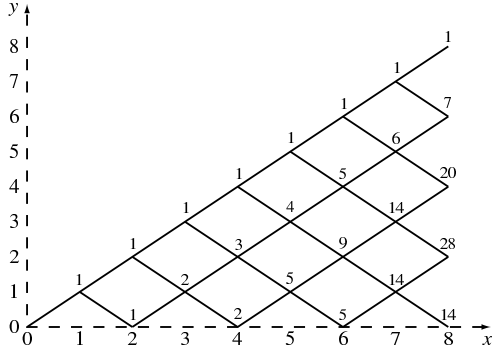}
    \caption{\label{Lattice1}  The lattice in the $x$-$y$ plane.  The 
    numbers above each intersection correspond to the number of paths 
    from the origin to that intersection.}
    \end{center}
\end{figure}

We wish to know, then, how many intertwiners exist with a given 
$n_1$, $n_2$, and $n_r$.  This is found in the following lemma:
\begin{lemma}  If $i$ is the number of edges (of colour 1) entering a 
branch, the number of distinct branches with internal edges 
$\{a_2, a_{3}, \ldots, a_{i}\}$ that end in a given edge label $a_{i}$ is
\begin{equation} \label{collnum}
Q(i, a_{i}) = \frac{a_{i}+1}{i+1} {i+1 \choose 
\frac{a_{i}+i}{2} + 1}
\end{equation}
\end{lemma}
\textbf{Proof:}  Consider the network depicted in Figure 
\ref{Lattice1}.  
Given a sequence of internal edges $\{a_{1}, a_{2}, 
\ldots, a_{n}\}$, we can define a path on this network which passes 
through the points $\{ (1, a_{1}), (2, a_{2}), \ldots, (n, a_{n})\}$.  
Thus, the number of distinct branches with internal edges $\{ a_{1}, 
a_{2}, \ldots, a_{i} \}$ is equal to the number of distinct paths on 
this lattice from the point $x = 1, y = 1$ to the point $x = i, 
y = a_{i}$ (which go strictly to the right at all times, not to the 
left.)  We also note that $Q(i, a_{i}) = Q(i-1, a_{i}-1) + Q(i-1, 
a_{i}+1)$;  in other words, the number of paths to a given point in 
the network is equal to the sum of the numbers of paths to the two 
points closest to it in the previous column.  Using this technique, 
the number of paths for small values of $x$ and $y$ can be easily 
calculated;  these numbers are shown in Figure \ref{Lattice1}, above 
each vertex.

\begin{figure}
    \begin{center}
    \includegraphics[scale=0.7]{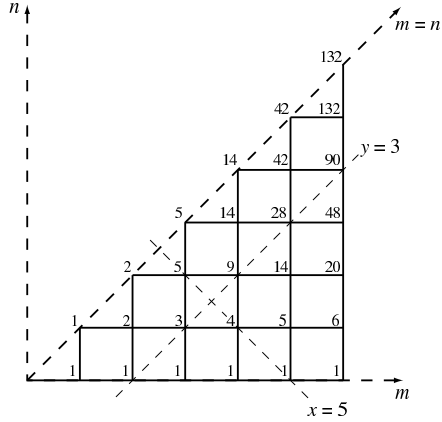}
    
    \caption{\label{Lattice2}  The lattice in the $m$-$n$ plane.  To make 
    the correspondence between this figure and Figure \ref{Lattice1} 
    clearer, we have included two representative lines of constant
    $x$ and $y$ in this diagram.}
    \end{center}
\end{figure}

We now make the substitutions
$$
m = \frac{x+y}{2} \qquad \mbox{and} \qquad
n = \frac{x-y}{2} \mbox{;}
$$
note that $m$ and $n$ are integers since $x \equiv y \pmod 2$ for any 
point on our original lattice.  If we redraw our lattice in the 
$m$-$n$ plane, we obtain Figure \ref{Lattice2}.  
Our paths are now on 
a conventional lattice;  however, the condition that $y \geq 0$ for 
the paths on our original lattice translates into the condition that 
$m \geq n$ (since $y = m - n$.)  Thus, we wish to find the number of 
lattice paths from $(0,0)$ to $(m_{0}, n_{0})$ (where $m_{0} = (i + 
a_{i})/2$ and $n_{0} = (i - a_{i})/2$) which never go above the line 
$m = n$.

This new problem is a classic problem in combinatorics;  it is 
discussed, along with other related problems, by Hilton and Pedersen 
\cite{HiltPed}.  The result for the lattice paths is then
$$
Q(m_{0}, n_{0}) = \frac{m_{0} - n_{0} + 1}{m_{0} + n_{0} + 1} {m_{0} + 
n_{0} + 1 \choose m_{0} + 1}
$$
and the solution to our original problem is then
$$
    Q(i,a_{i}) = \frac{a_{i} + 1}{i + 1} {i + 1 \choose \frac{i + 
    a_{i}}{2} + 1}
$$

If we wish to turn this into a normalized probability distribution, 
we can approximate the binomial coefficient in the above expression by 
a Gaussian distribution multiplied by a normalization factor $A$:
$$
{x \choose y} = A \exp \left[ 
\frac{-(x-\frac{y}{2})^{2}}{\frac{y}{2}} \right]
$$
Applying this to (\ref{collnum}), we have
\begin{equation}
P(i, a_{i}) = \frac{a_{i}+1}{i+1} \exp \left[ \frac{-(a_{i}^{2} + 
2a_{i})}{2(i+1)} \right]
\end{equation}
where $A$ has been subsumed into the exponential term.

\begin{figure}[p]
    \begin{center}
    \includegraphics[width=4in]{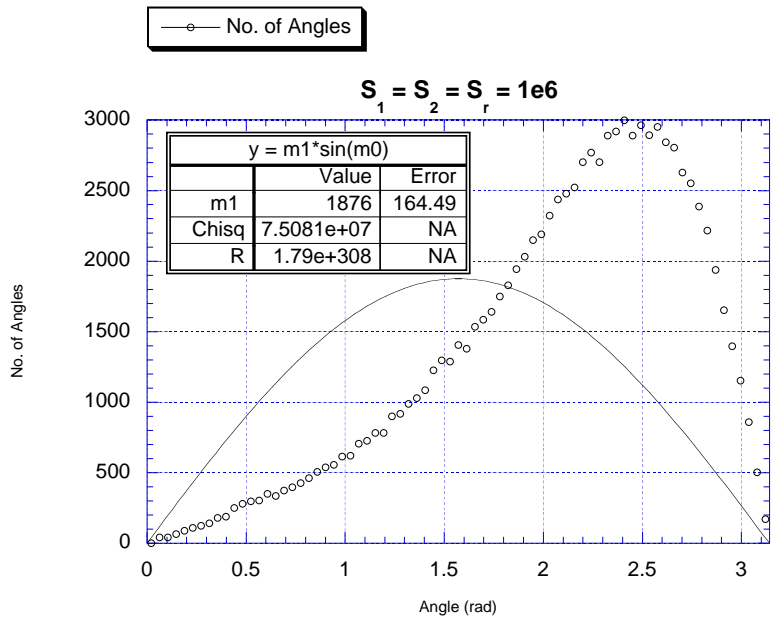}
    \caption{\label{dist1} Angle distribution for $s_{1} = s_{2} = s_{r}$.}
    \end{center}
\end{figure}

\begin{figure}[p]
    \begin{center}
    \includegraphics[width=4in]{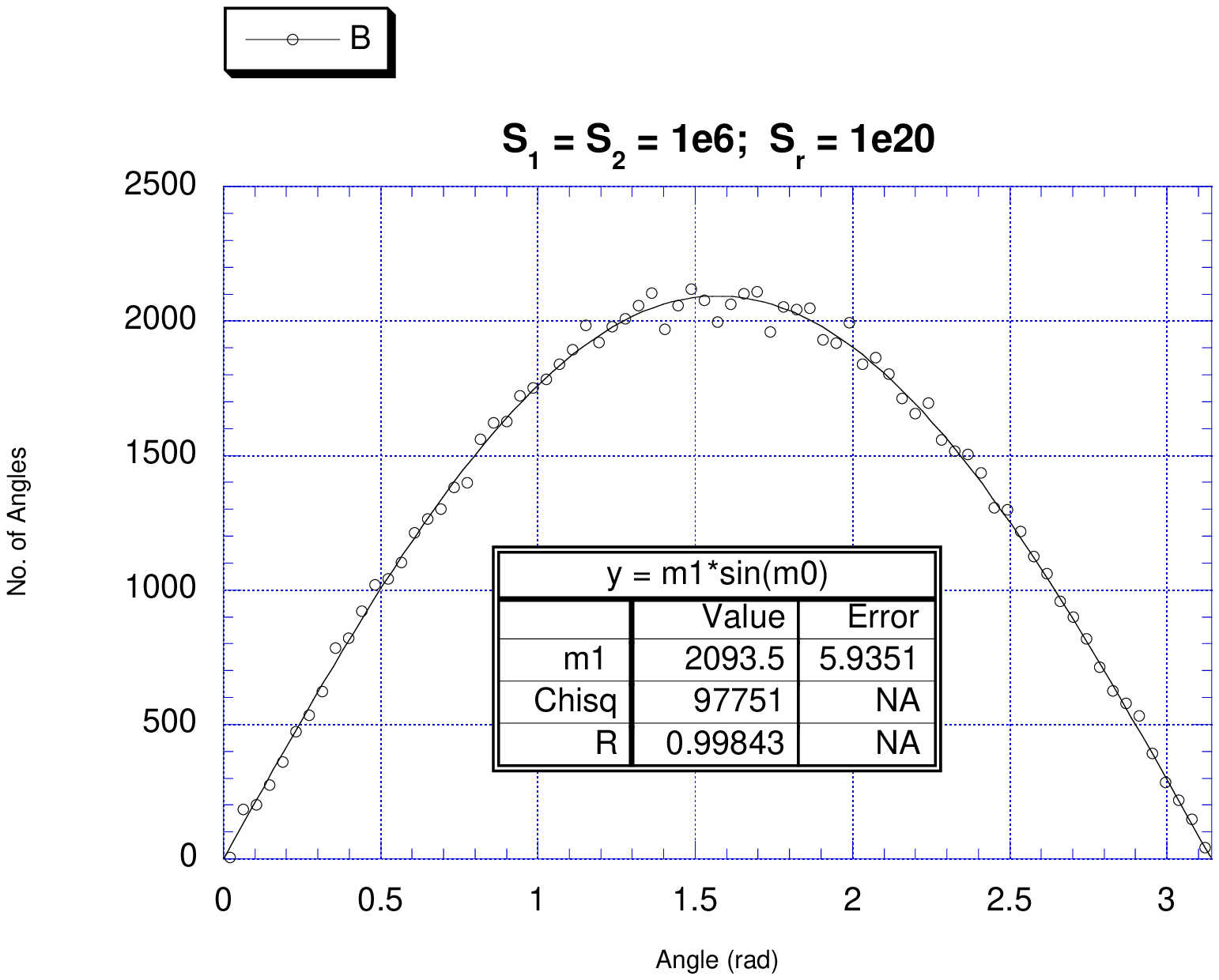}
    \caption{\label{dist2} Angle distribution for $s_{1} = s_{2} \ll s_{r}$.}
    \end{center}
\end{figure}

\begin{figure}[p] 
    \begin{center}
    \includegraphics[width=4in]{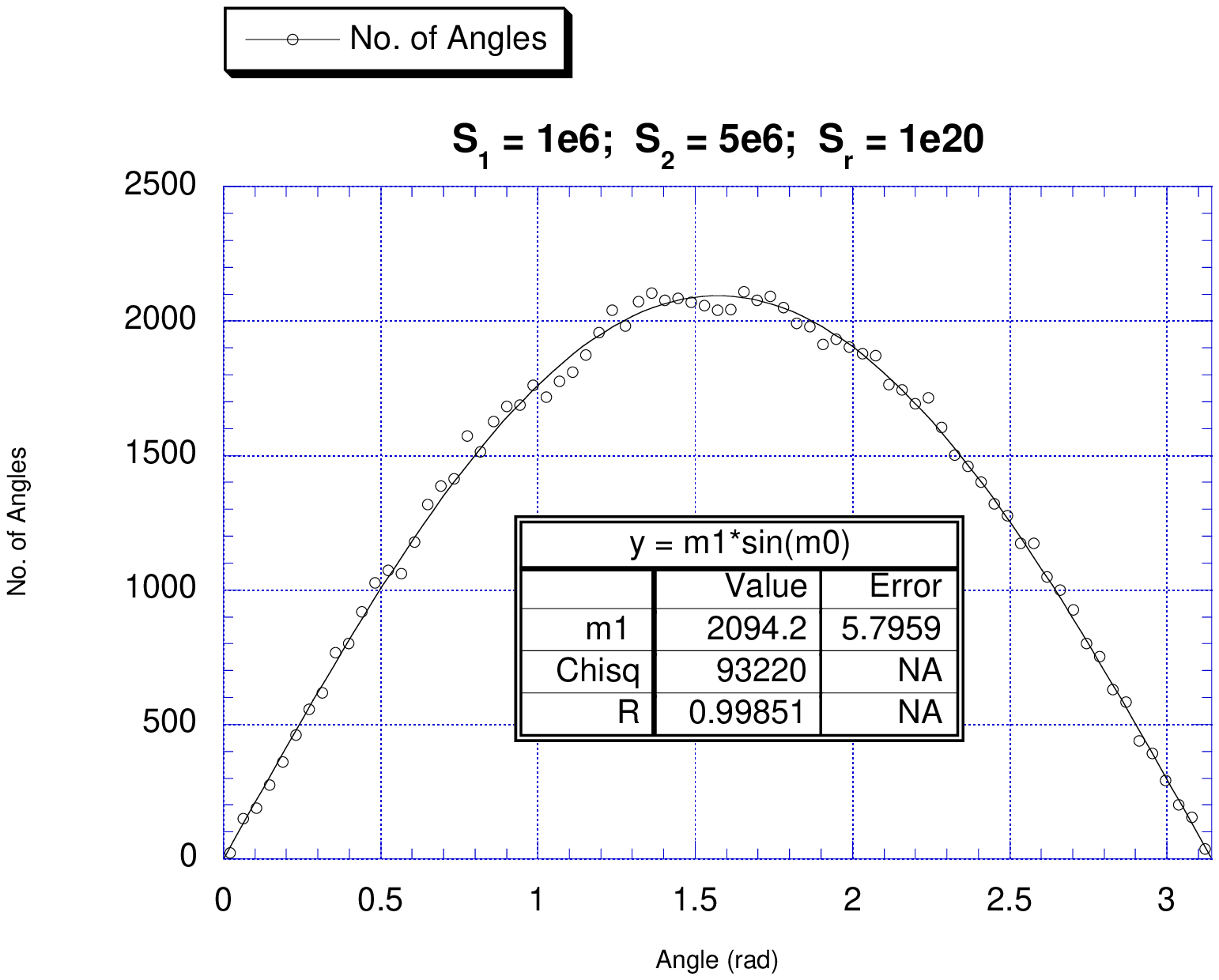}
    \caption{\label{dist3} Angle distribution for $s_{1} < s_{2} \ll s_{r}$.}
    \end{center}
\end{figure}

\begin{figure}[p] 
    \begin{center}
    \includegraphics[width=4in]{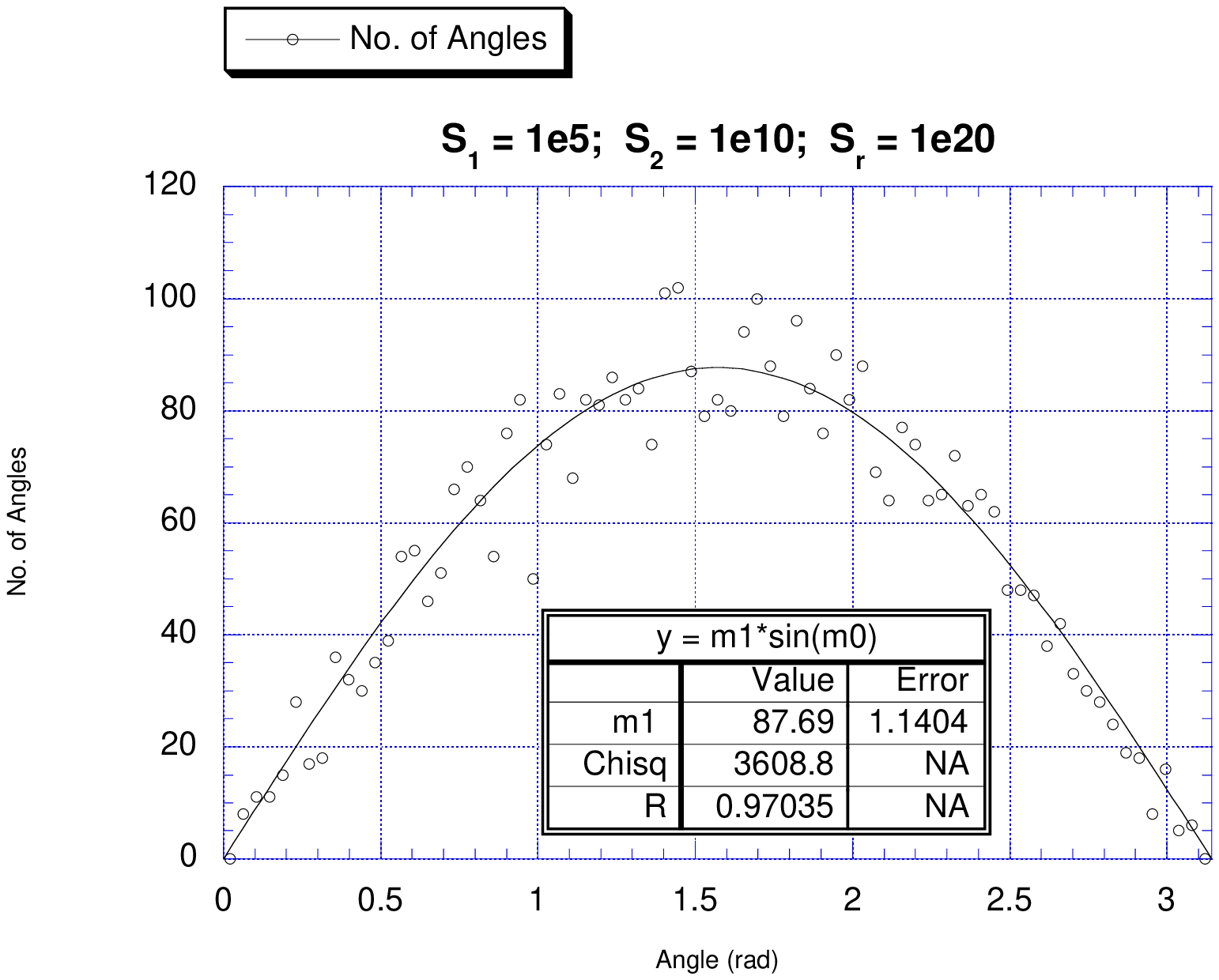}
    \caption{\label{dist4} Angle distribution for $s_{1} \ll s_{2} \ll s_{r}$.}
    \end{center}
\end{figure}

Using this probability distribution, we numerically computed the distribution of 
the angles in the interval $[0, \pi]$.  The program randomly selected
candidate values of $n_1$, $n_2$, and $n_r$, making sure that they satisfied
the vertex constraints;  it then accepted or rejected the angle using the
probability distribution derived above.  The results for various values 
of $s_{1}$, $s_{2}$, and $s_{r}$ are shown in Figures \ref{dist1} to 
\ref{dist4}.

Figure \ref{dist1} shows the distribution for $s_{1} = s_{2} = s_{r}$.  This 
distribution certainly does not match the expected classical 
distribution;  most notable is the fact that the distribution peaks 
at approximately $2.5 \approx 140^{\mathrm{o}}$.  

Figures \ref{dist2} and \ref{dist3} show the distribution for $s_{1} 
= s_{2} \ll s_{r}$ and $s_{1} < s_{2} \ll s_{r}$, respectively.  These have 
surprisingly good correspondence with the expected classical angle distribution.

Figure \ref{dist4} shows the case $s_{1} \ll s_{2} \ll s_{r}$.  This 
particular case required an extraordinary amount of computer time;  as 
a result, the deviation of the data points is much larger.  However, 
the data still matches the expected $\sin \theta$ distribution very 
well.

These data, while not conclusive, are certainly indicative that the 
angle operator can indeed reproduce the classical angle distribution 
in the case where $s_{1}, s_{2} \gg 1$ and $s_{1}, s_{2} \ll s_{r}$.  
This first condition corresponds to what normally thinks of as a 
``classical limit,'' well away from the regions where quantum effects 
dominate.  The second condition can be thought of as a requirement 
for a sufficient amount of ``background geometry'':  not only do we 
need to make sure that the angles we are measuring include a large 
amount of spin, but we also need to ensure that the background 
spacetime upon which we measure the angles is sufficiently classical. 
\setcounter{equation}{0}

\section{The Volume Operator}
\label{VolRes}
\setlength{\arraycolsep}{1mm}
\subsection{The $W$-matrix and Eigenvalue Bounds}

We recall from section \ref{VolOp} that the elements of the $W$-matrix are given by
\begin{eqnarray} \label{Weq3}
W^{(n)}_{[rst]} {}^{k_2 \cdots k_{n-2}}_{i_2 \cdots i_{n-2}} & = & - P_r P_s P_t 
\left\{  \begin{array}{ccc}
k_2 & P_t & k_3 \\
i_2 & P_t & i_3 \\
2 & 2 & 2
\end{array} \right\}  \lambda_{k_2}^{i_2 2} \delta_{i_4}^{k_4} 
\cdots \delta_{i_{n-2}}^{k_{n-2}} \nonumber \\
& & \, \times \frac{ \left\{ \begin{array}{ccc}
P_r & P_r & P_0 \\
k_2 & i_2 & 2 
\end{array} \right\} \left\{ \begin{array}{ccc}
P_s & P_s & k_4 \\
k_3 & i_3 & 2 
\end{array} \right\} }{\theta(k_2, k_3, P_t)}
\end{eqnarray}
where $k_2, \ldots, k_{n-2}$ and $i_2, \ldots, i_{n_2}$ are the internal 
spins in the  ``comb'' basis (see Figure \ref{SnowComb}.)

From this operator, we then define the ``volume-squared'' operator as 
\begin{equation} \label{GenVol}
\hat{V^2} = l_{0}^{6} \sum_{r < s < t} \left| \frac{i \hat{W}^{(n)}_{[rst]}}
{16} \right| \mbox{,}
\end{equation}
where $l_{0}$ is once again the Planck length.  The eigenvalues of this 
operator are the squares of the volume operator $\hat{V}$.

We can see that this expression is horrendously difficult to handle in all 
its generality, for three main reasons.  First, for an $n$-valent vertex with 
arbitrary edge colours, there are ${n \choose 3}$ possible choices for $r$, 
$s$, and $t$;  each of these choices could conceivably produce a 
different $W^{(n)}_{[rst]}$.  Second, the eigenvalues of the $\hat{V^2}$
operator are difficult to find;  even those of the $W$-matrix are 
distinctly non-trivial, since all that (\ref{Weq3}) gives us is the matrix 
entries.  Third, the matrix entries themselves, as given by (\ref{Weq3}),
are not easy to calculate analytically.  These three factors make the 
overall problem rather intractable, either analytically or numerically.

With these problems, it is to our advantage to make some simplifications.  Instead 
of considering an arbitrary vertex, we will examine the case of a 
monochromatic vertex;  in this case, all of the matrices $W^{(n)}_{[rst]}$ 
are identical (since our ``graspings'' are always on edges of the same 
colour.)\footnote
{We will often suppress the indices $[rst]$, since 
they are always the same.  In other words, when referring to the 
monochromatic case, we define $W^{(n)} = W^{(n)}_{[rst]}$.}
The formula (\ref{GenVol}) then reduces to
\begin{equation} \label{V2andW}
\hat{V^2} = l_{0}^{6} {n \choose 3} \left|\frac{iW^{(n)}}{16}\right| = 
\frac{n(n-1)(n-2)}{96} \left|iW^{(n)}\right|
\end{equation}

This solves our first problem.  The second problem, however, is somewhat 
less tractable;  analytically finding the eigenvalues of an $n \times n$ 
matrix is equivalent to finding the roots of a $n$th degree polynomial,
for which no analytic solution exists for $n > 4$.  The matrix $W^{(n)}$, 
on the other hand, has a row for each possible intertwiner core - a number 
that is, in general, much greater than 4.
To get around this, we will examine bounds on the eigenvalues instead of 
the eigenvalues.  For any $n \times n$ matrix with entries $a_{ij}$, any 
given eigenvalue $\lambda$ of this matrix satisfies
\begin{equation} \label{eiglim}
| \lambda | \leq \max_i \sum_{j=1}^n |a_{ij}| \quad \mbox{and} \quad 
| \lambda | \leq \max_j \sum_{i=1}^n |a_{ij}|
\end{equation}
In other words, the magnitude of every eigenvalue must less than the sum 
of the absolute values of some column and of some row.\footnote{
The other two well-known results for eigenvalue bounds --- namely, 
Gershgorin's theorem and Schur's inequality --- turn out to be much less 
useful to us.  Gershgorin's theorem states that for some $i$,
$$
| \lambda - a_{ii} | \leq \sum_{j \neq i} |a_{ij}|
$$
and a similar result for the rows of the matrix;  however, the 
diagonal elements of the $W^{(n)}$-matrix are all zero, which reduces this 
formula to (\ref{eiglim}).  Schur's inequality states that
$$
\sum_{\alpha = 1}^n |\lambda_{\alpha}|^2 \leq \sum_{i,j = 1}^n |a_{ij}|^2
$$
where $\lambda_{\alpha}$ are the eigenvalues of the matrix.  However, it 
is much more difficult to obtain an analytic expression for the RHS of 
this equation, as it includes all of the elements of the matrix.  For more 
information on these bounds, consult Bell \cite{Bell}.}
Using this, we can find an upper bound $M$ on the 
magnitudes of the eigenvalues of $W^{(n)}$.  In the monochromatic case, however, 
the eigenvalues $\lambda_{W\alpha}$ of the $\hat{W}^{(n)}$ operator are related to 
the eigenvalues $\lambda_{V^2\alpha}$ of the $\hat{V^2}$ operator by
\begin{equation}
\lambda_{V^2\alpha} = \frac{n(n-1)(n-2)}{96} \left| \lambda_{W\alpha} 
\right|
\end{equation}
as can be seen from (\ref{V2andW}).  Hence, we can place a limit on the 
eigenvalues of the volume operator:
\begin{eqnarray}
\lambda_{V\alpha} & = & \sqrt{\lambda_{V^2\alpha}} \nonumber \\
& = & \sqrt{ \frac{n(n-1)(n-2)}{96} \left| \lambda_{W\alpha} \right| } 
\nonumber \\
& \leq & \sqrt{ \frac{n(n-1)(n-2)}{96} M} \label{Vlim}
\end{eqnarray}

The determination of $M$ --- the last of our three problems above --- is, 
unfortunately,  rather intractable.  While explicit expressions for 
the 6-$j$ and 9-$j$ symbols in (\ref{Weq3}) exist, they are rather 
complicated.  Simpler expressions do exist for the case of a 4-valent vertex;  
in this case, we can easily calculate the matrix elements explicitly.  However, 
in general, we will have great difficulty doing so.

\subsection{Volume Eigenvalue Bounds for 4-valent Vertices}
\label{4vbounds}

In the case of a 4-valent vertex, De Pietri \cite{DPVol2} has shown that the 
entries of the $W^{(4)}$ matrix for a 4-valent vertex with edge colours $a$, 
$b$, $c$, and $d$ are given explicitly by the formula
\begin{eqnarray}
W^{(4)}_{[012]} {}^{t+\epsilon}_{t-\epsilon} & = & \frac{-\epsilon 
(-1)^{(a+b+c+d)/2}}{32\sqrt{t(t+2)}} \nonumber \\
& & \quad \left[ (a+b+t+3)(c+d+t+3)(1+a+b-t) \right. \nonumber \\
& & \qquad (1+c+d-t)(1+a+t-b)(1+b+t-a) \nonumber \\
& & \qquad \left. (1+c+t-d)(1+d+t-c) \right]^{\frac{1}{2}} \mbox{.} 
\label{4val2}
\end{eqnarray}
where $t+\epsilon$ and $t-\epsilon$ correspond to the internal edge of the 
vertex.  It can be shown that these elements are zero unless $\epsilon = 
\pm 1$;  hence, each row or column will have at most two non-zero 
elements.  To find $M$ in this case, then, all we have to do is to find the 
maximum value of the polynomial on the RHS of (\ref{4val2}).  

In the monochromatic case, $a=b=c=d$;  we will call this edge colour $m$.  
The polynomial in question then becomes
\begin{equation}  \label{4lim1}
\left|W^{(4)}_{[012]} {}^{t+\epsilon}_{t-\epsilon} \right| = 
\frac{(2m+t+3)(2m-t+1)(t+1)^2}{32\sqrt{t(t+2)}}
\end{equation}
If we think of this as a continuous function $W(t)$, we can show that this 
expression is maximized at
\begin{equation}  \label{4lim2}
t_{\mathit{max}} = \sqrt{\frac{(2m+2)^2 + 4}{6} + \frac{1}{6} \sqrt{(2m+2)^4 - 
16(2m+2)^2 + 16}} - 1 \mbox{,}
\end{equation}
and the maximum value is, of course, $W(t_{\mathit{max}})$.  In the 
large-spin limit, where $m \gg 1$, this maximum value scales as
\begin{equation}
\left| W_{\mathit{max}} \right| \approx \frac{m^3}{6\sqrt{3}}
\end{equation}
The maximum absolute row or column sum (i.e. the quantity $M$) is at most 
twice this number; therefore, (\ref{Vlim}) tells us that for a 4-valent 
monochromatic vertex with edge colour $m \gg 1$, any eigenvalue $\lambda_V$ of 
the volume operator on this vertex will satisfy the 
inequality
\begin{equation} \label{large4lim}
\lambda_V \leq l_0^3 \frac{m^{3/2}}{\sqrt{12\sqrt{3}}}
\end{equation}

This is a very encouraging result when we consider it in relation to the 
area operator.  The eigenvalues of the area operator are given by
\begin{equation}
\lambda_A = l_0^2 \sum_{i} \sqrt{m_i(m_i+2)} 
\end{equation}
where the values $m_i$ are all the intersections of edges with the surface 
in question.  For a 4-valent vertex with large edge colour $m$, we have 
\begin{equation}
\lambda_A \propto m \mbox{.}
\end{equation}
Together with (\ref{large4lim}), this implies that the volume scales 
no faster than $A^{3/2}$, which is what we would expect from 
classical spatial geometry.

\begin{figure} 
    \begin{center}
    \includegraphics[width=4in]{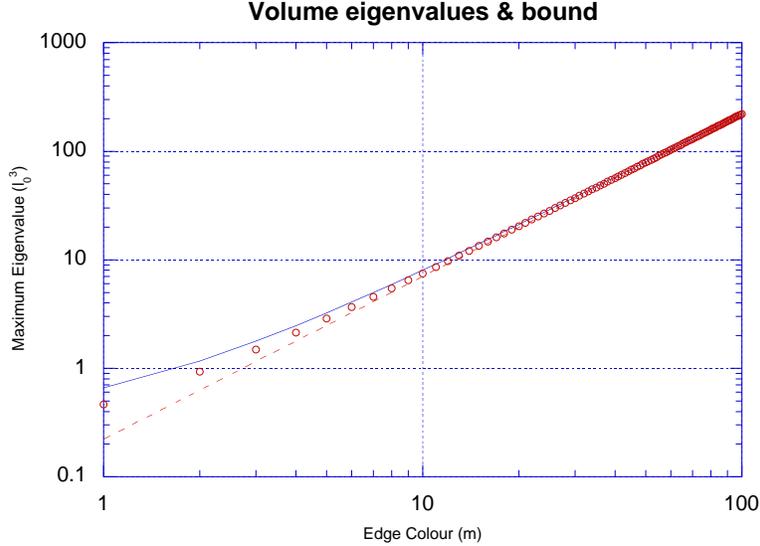}
    \caption{\label{4eigsandlim} The eigenvalue limit (shown as a solid 
    line) obtained in 
    (\ref{4lim1}) and (\ref{4lim2}), compared with the maximum volume 
    eigenvalue for $1 \leq m \leq 100$ (shown as points).  The lower dotted 
    line is the 
    result of a curve fit of the maximum eigenvalues to a curve of 
    the form $\lambda = k m^{3/2}$, where $k$ is the only free 
    parameter of the fit.  This fit yields a value of $k = 0.22155$.}
    \end{center}
\end{figure}
\begin{figure} 
    \begin{center}
    \includegraphics[width=4in]{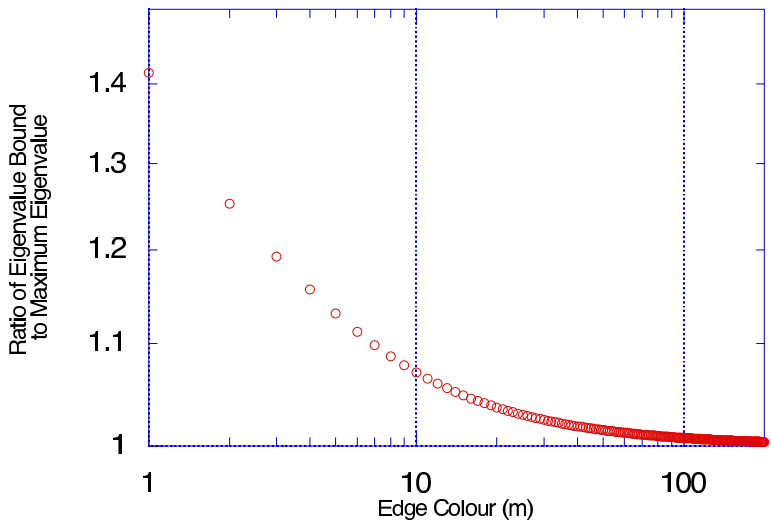}
    \caption{\label{limeigsrat} The ratio of the eigenvalue limit to the 
    maximum volume eigenvalue, for $1 \leq m \leq 200$.}
    \end{center}
\end{figure}

We must ask ourselves, however, how ``good'' this bound is:  do the 
eigenvalues actually scale proportionally to $m^{3/2}$, and if so, is the 
proportionality coefficient we have found (i.e. $1/\sqrt{12\sqrt{3}}$) close 
to the actual proportionality coefficient?  Figure \ref{4eigsandlim} shows 
the largest eigenvalue of the volume operator (found by numerical 
calculation) compared to the bound stated in (\ref{4lim1}) and (\ref{4lim2}).
As we can see, the bound fits the volume eigenvalues very tightly, diverging 
significantly from them for $n<10$ and becoming practically indistinguishable 
from the maximum eigenvalue for $n>50$.  We also show, in Figure 
\ref{limeigsrat}, the ratio between the eigenvalue bound we have found and 
the maximum eigenvalue.  Note that this converges rapidly to 1.

\subsection{Eigenvalue Bounds for $n$-Valent Vertices}

We now turn our attention to the more general case of the $n$-valent 
monochromatic vertex.  In this case, we cannot use the simple formula 
in (\ref{4val}), but must instead use the more general form in 
(\ref{Weq}) or (\ref{Weq2}).  There are some simplifications that can 
be made which aid us in computing these matrix entries and their 
bounds;  however, a general analytic result is not forthcoming.  
Numerical methods, however, will prove fruitful.

First, we recall that we are examining the special case where 
$P_{r} = P_{s} = P_{t} = P_{0} = m$.  We the note that the expression for 
the 9-$j$ symbol in (\ref{Weq}) and (\ref{Weq2}) is fairly symmetric, and can 
be simplified.  This symbol corresponds to the spin network
\begin{equation}
\left\{  \begin{array}{ccc}
k_2 & m & k_3 \\
i_2 & m & i_3 \\
2 & 2 & 2
\end{array} \right\} = \eqngraph[0.7]{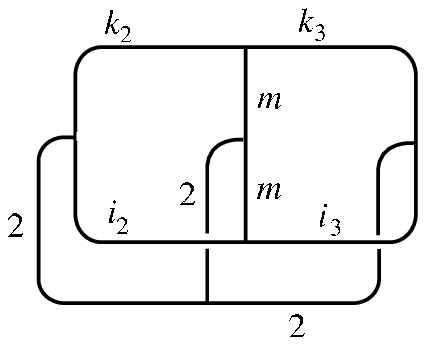}
\end{equation}
We can, however, move the graspings of the 2-edges around using 
equation (\ref{Slide}), and then close the resulting ``triangles'' using 
equation (\ref{Triangle}).  Doing this, we obtain
\begin{eqnarray*}
\left\{
\begin{array}{ccc}
k_2 & m & k_3 \\
i_2 & m & i_3 \\
2 & 2 & 2
\end{array} \right\} & = & \frac{{\lambda}_{k_3}^{i_3 2}}{m} \left(
\frac{-i_2 \, \mbox{Tet} \left[
\begin{array}{ccc}
k_2 & i_2 & i_2 \\
2 & 2 & 2
\end{array} \right]}{\theta(k_2,i_2,2)} \right. \\
 & & \quad + \left. \frac{i_3 \, \mbox{Tet} \left[
\begin{array}{ccc}
k_3 & i_3 & i_3 \\
2 & 2 & 2
\end{array} \right]}{\theta(k_3,i_3,2)} \right) \mbox{Tet} \left[
\begin{array}{ccc}
k_2 & k_3 & 2 \\
i_3 & i_2 & m
\end{array} \right]
\end{eqnarray*}
Plugging this back in, (\ref{Weq}) becomes
\begin{eqnarray*}
\lefteqn{W^{(n)}_{[rst]} {}^{k_2 \cdots k_{n-2}}_{i_2 \cdots 
i_{n-2}} = {}} \\
& & \quad - m^{3} 
\mathcal{Q} \lambda_{k_2}^{i_2 2} \delta_{i_4}^{k_4} 
\cdots \delta_{i_{n-2}}^{k_{n-2}}\nonumber \\
& & \quad \, \times \frac{\mbox{Tet} \left[ \begin{array}{ccc}
m & m & m \\
k_2 & i_2 & 2 
\end{array} \right] \mbox{Tet} \left[ \begin{array}{ccc}
m & m & k_4 \\
k_3 & i_3 & 2 
\end{array} \right] \mbox{Tet} \left[
\begin{array}{ccc}
k_2 & k_3 & 2 \\
i_3 & i_2 & m
\end{array} \right]\Delta_{k_2} \Delta_{k_3}}
{\theta(k_2, i_2, 2) \theta(k_3, i_3, 2) \theta(m, m, k_2) 
\theta(k_2, k_3, m) \theta(k_3, k_4, m)}
\end{eqnarray*}
where
\begin{equation}
\mathcal{Q} = \frac{{\lambda}_{k_3}^{i_3 2}}{m} \left(
\frac{-i_2 \, \mbox{Tet} \left[
\begin{array}{ccc}
k_2 & i_2 & i_2 \\
2 & 2 & 2
\end{array} \right]}{\theta(k_2,i_2,2)} +  \frac{i_3 \, \mbox{Tet} \left[
\begin{array}{ccc}
k_3 & i_3 & i_3 \\
2 & 2 & 2
\end{array} \right]}{\theta(k_3,i_3,2)} \right) \label{Qdef}
\end{equation}

We can also put the Tets in this resulting equation in terms of 
Wigner 6-$j$ symbols (see Appendix \ref{KLvsW});  at 
least one of the arguments of the resulting Wigner 6-$j$ symbols will 
be 1, and analytic formulas for this case have been compiled (see, 
for example, Varshalovich \cite{Varsh}.)  Equation (\ref{Weq}) then 
becomes
\begin{equation}
W^{(n)}_{[rst]} {}^{k_2 \cdots k_{n-2}}_{i_2 \cdots 
i_{n-2}} = - m^{3} 
\mathcal{Q R S} \lambda_{k_2}^{i_2 2} \delta_{i_4}^{k_4} 
\cdots \delta_{i_{n-2}}^{k_{n-2}}
\end{equation}
where $\mathcal{Q}$ is given in (\ref{Qdef}), and $\mathcal{R}$ and 
$\mathcal{S}$ are given by
\begin{equation}
    \mathcal{R} = \left\{ \begin{array}{ccc}  m/2 & k_{2}/2 & k_{3}/2 
    \\ 1 & i_{3}/2 & i_{2}/2 \end{array} \right\}_{W} \left\{ \begin{array}{ccc} 
    m/2 & m/2 & k_{2}/2 \\ 1 & i_{2}/2 & m/2 \end{array} \right\}_{W} \left\{ 
    \begin{array}{ccc} k_{4}/2 & m/2 & k_{3}/2 \\ 1 & i_{3}/2 & m/2 \end{array} 
    \right\}_{W}
\end{equation}
\begin{equation}
    \mathcal{S} = \theta(m,m,2) \sqrt{\frac{\theta(i_{2}, i_{3}, m) 
    \theta(m, m, i_{2}) \theta(i_{3}, k_{4}, m)}{\theta(k_{2}, k_{3}, m) 
    \theta(m, m, k_{2}) \theta(k_{3}, k_{4}, m)}}
\end{equation}

Finally, we note that there are certain ``selection rules'' for finding 
the non-zero elements of $W$ in this basis.  The presence of a series of 
Kronecker deltas in (\ref{Weq}) gives us the obvious selection rules that
\begin{equation} \label{latersame}
k_4 = i_4, \quad k_5 = i_5, \quad \cdots, \quad k_{n-2} = i_{n-2}
\end{equation}
It is also evident from the spin 
diagram in (\ref{9jdef}) that for the 9-$j$ symbol in (\ref{Weq}) to be 
non-zero, we must have
\begin{equation} \label{2and3}
 k_2 - i_2 = 0, \pm 2 \quad \mbox{and} \quad  k_3 - i_3 = 0, \pm 2
\end{equation}
in order to satisfy the triangle inequalities.  Finally, it can be shown that 
\begin{equation}
\frac{a}{\theta(a,b,2)} \mbox{Tet} \left[ \begin{array}{ccc} 
a & a & b \\ 
2 & 2 & 2 \end{array} \right] = \left\{ \begin{array}{l @{\quad} c} 
(-1)^{a+1} \cdot \frac{a+2}{2} & b = a-2 \\
-1 & b=a \\
(-1)^a \cdot \frac{a}{2} & b=a+2 \\
0 & \mbox{otherwise}
\end{array} \right.
\end{equation}
Hence, if $k_2 = i_2$ and $k_3 = i_3$, the quantity in parentheses 
in (\ref{Qdef}) will be $(-1 + 1)$, with the result that $\mathcal{Q} = 0$.  
We can then rewrite (\ref{2and3}) as 
\begin{equation} \label{2and3rev}
\begin{array}{c}
k_2 - i_2 = 0, \pm 2, \quad  k_3 - i_3 = 0, \pm 2, \quad \mbox{and} \\ 
k_2 - i_2 \mbox{ \& } k_3 - i_3 \mbox{ are not both zero.}
\end{array}
\end{equation}
We conclude that for a given column of the $W$ matrix 
(i.e. fixed $k_2, k_3, \ldots,$ $k_{n-2}$), there will be at most eight 
non-zero elements:
\begin{equation}
    \begin{array}{c @{\qquad} c @{\qquad} c}
    i_{2} = k_{2} - 2 & i_{2} = k_{2} - 2 & i_{2} = k_{2} - 2 \\
    i_{3} = k_{3} - 2 & i_{3} = k_{3} & i_{3} = k_{3} + 2 \\[10pt]  
    i_{2} = k_{2} & & i_{2} = k_{2} \\
    i_{3} = k_{3} - 2 & & i_{3} = k_{3} + 2 \\[10pt]
    i_{2} = k_{2} + 2 & i_{2} = k_{2} + 2 & i_{2} = k_{2} + 2 \\
    i_{3} = k_{3} - 2 & i_{3} = k_{3} & i_{3} = k_{3} + 2
    \end{array}
\end{equation}

Unfortunately, this is about as far as analysis will easily take us.  
The quantities $\mathcal{Q}$, $\mathcal{R}$, and $\mathcal{S}$ are 
fairly simple to calculate for a given set of arguments;  
however, finding the maximum product of all three, over all possible values 
of $\{k_2, k_3, k_4, i_2, i_3, i_4\}$ is a much more daunting 
task.\footnote{
Note that, in a small mercy granted to us by equation (\ref{Weq3}), the entries of this 
form of the $W$-matrix depend only on these six intertwiner strands (and 
the edge colour $m$), and not on the entire set.}
Things are complicated by the fact that both the quantities $\mathcal{Q}$ and 
$\mathcal{R}$ have slightly different analytic forms for each of the eight 
entries in a given column --- one form for $k_2 - i_2 = 2$ and $k_3 - 
i_3 = 2$, one form for $k_2 - i_2 = 2$ and $k_3 - i_3 = 0$, and so 
forth (see Varshalovich \cite{Varsh}.)  
We see, then, that an analytic answer is difficult to find.

Despite this difficulty, we can get an idea of the scaling properties of the 
absolute row and column sums by numerical analysis.  By a happy coincidence, 
we do not even need to look at all of the rows and columns of the matrix --- 
only a small subset of them.  This is because (as mentioned above) the entries 
of the $W$-matrix are only dependent on the six values $\{k_2, k_3, k_4, i_2, 
i_3, i_4\}$;  all other pairs of corresponding indices must be equal, or 
the matrix entry will be zero.  These facts imply that the $W$-matrix must 
be of block-diagonal form, and that (more importantly) each one of the 
blocks is identical:
\begin{equation}
W = \left( \begin{array}{ccccc}
A & 0 & 0 &  & 0 \\
0 & A & 0 & \cdots & 0 \\
0 & 0 & A &  & 0 \\
& \vdots & & \ddots & \vdots \\
0 & 0 & 0 & \cdots & A
\end{array} \right)
\end{equation}
where each block corresponds to a different choice of $k_5, \ldots, 
k_{n-2}$, and the indices of the matrix $A$ are all possible choices for 
$k_2$, $k_3$, and $k_4$.  We see that we need only compute the absolute 
row and column sums for the (much smaller) matrix A, instead of the full 
$W$-matrix.  This is still a non-trivial calculation, as the size of the 
matrix $A$ scales fairly quickly (for $m = 10$, the matrix has already 
grown to $891 \times 891$,) but it does simplify the problem greatly.

\begin{figure} 
    \begin{center}
    \includegraphics[width=4in]{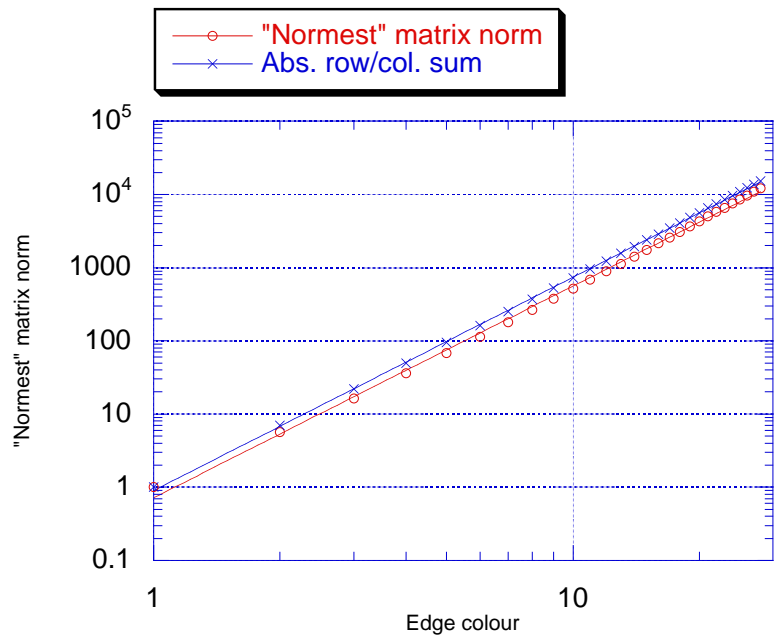}
    \caption{\label{Mvals} The eigenvalue limits found by the absolute row 
    and column sums and by MATLAB's internal \ttfamily normest \rmfamily 
    function.  The curve fits are for a power law relation.  For the 
    absolute row/column sum method, the best fit is $M = 0.90432 \cdot 
    m^{2.9126}$;  for the \ttfamily normest \rmfamily method, it is $M = 
    0.71884 \cdot m^{2.8921}$.}
    \end{center}
\end{figure}

Using this trick, combined with the selection rules in (\ref{latersame}) 
and (\ref{2and3rev}), the bounding quantity $M$ (as defined above in 
section \ref{4vbounds}) was calculated with MATLAB.  Two methods were 
used.  The first was the absolute row and column sum discussed 
previously.  The second was MATLAB's internal \ttfamily normest \rmfamily
function, which estimates the largest eigenvalue of a sparse matrix.\footnote{
For the mathematically inclined reader, we note that the absolute row and 
column sums are known as the 1-norm and the $\infty$-norm of a given 
matrix, respectively;
that which \ttfamily normest \rmfamily finds is the 2-norm.}
These results are shown in Figure \ref{Mvals}.  Each of these data sets is 
fitted to a power law relation.  We see that for both of these data sets, 
we have an approximate relation $M \propto m^3$.  

There is good reason to believe that this power dependence should tend 
towards $m^3$ (exactly) as $m \to \infty$:  if it does, then we have the 
relation
\begin{eqnarray}
\left| \lambda_V \right| & \approx & \sqrt{M \frac{n(n-1)(n-2)}{96}} 
\nonumber \\
& \approx & C \sqrt{m^3 n^3} \nonumber \\
& = & C \mbox{(total spin)}^{3/2}
\end{eqnarray}
where $C$ is a proportionality constant.  As noted in the previous 
section, this is the dependence expected from classical spatial geometry.  
While our data are not conclusive evidence of this relation, they are 
highly suggestive that this relation holds for a monochromatic 
vertex of any valence.

Thus, we have placed limits on the volume eigenvalues of 
monochromatic vertices;  in both cases, we find that volume of a 
region scales as the $3/2$ power of the area of its bounding 
surface.  This result was found analytically in the 4-valent case;  
numerical results strongly suggest that this relation holds for the 
general case as well. 
\setcounter{equation}{0}

\section{Discussion}
\label{Disc}

We have seen some interesting consequences stemming from the 
discreteness of the angle operator (see Section \ref{AngRes});  these 
results will be referred to as the \emph{small angle} property (Equation 
\ref{SmallAng}), the \emph{angular resolution} property (Equation 
\ref{AngResol}), and the \emph{angular distribution} property 
(Figures \ref{dist1}--\ref{dist4}.)  We now wish to assign some 
scaling behaviour to these predicted phenomena;  in other words, on 
what length scales do we expect these phenomena to be observable?

The small angle property is the best-quantified of these three 
properties;  we have the equation
\begin{equation}
\epsilon \geq \frac{4}{\sqrt{n_T + 8}} \geq \frac{4}{\sqrt{s_{T} + 8}} 
\mbox{,}
\end{equation} 
where $s_{T}$ is the total surface flux.
From our numerical studies of the volume operator, we also have the 
result that 
\begin{equation} \label{len2spin}
V \propto \ell^{3} \propto s_{T}^{3/2}
\end{equation}
where $\ell$ is the length scale (as defined by either the volume or 
area operator --- we have shown that these are equivalent.)
Hence, if we combine these results, we obtain the result that for 
large $s_{T}$, the 
minimum observable angle is related very simply to the length scale 
observed:
\begin{equation}
\epsilon \propto \ell^{-1} \propto s_{T}^{-1/2}
\end{equation}
Moreover, from our discussion in Section \ref{AngSpectrum}, we recall 
that for a vertex with large amounts of ``background geometry'' 
(i.e.~$s_{r} \gg s_{1}, s_{2}$), the primary limit on $n_{T}$ is not 
$s_{T}$ but $s_{1}+s_{2}$.  Hence, for these vertices we have the 
result that
\begin{equation}
    \epsilon \propto (s_{1} + s_{2})^{-1/2}
\end{equation}

Using these relations, we can also gain an idea of the absolute length 
scale on which this phenomenon might be viewed.  Recalling that
\begin{equation}
    V = l_{0}^{3} \sqrt{\frac{M}{96}} s_{T}^{3/2}
\end{equation}
where $M$ was found to be equal to $0.71884$ (numerically), we can 
state an absolute relation between the length scale and the smallest 
observable angle:
\begin{equation}
    \epsilon = k_{\epsilon} \left( \frac{\ell}{l_{0}} \right)^{-1}
\end{equation}
where $k_{\epsilon} = 4 \sqrt[6]{M/96} \approx 1.7692$.  This implies 
that if we observe angles on a length scale of $\ell \approx 10^{-15} 
\mbox{ m}$ (i.e.~ the radius of a proton), the minimum observable 
angle is then approximately $10^{-20}$ radians --- a miniscule angle 
by any account.

In Section \ref{AngResolution}, we found results for the mean
angular resolution associated with a vertex.  The most important of 
these was the average spacing between expectation values, given by
\begin{equation}
\delta \geq \frac{4 \pi}{n^2 + 3n + 6} \mbox{,}
\end{equation}
Again, combining these results with the result in (\ref{len2spin}) 
allows us to associate a scaling behaviour with the mean angular 
separation and resolution.  For large $n$, the angular separation 
scales as
\begin{equation}
\delta \propto n^{-2} \propto \ell^{-4} \mbox{,}
\end{equation}
This scaling property of the angular resolution is also somewhat striking, 
since it scales proportionally to such a high value of the total spin 
(i.e.\ inversely proportional to the fourth power of the length.)  

Using similar techniques to those found above, we can find absolute 
length scales for $\delta$ as well;  these can be shown to be
\begin{equation}
    \delta = k_{\delta} \left( \frac{\ell}{l_{0}} \right)^{-4} 
\end{equation}
where $k_{\delta} = 4 \pi (M/96)^{2/3} \approx 0.4810$.  On the length 
scale of a proton, this implies a resolution of roughly $10^{-80}$ 
radians!

Finally, we found that we could reproduce a classical angle distribution 
under the condition that
\begin{equation}
1 \ll \{ s_1, s_2 \} \ll s_r \mbox{.}
\end{equation}
This is somewhat harder to quantify;  however, we \emph{can} associate
lengths with the distributions shown in Figures \ref{dist1}--\ref{dist4}.
For Figure \ref{dist1}, we have $n_T \approx 10^6$;  hence, the length 
scale associated with such a vertex is on the order of $10^3 \: l_0 \approx
10^{-32}$ m;  for Figures \ref{dist2}--\ref{dist4}, the length scale would 
be on the order of $10^{10} \: l_0 \approx 10^{-25}$ m.  For reference, 
the diameter of an atomic nucleus is on the order of $10^{-15}$ m, a full 
17 orders of magnitude larger than the length scale associated with the 
``drastically altered'' distribution of Figure \ref{dist1}.  Even on a 
length scale ten orders of magnitude less than the size of a nucleus (as 
in Figures \ref{dist2}--\ref{dist4}), the classical angle spectrum is 
reproduced.

We must be careful, however, that we do not read too much into these diagrams.  The 
main difference between the conditions for the normal and skewed angle 
distributions is that $s_{1}, s_{2} \ll s_{r}$ for the former while 
$s_{1}, s_{2} \approx s_{r}$ for the latter.  To gain a better 
understanding of where the transition between these two types of 
distribution actually occurs, we would have to examine the 
distributions over a wide range of values of $s_{1}/s_{r}$ and 
$s_{2}/s_{r}$;  the results presented here only show two values of 
$s_{1}/s_{r}$ and three of $s_{2}/s_{r}$ (not counting the minor difference 
between Figures \ref{dist2} and \ref{dist3}).

Moreover, while the skewed angle distribution in Figure \ref{dist1}
certainly occurs for $s_{1}, s_{2}, s_{r} \approx 
10^{5}$, there is no reason to believe that this behaviour will go 
away for larger values of the spins.  If this behaviour does indeed 
persist at higher spins, our interpretation of the angle operator may 
be incorrect.  We recall that the angle operator finds the angle between
the two patches $S_{1}$ and $S_{2}$;  perhaps all this distribution 
is telling us is that these patches don't ``want'' to overlap;  as 
they get larger, they will have to spread themselves out to avoid each 
other.  However, this is purely conjectural;  more research is needed 
to find out whether this problem even arises.

The true importance of our angle distribution results becomes clearer 
when we consider them in light of the small-angle phenomenon.  We 
recall that the smallest angle observable about a vertex depends on 
$s_{T}$ if $s_{1}, s_{2} \approx s_{r}$, but on $s_{1} + s_{2}$ if 
$s_{1}, s_{2} \ll s_{r}$.  To obtain a classical angle distribution, 
the latter of these conditions must hold;  the magnitude of $s_{1}$ and 
$s_{2}$ then determines the minimum observable angle.  Thus, if we 
have a vertex of a given spin, we have to balance two factors to 
obtain a classical angle distribution:  the need to assign enough 
spin to $s_{r}$ to obtain the correct distribution, along with the 
need to assign enough spin to $s_{1}$ and $s_{2}$.  Hence, to 
approximate classical angles in both of these ways, the total spin of the 
vertex must be sufficiently high. 
\setcounter{equation}{0}

\section{Conclusion}

The search for a quantum theory of gravity is one of the major 
challenges facing modern physics today.  One of the more promising 
approaches in this search is the canonical or \emph{loop quantum 
gravity} approach;  in this theory, the fundamental variables of the 
theory are not field variables, but path integrals of field variables around 
closed loops in space.  A difficulty arises from the properties of 
these loops:  a given set of loops may satisfy any number of linear 
relations between themselves, making it difficult to compare two 
arbitrary states.  This difficulty can be resolved through the use 
of \emph{spin networks}, whose properties were described in Section 
\ref{SpinNet}.

Since general relativity predicts the structure of space, these spin 
networks must do so as well.  This is done through Hermitian operators 
on spin network states, corresponding to ``spatial observables'' such 
as area, volume, and angle.  While the forms of these last two 
operators are known, their properties have not been extensively 
studied.  

Our research into the properties of these operators has revealed 
several peculiarities.  In particular, we have shown that the minimum 
observable angle about a vertex with total spin $s_{T}$ scales 
roughly as $\epsilon \propto s_{T}^{-1/2}$, a fairly slow decrease to 
zero.  Moreover, since the angle operator has a discrete spectrum, 
the possible observable angles will be separated from each other.  
We have shown that the average separation between adjacent angles in the 
spectrum decreases as $\delta \propto s_{T}^{-2}$.  Finally, we have 
shown that for ``sufficiently classical'' vertices (i.e.~$1 \ll 
\{s_{1}, s_{2}\} \ll s_{r}$), the classical angle distribution of 
$\mathcal{P} (\theta) \, \dif \theta = \sin \theta \, \dif \theta$ is 
reproduced.  However, in the case where $s_{1}, s_{2} \approx s_{r}$, 
the angle distribution is vastly skewed from its normal behaviour.

Our studies of the volume operator were hindered by its complexity.  
However, we have found (through a combination of analytical and numerical
methods) evidence suggesting that the volume associated with 
a vertex is proportional to the $3/2$ power of a surface surrounding 
that vertex.  This is identical to the scaling properties of these 
quantities in the classical model of space;  hence, this result is a 
major step towards showing that spin networks can fully approximate 
classical continuum space.

The true measure of any theory, of course, is not how well it predicts 
already-observed effects but whether it predicts any new, hitherto 
unexpected phenomena.  In this sense, our angle results are more 
important than our volume result.  The latter result is still 
important, of course;  if we had found any other dependence than 
$(\mathrm{volume}) \propto (\mathrm{area})^{3/2}$, we would have had 
to discard our current volume or area operator --- if not the entire 
theory --- as untenable.  However, the fact that our angle results predict 
unexpected phenomena (the small-angle phenomenon and the skewed angle 
distribution) makes these results rather more interesting.

These phenomena imply that the angle operator predicts a drastically 
different structure of space for low-valence vertices.  Unfortunately, 
such vertices are very far removed from our current realm of 
observation, even if we consider the next generations of 
observational equipment.  To get a rough idea of how far away we are 
from experimentally probing the predicted properties of spin 
networks, we can note that to do so would require a particle with a 
de Broglie wavelength on the order of the Planck length.  
Hence, such a particle would have a momentum
\begin{equation}
    p_{0} = \frac{h}{l_{0}} \approx 3 \times 10^{28} \: \mathrm{eV/c}
    \approx 16 \: \mathrm{kg \cdot m/s}.
\end{equation}
While this may not seem like terribly much momentum on a macroscopic 
scale, it is a tremendous amount of energy for an elementary particle 
to have.   If we make the assumption that the rest mass of such a particle 
is very small (i.e.~$m_{0} \ll p_{0}/c \approx 50 \: \mu 
\mathrm{g}$), its energy will be (to first order) determined by its 
momentum only:  
\begin{equation}
    E = p_{0} c \approx 3 \times 10^{28} \: \mbox{eV} \approx 5 \: \mbox{GJ.}
\end{equation}
This is a stupendous amount of energy;  for comparison, the kinetic 
energy of a cruising Boeing 747 is roughly 12 gigajoules. We cannot 
possibly hope, then, to probe these distances directly at any point in 
the near future.

Because of this, the skewed angle distribution result is rather 
interesting:  it predicts a change in angle distributions at a scale 
three orders of magnitude larger than the Planck length.  This would 
bring down the required energy of a particle required to see this 
phenomenon to approximately $10^{-25} \: \mathrm{eV}$ --- still well beyond our 
observational capability, but a good deal closer.  However, as 
mentioned in Section \ref{Disc}, this result is still open to 
interpretation.  The normal distributions shown in Figures 
\ref{dist2}--\ref{dist4}, however, do place an upper bound on the 
length scales on which quantum angle effects can be observed --- by 
the time we have reached a length scale of $10^{-25}$ m, the quantum angle 
spectrum approximates the classical angle spectrum very well.

In the face of this, the small-angle property may end up being the 
``best'' prediction stemming from our results.  This is not to 
denigrate the other results we have found;  the angle distribution and 
volume results show that the angle and volume operators do satisfy 
``classical'' relations on a large scale.   These are important 
properties for an quantum theory of general relativity to have, and 
are important steps towards showing that loop quantum gravity is a 
self-consistent example of such a theory. 
\setcounter{equation}{0}

\appendix
\renewcommand{\theequation}{\Alph{section}.\arabic{equation}}

\section*{Appendices}
\addcontentsline{toc}{section}{Appendices}

\section{Dirac-Bergmann Constraint Analysis}
\label{BDCA}

\subsection{The Generalized Hamiltonian}
\label{GenHam}

To canonically quantize a classical theory, we need first to find a 
general expression for its Hamiltonian.  It is fairly simple to find one 
Hamiltonian for a given system;  however, our objective is to find an 
expression for all the possible Hamiltonian of a given theory.  Such an 
expression should permit arbitrary variations that are implicit in the 
theory (e.g.\ gauge freedom in classical electrodynamics.)  The following 
derivation of the generalized Hamiltonian follows from 
Dirac~\cite{DiracLec}.

To construct the generalized Hamiltonian, it is easiest to start from the 
point of the action integral:
\begin{equation}
S = \int L(q_{1}, q_{2}, \ldots, q_{N}, {\dot{q}}_{1}, {\dot{q}}_{2}, 
\ldots, {\dot{q}}_{N} ) \,dt
\end{equation}
where $\{q_{1}, q_{2}, \ldots, q_{N}\}$ are the coordinates of the 
configuration space of the system, and $\{{\dot{q}}_{1}, {\dot{q}}_{2}, 
\ldots, {\dot{q}}_{N}\}$ are their time derivatives.  By applying the 
calculus of variations to this integral, we obtain the Lagrangian 
equations of motion:
\begin{equation} \label{Leom}
\frac{\partial L}{\partial q_{n}} - \frac{d}{dt}\Big(\frac{\partial 
L}{\partial {\dot{q}}_{n}}\Big) = 0
\end{equation}
Note that there are $N$ such equations, one for each coordinate of the 
configuration space.  

The Hamiltonian formalism first requires us to change variables from the 
velocities ${\dot{q}}_{n}$ to the conjugate momenta $p_{n}$:
\begin{equation} \label{pDef}
p_{n} = \frac{\partial L}{\partial {\dot{q}}_{n}}
\end{equation}
These momenta will satisfy the Poisson bracket relation
\begin{equation}
\{ q_{n}, p_{m} \} = \delta_{nm},
\end{equation}
where the \emph{Poisson bracket} of two quantities $f$ and $g$ is defined as
\begin{equation}
\{ f, g \}  = \frac{\partial f}{\partial q_{n}} \frac{\partial g}{\partial 
p_{n}} - \frac{\partial f}{\partial p_{n}} \frac{\partial g}{\partial 
q_{n}} \mbox{.}
\end{equation}
It is possible that these momenta are all independent of one another;  
however, this is not required in general.  As a simple example, consider 
a disc rolling without slipping on a plane.  The disc's 
linear momentum $p_{x}$ is not independent of its angular momentum 
$p_{\theta}$.

There will therefore be a certain number of constraints on the coordinates 
and momenta.  These constraints can be written in the form
\begin{equation} \label{PrimCons}
{\phi}_{m}(q, p) = 0 \qquad m = 1, 2, \ldots, \mathcal{M}
\end{equation}
where $q$ and $p$ denote any combination of the coordinates and the 
conjugate momenta.  The expressions in (\ref{PrimCons}) are called the 
\emph{primary constraints} of the formalism.

Let us now consider the quantity $H = p_{n} {\dot{q}}_{n} - L$.  We first 
note that an arbitrary variation in this quantity can be expressed as
\begin{eqnarray} \label{Hamvar}
\nonumber \delta H & = & \delta p_{n} {\dot{q}}_{n} + p_{n} \delta 
{\dot{q}}_{n} - \frac{\partial L}{\partial q_{n}} \delta q_{n} - 
\frac{\partial L}{\partial {\dot{q}}_{n}} \delta {\dot{q}}_{n} \\
& = & \delta p_{n} {\dot{q}}_{n} - \frac{\partial L}{\partial q_{n}} 
\delta q_{n} \mbox{,}
\end{eqnarray}
Thus, a variation in $H$ can be expressed solely in terms of variations in 
$p_{n}$ and $q_{n}$;  this implies that $H$ is solely a function of the 
coordinates and their conjugate momenta, and not of their velocities 
${\dot{q}}_{n}$.

This quantity $H$ is called the \emph{Hamiltonian} of the system.  
However, it is not uniquely determined.  Since the primary constraints are 
identically zero, we can add any combination of them to the Hamiltonian 
without changing the Hamiltonian's value:
\begin{equation} \label{Ham+phi}
H^{*} = H - u_{m} {\phi}_{m}
\end{equation}
where the coefficients $u_{m}$ are arbitrary.  (The negative sign is 
chosen merely for notational convenience later.)  Using this new 
Hamiltonian, we can then deduce the generalized Hamiltonian equations of 
motion from (\ref{Hamvar}) and (\ref{Ham+phi}):
\begin{eqnarray*}
\frac{\partial H^{*}}{\partial p_{n}} & = & {\dot{q}}_{n} - u_{m} 
\frac{\partial {\phi}_{m}}{\partial p_{n}} - \frac{\partial 
u_{m}}{\partial p_{n}} {\phi}_{m} \\
\frac{\partial H^{*}}{\partial q_{n}} & = & -\frac{\partial L}{\partial q_{n}} 
- u_{m} \frac{\partial {\phi}_{m}}{\partial q_{n}} - \frac{\partial 
u_{m}}{\partial q_{n}} {\phi}_{m}
\end{eqnarray*} 
Noting that the primary constraints vanish and applying (\ref{Leom}) to 
the second equation, these equations lead to the \emph{Hamiltonian 
equations of motion}:
\begin{eqnarray}
\label{qdot} {\dot{q}}_{n} & = & \frac{\partial H}{\partial p_{n}} + u_{m} 
\frac{\partial {\phi}_{m}}{\partial p_{n}} \\
\label{pdot} {\dot{p}}_{n} & = & -\frac{\partial H}{\partial q_{n}} - 
u_{m} 
\frac{\partial {\phi}_{m}}{\partial q_{n}}
\end{eqnarray}

We can rewrite these equations in a fairly simple form using Poisson 
bracket notation.  The time evolution of any quantity $g(q,p)$ is given by
\begin{equation}
\dot{g} = \frac{\partial g}{\partial q_{n}} {\dot{q}}_{n} + \frac{\partial 
g}{\partial p_{n}} {\dot{p}}_{n}
\end{equation}
Applying (\ref{qdot}) and (\ref{pdot}), we see that
\begin{eqnarray}
\nonumber \dot{g} & = & -\frac{\partial g}{\partial 
p_{n}}\Big(\frac{\partial 
H}{\partial q_{n}} + u_{m}\frac{\partial {\phi}_{m}}{\partial q_{n}}\Big) 
+ 
\frac{\partial g}{\partial q_{n}}\Big(\frac{\partial  H}{\partial p_{n}} + 
u_{m}\frac{\partial {\phi}_{m}}{\partial p_{n}}\Big) \\
\nonumber & = & \bigg(\frac{\partial g}{\partial q_{n}}\frac{\partial 
H}{\partial p_{n}} - \frac{\partial g}{\partial p_{n}}\frac{\partial 
H}{\partial q_{n}}\bigg) + \bigg[\frac{\partial g}{\partial 
q_{n}}\Big(u_{m}\frac{\partial {\phi}_{m}}{\partial p_{n}} + {\phi}_{m}
\frac{\partial u_{m}}{\partial p_{n}}\Big) \nopagebreak \\
\nonumber & & \qquad + \frac{\partial g}{\partial p_{n}}\Big(u_{m}
\frac{\partial {\phi}_{m}}{\partial q_{n}} + {\phi}_{m} \frac{\partial 
u_{m}}
{\partial q_{n}}\Big)\bigg] \\
\nonumber & = & \frac{\partial g}{\partial q_{n}}\frac{\partial (H + u_{m} 
{\phi}_{m})}{\partial p_{n}} - \frac{\partial g}{\partial 
p_{n}}\frac{\partial 
(H + u_{m} {\phi}_{m})}{\partial q_{n}} \\
\label{Hcom1} & = & \{ g, H_{T} \}
\end{eqnarray}
where $H_{T} = H + u_{m} {\phi}_{m}$ is the \emph{total 
Hamiltonian}.  Note that the extra terms introduced in the second step 
are identically zero.

This formulation of the Hamiltonian equations of motion is concise;  
however, it does pose a potential pitfall.  Our primary constraints 
${\phi}_{m}$ are, by definition, equal to zero;  however, if we were to 
apply this equality immediately, before working out the Poisson brackets 
in (\ref{Hcom1}),  we would lose the generality that they allow.  To 
clarify this distinction, we introduce the concept of ``weak equality'', 
denoted
\begin{equation}
{\phi}_{m} \approx 0
\end{equation}
This notation signifies that one may only make use of this relation after 
the Poisson brackets have been worked out.  Using this notation, then, 
(\ref{Hcom1}) becomes 
\begin{equation}
\label {Hcom2} \dot{g} \approx \{ g, H_{T} \} \mbox{.}
\end{equation}

We can use (\ref{Hcom2}) to check for further consistency conditions of 
the theory.  If we apply (\ref{Hcom2}) to one of the primary constraints 
${\phi}_{m}$, we have
\begin{equation} \label{conscond} 
{\dot{\phi}}_{m} \approx 0 \approx \{ {\phi}_{m}, H \} + u_{m'} \{ 
{\phi}_{m}, {\phi}_{m'} \}
\end{equation}
Thus, for each value of $m$, we have a consistency condition.  Each one of 
these conditions leads to one of three results:\footnote
{It is possible that one or more of these equations lead directly to an 
inconsistency (i.e.\ an equation of the form $1=0$);  in this case, our 
original Lagrangian equations of motion were inconsistent, and the theory 
must be modified or rejected.}
\begin{itemize}
\item The equations in (\ref{conscond}) could lead to an equation that is 
trivially satisfied, i.e. of the form $ 0 = 0$.
\item The consistency conditions could reduce to an equation that contains 
only coordinates and momenta.  Such a condition must be independent of the 
primary constraints (otherwise, it would be identically zero and therefore 
be of the first type.)  We therefore have a new set of \emph{secondary 
constraints}:
\begin{equation}
{\phi}_{k}(q, p) \approx 0 \qquad k = \mathcal{M} +1, \mathcal{M} +2, 
\ldots, 
\mathcal{K}
\end{equation}
\item  Finally, the consistency conditions may not reduce in either of the 
above ways;  in this case, they impose a constraint on the unknown 
functions $u_{m}$.
\end{itemize}

Secondary constraints may themselves generate other constraints by further 
applications of (\ref{Hcom2}), i.e.
\begin{equation}
\{ {\phi}_{k}, H\} + u_{m} \{ {\phi}_{k}, {\phi}_{m}\} \approx 0,
\end{equation}
where $ k > \mathcal{M}$.  These new constraints may be of any one of the
three types listed above.  
If they turn out to be new secondary constraints, we can continue to apply 
(\ref{Hcom2}) recursively to them.  In the end, we are left with our 
original set of primary constraints ${\phi}_{m}$, a new set of secondary 
constraints ${\phi}_{k}$, and a set of equations of the third type 
relating the $u_{m}$'s.

Let us further examine these equations of the third type.  Each of these 
equations is of the form
\begin{equation} \label{3rdType}
\{ {\phi}_{j}, H \} + u_{m} \{ {\phi}_{j}, {\phi}_{m} \} \approx 0 \qquad 
j = 1, 2, \ldots, \mathcal{K}
\end{equation}
We can consider (\ref{3rdType}) to be a system of $\mathcal{K}$ linear 
non-homogeneous equations, with the $u_{m}$'s as variables and 
coefficients in terms of the commutators of the primary and secondary 
constraints.  Since these coefficients are in terms of $p$'s and $q$'s, we 
conclude that the $u_{m}$'s must also be functions of the coordinates and 
the momenta:
\begin{equation}
u_{m} = U_{m}(p,q)
\end{equation}
A complete solution for these variables also includes an arbitrary linear 
combination of the independent solutions (if any) to the homogeneous 
system 
of equations
\begin{equation} \label{VConsts}
V_{m} \{ {\phi}_{j}, {\phi}_{m} \} = 0\mbox{.}
\end{equation}
Hence, the general solution to this system of equations is
\begin{equation}
u_{m} = U_{m} + v_{a}V_{am}\mbox{,}
\end{equation}
where $a$ sums over the solutions found in (\ref{VConsts}).  
The total Hamiltonian can then be written as 
\begin{equation}
\label{RealH} H_{T} = H' + v_{a}{\phi}_{a}
\end{equation}
where
\begin{equation}
\label{HPrime} H' = H + U_{m}{\phi}_{m}
\end{equation}
and
\begin{equation}
{\phi}_{a} = V_{am}{\phi}_{m}\mbox{.}
\end{equation}
It is important to note that while the original $u_{m}$ coefficients were 
not necessarily independent of one another, the new coefficients $v_{m}$ 
in (\ref{RealH}) are totally independent.\footnote
{In general, there will be fewer of the $v_{m}$'s than of the $u_{m}$'s.}

At this point, we introduce a bit of nomenclature.  We call a dynamical 
variable $R$ \emph{first-class} if its Poisson brackets with every 
constraint (primary and secondary) are weakly zero:
\begin{equation} \label{1stClass}
\{ R, {\phi}_{j} \} \approx 0
\end{equation}
An expression that is weakly zero must be equal to some linear combination 
of the constraints;  hence, we have
\begin{equation}
\{ R, {\phi}_{j} \} = c_{jj'} {\phi}_{j'}
\end{equation}
If $R$ does not satisfy (\ref{1stClass}) for each ${\phi}_{j}$, we say 
that $R$ is \emph{second-class}.  The ``class'' of a constraint is 
independent 
of whether it is a primary or secondary constraint;  hence, we now have a 
total 
of four types of constraints.

Let us again consider the arbitrary coefficients $v_{a}$ in 
(\ref{RealH}).  These new coefficients enter into the equations of motion:
\begin{equation}
\dot{g} \approx \{ g, H' + v_{a}{\phi}_{a} \}
\end{equation}
However, a classical theory such as ours should be deterministic;  we must 
therefore inquire as to the meaning of these arbitrary coefficients in the 
equations of motion.  

Initially, the state of the system is determined solely by the coordinates 
and momenta, not by the $v_{a}$'s.  If the theory is deterministic, then 
this state uniquely determines the system's later state.  However, since 
the evolution of the coordinates and momenta depends on the values of the 
coefficients $v_{a}$, certain values of coordinates and momenta must 
correspond to the same physical state.  To determine the relations between 
these values, we consider an infinitesimal time evolution in any dynamical 
variable $g$:
\begin{eqnarray}
\nonumber {\delta}_{t} g & = & g_{0} + \dot{g} \delta t \\
\nonumber & = & g_{0} + \{ g, H_{T} \} \delta t \\
& = & g_{0} + \delta t (\{ g, H'\,\} + v_{a}\{ g, {\phi}_{a} \})
\end{eqnarray}
The $v_{a}$ coefficients are completely arbitrary;  thus, we change them 
to obtain a new ${\delta}_{t} g$.  Since our theory is deterministic, 
these two new $g$'s must correspond to the same state;  therefore, the 
difference between them must correspond to a transformation that does not 
change the state:
\begin{eqnarray}
\nonumber \Delta g & = & \delta t(v_{a} - v'_{a}) \{ g, {\phi}_{a} \} \\
& \equiv & {\epsilon}_{a} \{ g, {\phi}_{a} \}
\end{eqnarray}

We can therefore change $g$ using this rule, and the physical state will 
remain the same.  These are known as {\em infinitesimal canonical 
transformations}, and the ${\phi}_{a}$'s are their generators.\footnote 
{For our purposes, we will consider an infinitesimal canonical 
transformation 
to be a transformation that adds an arbitrary infinitesimal (possibly a 
function of the original variable) to $g$, e.g.
\begin{equation}
g \mapsto g + {\epsilon}_{a} \{ g, {\phi}_{a} \}
\end{equation}
In the above equation, ${\phi}_{a}$ is called the \emph{generator} of the 
infinitesimal contact transformation.  For further explanation of 
canonical 
transformations, consult Goldstein~\cite{Goldstein}.}

Suppose we apply two such transformations successively.  We obtain
\begin{eqnarray}
\nonumber g & \mapsto & g + {\epsilon}_{a} \{ g, {\phi}_{a} \} \\
& \mapsto & g + {\epsilon}_{a} \{ g, {\phi}_{a} \} + \gamma_{a'} \{ g + 
{\epsilon}_{a} \{ g, {\phi}_{a} \}, {\phi}_{a'} \} \equiv g'
\end{eqnarray}
If we apply them in the opposite order instead, we get
\begin{equation}
g \mapsto g + {\gamma}_{a'} \{ g, {\phi}_{a'} \} + {\epsilon}_{a} \{ g + 
{\gamma}_{a'} \{ g, {\phi}_{a'} \}, {\phi}_{a} \} \equiv g''
\end{equation}
Since these two values of $g$ correspond to the same physical state, a 
transformation corresponding to their difference must also be an 
infinitesimal canonical transformation that is not associated with a 
change 
of state:
\begin{eqnarray}
\nonumber g' - g'' & = & {\epsilon}_{a} \{ g, {\phi}_{a} \} + 
{\gamma}_{a'} \{ g, 
{\phi}_{a'} \} + {\gamma}_{a'}{\epsilon}_{a} \{ \{ g, {\phi}_{a} \} , 
{\phi}_{a'} \} \\
\nonumber & & \qquad {} - {\gamma}_{a'} \{ g, {\phi}_{a'} \} - 
{\epsilon}_{a} \{ g, {\phi}_{a} \} - {\epsilon}_{a}{\gamma}_{a'} \{ \{ g, 
{\phi}_{a'} \},
{\phi}_{a} \} \\
\nonumber & = & {\epsilon}_{a}{\gamma}_{a'}( \{ \{ g, {\phi}_{a} \}, 
{\phi}_{a'} \} + 
 \{ \{{\phi}_{a'}, g \}, {\phi}_{a} \}) \\
& = & {\epsilon}_{a}{\gamma}_{a'}( \{g, \{{\phi}_{a}, {\phi}_{a'} \} \})
\end{eqnarray}

We see that the commutators of the ${\phi}_{a}$'s can also generate 
infinitesimal canonical transformations.  However, it is important to note 
that while the original set of canonical transformations were generated by 
primary constraints, the new transformations generated by the commutators 
are not necessarily primary constraints.

This leads us to the \emph{extended Hamiltonian} $H_{E}$, which consists 
of 
the total Hamiltonian $H_{T}$ plus any linear combination of expressions 
$\phi_{a'}$ which generate 
infinitesimal canonical transformations not associated with a change of 
state:
\begin{equation}
H_{E} = H_{T} + v'_{a'}{\phi}_{a'}
\end{equation}
We will find in the next section that in the case of electrodynamics, 
these canonical transformations will correspond to gauge 
transformations.  This last equation, then, is the most general 
expression for the Hamiltonian of a system.

%
%   SUBSECTION BEGINS HERE
%

\subsection{Dirac-Bergmann Analysis \& Electrodynamics}

As an example of Dirac-Bergmann analysis, let us attempt to 
construct the generalized Hamiltonian for classical electrodynamics.  We 
saw in Section \ref{CanElec} that the \emph{field tensor} is given by 
\begin{equation}
F_{\mu \nu} = {\partial}_{\mu} A_{\nu} - {\partial}_{\nu} A_{\mu}\mbox{.}
\end{equation}
In terms of this tensor, the action can be shown to be
\begin{equation} \label{1stAction}
S[A_{\mu}] = -\frac{1}{4} \int F^{\mu \nu} F_{\mu \nu} \, \dif^{4}\mathbf{x}
\end{equation}
where the integral is taken over a region of spacetime.

Constructing the momenta from the action requires a Lagrangian;  
normally this Lagrangian is related to the action by the relation 
$S = \int L \, \dif t$.  To obtain our Lagrangian, then, we must define 
a reference frame;  this defines a $t$-direction on spacetime.
We can then split the four-dimensional integral in 
(\ref{1stAction}) into a double integral over space and time:
\begin{equation}
S = -\frac{1}{4} \int \!\!\! \int F^{\mu \nu} F_{\mu \nu} \, 
\dif^{3}\mathbf{x}\,\dif t
\end{equation}
The Lagrangian is then the spatial part of this equation:
\begin{equation} \label{MaxLag}
L = -\frac{1}{4} \int F^{\mu \nu} F_{\mu \nu} \, \dif^{3}\mathbf{x}
\end{equation}
In a simple particle-dynamics system, we find the momenta by taking the 
derivative of the Lagrangian with respect to velocities, as in 
(\ref{pDef}).  Taking the derivative with respect to a field, however, is 
less straightforward.  

To find these momenta, we first define the \emph{Lagrangian density} as a scalar 
field over all space:
\begin{equation}
\mathcal{L}(\mathbf{x}) = -\frac{1}{4}F^{\mu \nu}(\mathbf{x}) F_{\mu \nu}
(\mathbf{x})
\end{equation}
Note that this is simply the integrand of (\ref{MaxLag}).  The momenta, which 
we rather suggestively label $E^{\mu}$, are then defined by
\begin{equation}
E^{\mu}(\mathbf{x}) = \frac{\partial \mathcal{L}(\mathbf{x})}{\partial 
{\dot{A}}_{\mu}(\mathbf{x})}
\end{equation}
To find these explicitly, we split the Lagrangian density into 
time-dependent and non-time-dependent parts:
\begin{equation}
\mathcal{L} = -\frac{1}{4}(F^{00} F_{00} + F^{a0} F_{a0} + F^{0a} F_{0a} + 
F^{ab} F_{ab})
\end{equation}
The first term of this expression vanishes, since $F_{\mu\nu}$ is 
anti-symmetric.  Moreover, the last term does not contain any terms 
containing the velocities (i.e.\ terms of the form ${\partial}_{0} 
A_{\mu}$), so we can discard it when taking the derivatives:
\begin{eqnarray*}
E^{\mu} & = & \frac{\partial \mathcal{L}}{\partial {\dot{A}}_{\mu}} \\
& = & -\frac{1}{4} \frac{\partial}{\partial {\dot{A}}_{\mu}} (F^{a0} 
F_{a0} + F^{0a} F_{0a}) \\
& = & -\frac{1}{2} \frac{\partial}{\partial {\dot{A}}_{\mu}} (F^{a0}
({\partial}_{a} A_{0} - {\partial}_{0} A_{a}) + F^{0a} ({\partial}_{0} 
A_{a} - {\partial}_{a} A_{0}))
\end{eqnarray*}
Recalling that ${\dot{A}}_{\mu} = {\partial}_{0} A_{\mu}$ and that 
$F^{0a} = -F^{a0}$, this reduces to
\begin{equation}
E^{\mu} = F^{\mu 0}\mbox{.}
\end{equation}

We see now why the symbol $E^{\mu}$ was used to denote the momenta:  the 
momenta turn out to be the components of the electric field.  
Moreover, since $E^{0} = F^{00} = 0$, we can discard the 
$t$-component of this vector;  hence, we will also denote the 
electric field as $E^{a}$.  Moreover, since $E^{0} = 0$, its conjugate 
variable $A_{0}$ is not physically significant (i.e.~$A_{0}$ is a 
cyclic variable);  hence, we will denote the configuration variable 
as $A_{a}$ as well.  However, it is important to note that these 
constraints are only weakly equal to 0;  to retain all generality, we 
will have to retain $E^{0}$ and $A_{0}$ until we have found the 
secondary constraints.

The Poisson brackets of the configuration variables $A_{a}$ and 
$E^{a}$ will then satisfy the relations
\begin{equation}
    \begin{split}
    \left\{E^a(x), E^b(y) \right\} & = 0 \\
    \left\{A_a(x), A_b(y) \right\} & = 0 \\
    \left\{A_a(x), E^b(y) \right\} & = \delta_a^b \delta^3(x-y)
    \end{split}
\end{equation}

Now that we have the momenta, we can construct the Hamiltonian for the 
electromagnetic field:
\begin{eqnarray*}
H & = & \int E^{\mu} {\partial}_{0} A_{\mu} \, \dif^{3}\mathbf{x} - 
L \\
& = & \int F^{a0} {\partial}_{0} A_{a} + \frac{1}{4}(F^{00} 
F_{00} + 
F^{a0} F_{a0} + F^{0a} F_{0a} + F^{ab} F_{ab}) \, \dif^{3}\mathbf{x}) \\
& = & \int F^{a0} ({\partial}_{a} A_{0} - F_{a0}) + 
\frac{1}{2} F^{a0} F_{a0} + \frac{1}{4} F^{ab} F_{ab} \, \dif^{3}\mathbf{x}
\end{eqnarray*}
where we have used the relation $\partial_{0} A_{a} = \partial_{a} 
A_{0} - F_{a0}$ in the first term and collected two terms together 
to make the second (through the symmetry of $F$.)  Rearranging and 
combining terms, we have
\begin{eqnarray}
\nonumber H & = & \int \frac{1}{4} F^{ab} F_{ab} - \frac{1}{2} F^{a0} 
F_{a0} + F^{a0} {\partial}_{a} A_{0} \, \dif^{3}\mathbf{x} \\
\label{EMHamil} & = & \int \frac{1}{4} F^{ab} F_{ab} + \frac{1}{2} E^{a} 
E_{a} 
- A_{0} {\partial}_{a} E^{a} \, \dif^{3}\mathbf{x} + \int A_{0} E^{a} n_{a} \,
\dif^{2}\mathbf{x}
\end{eqnarray}
Note that we have applied an integration by parts in this last step, 
creating the last term of the three-dimensional integral and the 
surface integral;  the sign of the second term also changes since 
$F^{a0} = - E^{a}$.  This shows that the Hamiltonian is not dependent on the velocities 
${\dot{A}}_{\mu}$, as desired.  The final term, in which $n_{a}$ is a 
vector normal to the surface enclosing the region being integrated, can be ignored if we 
assume that the strength of the electric field falls off at least as fast 
as $1/r^{2}$.

Our next step is to examine the constraints.  We have already found 
one constraint from the antisymmetry of the field tensor, namely that
\begin{equation} \label{EMPrimC}
E^{0}(\mathbf{x}) \approx 0\mbox{.}
\end{equation}
We note that this is a constraint on $E^{0}(\mathbf{x})$ at every point in 
the 
three-space under consideration.  To find the secondary constraint 
associated 
with this primary constraint, we examine the time evolution of $E^{0}
(\mathbf{x})$:
\begin{eqnarray*}
{\dot{E}}^{0}(\mathbf{x}) & \approx & 0 \\
{[E^{0}, H]} & \approx & 0 \\
\frac{\partial E^{0}(\mathbf{x})}{\partial A_{0}(\mathbf{x})} \frac 
{\partial H}{\partial E^{0}(\mathbf{x})} - \frac{\partial 
E^{0}(\mathbf{x})}
{\partial E^{0}(\mathbf{x})} \frac {\partial H}{\partial 
A_{0}(\mathbf{x})} 
& = & 0 \\
\frac {\partial}{\partial A_{0}(\mathbf{x})} \Big(\int A_{0}(\mathbf{x}) 
{\partial}_{a} E^{a}(\mathbf{x}) \, \dif^{3}\mathbf{x}\Big) & = & 0 \\
\int \Big(\frac{\partial A_{0}(\mathbf{x})}{\partial A_{0}(\mathbf{x})}
{\partial}_{a} E^{a}(\mathbf{x}) + A_{0}(\mathbf{x}) \frac{\partial} 
{\partial 
A_{0}(\mathbf{x})}({\partial}_{a} E^{a}(\mathbf{x}))\Big)\, 
\dif^{3}\mathbf{x} & = & 0
\end{eqnarray*}
The second term is zero, since the momenta are totally independent of the 
coordinates.  Since the integral of the first term must be zero for a 
variation of the field at any point in space $\mathbf{x}$, it follows that 
the secondary constraint associated with (\ref{EMPrimC}) is
\begin{equation}
{\partial}_{a} E^{a}(\mathbf{x}) \approx 0 \mbox{.}
\end{equation}
The reader may note that this is Gauss' law in the absence of 
charge;  this makes sense, since we have been examining the 
source-free versions of the Maxwell equations.
If we examine the time evolution of this new constraint, we find that the 
equation reduces identically to zero;  hence, we have one primary 
constraint and one secondary constraint for this Hamiltonian.  It can be 
shown that these constraints are both first-class.

The total Hamiltonian of the electromagnetic field is then given by
\begin{equation}
H_{T} = \int \frac{1}{4} F^{ab} F_{ab} + \frac{1}{2} E^{a} E_{a} \, 
\dif^3 \mathbf{x} - \int A_{0} {\partial}_{a} E^{a} \, \dif^3 \mathbf{x} + 
\int v(\mathbf{x}) E^{0} \, \dif^3 \mathbf{x}\mbox{,}
\end{equation}
where $v(\mathbf{x})$ is an arbitrary scalar function over all of the 
region of integration.  To obtain the extended Hamiltonian, we add in the 
secondary constraints:
\begin{equation}
H_{E} = H_{T} + \int u(\mathbf{x}) {\partial}_{a} E_{a} \, 
\dif^3 \mathbf{x}
\end{equation}
where $u(\mathbf{x})$ is another arbitrary scalar function.  We note the 
similarity of this term to the third term of the total Hamiltonian;  this 
suggests that $A_{0}$ does not have any physical significance, since it 
can 
have an arbitrary coefficient added to it.  If we wish to discard all 
variables 
without physical significance, we can also drop the final term of the 
total 
Hamiltonian (involving the term $E^{0} = 0$), to obtain a new Hamiltonian
\begin{equation} \label{FinalHam}
H = \int \frac{1}{4} F^{ab} F_{ab} + \frac{1}{2} E^{a} E_{a} + 
u(\mathbf{x})
{\partial}_{a} E_{a} \, \dif^3 \mathbf{x}\mbox{.}
\end{equation}
This Hamiltonian has the advantage of containing no reference to the 
unphysical variables $A_{0}$ and $E^{0}$, while still retaining some gauge 
freedom (in terms of the function $u(\mathbf{x})$.)

Finally, we note that this expression is not as unfamiliar as it might 
seem.  The first term of (\ref{FinalHam}) can be written as
\begin{equation}
    \frac{1}{4} F^{ab} F_{ab} = \frac{1}{2} (B_{x}^{2} + B_{y}^{2} + B_{z}^{2})
\end{equation}
(where the extra factor of two comes from the fact that we have 
summed both the term $F^{ab} F_{ab}$ and $F^{ba} F_{ba}$.)  
Similarly, the second term can be written as
\begin{equation}
    \frac{1}{2} E^{a} E_{a} = \frac{1}{2} (E_{x}^{2} + E_{y}^{2} + E_{z}^{2})
\end{equation}
Noting that $\partial_{a} E^{a} = 0$ in free space, the last term goes 
away, and we are left with the familiar expression
\begin{equation}
    H = \frac{1}{2} \int \left( \mathbf{B}^{2} + \mathbf{E}^{2} 
    \right) \, \dif^{3} \mathbf{x} \mbox{,}
\end{equation}
which is quite simply the energy stored in an electromagnetic field.

\setcounter{equation}{0}

\section{Simple Spin Networks and their Values}
\label{zoo}

With the definitions and identities in Section \ref{SpinNet}, we can create 
a veritable panoply of 
spin networks.  More importantly, we can assign numerical values to any 
closed spin network (since if a spin network is closed, we are effectively 
summing over all the indices of its component matrices.)  Some important 
strand identities and 
closed spin networks, along with their values, are given 
in~\cite{KaufLins};  we cite these results here:

\begin{itemize}
\item The $n$-loop, denoted $\Delta_{n}$:
\begin{equation}
\eqngraph[0.75]{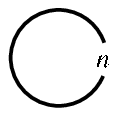} = (-1)^n (n+1)
\end{equation}

\item  The ``theta'', denoted $\theta(a,b,c)$:
\begin{equation}
\eqngraph[0.75]{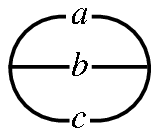} = (-1)^{(a+b+c)/2} \frac{(\frac{a+b+c}{2} + 
 1)!(\frac{a+b-c}{2})!(\frac{b+c-a}{2})!(\frac{c+a-b}{2})!}{a! b! c!}
\end{equation}
Often, we will encounter a theta-diagram with two identical strands, 
and the third strand equal to 2.  In this case, the value of the theta 
diagram becomes
\begin{equation} \label{mm2id}
    \theta(m,m,2) = (-1)^{m+1} \frac{(m+2)(m+1)}{2m}
\end{equation}

\item  The ``Tet'', denoted $\mbox{Tet} \left[ \begin{array}{ccc}
a & b & e \\
c & d & f 
\end{array} \right]$, whose value is given by
\begin{eqnarray}
\nonumber \mbox{Tet} \left[ \begin{array}{ccc}
a & b & e \\
c & d & f 
\end{array} \right]
 & = & \eqngraph[0.75]{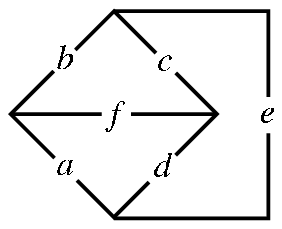}\\ 
& = & N \sum_{m \leq s \leq M} \frac{ (-1)^s (s+1)! }{\prod_{i} (s-a_i)! 
\prod_j (b_j-s)!}
\end{eqnarray}
where
$$
N = \frac{\prod_{i,j} (b_j - a_i)!}{a! b! c! d! e! f!}
$$
$$  \begin{array}{cc}
a_1 = \frac{1}{2}(a + d + e) & b_1 = \frac{1}{2}(b+d+e+f) \\[4pt]
a_2 = \frac{1}{2}(b+c+e) & b_2 = \frac{1}{2}(a+c+e+f)\\[4pt]
a_3 = \frac{1}{2}(a+b+f) & b_3 = \frac{1}{2}(a+b+c+d)\\[4pt]
a_4 = \frac{1}{2}(c+d+f) &
\end{array}
$$
$$ \begin{array}{cc}
m = max\{a_i\} & M = min\{b_j\}
\end{array}
$$

\item  The ``bubble'' diagram, which is proportional to a single strand:
\begin{equation} \label{bubble}
\eqngraph[0.75]{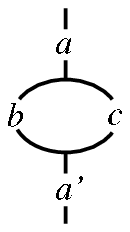} = \delta_{a a'} \frac{\theta(a,b,c)}{\Delta_a} \eqngraph[0.75]{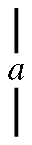}
\end{equation}

\item The lambda-move:
\begin{equation} \label{Lambda}
\eqngraph[0.75]{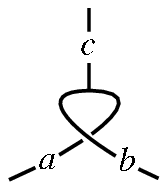} = \lambda_{c}^{ab} \eqngraph[0.75]{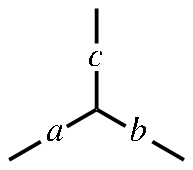} 
\end{equation}
where
\begin{equation}
\lambda_{c}^{ab} = (-1)^{(a(a+1) + b(b+1) - c(c+1))/2}
\end{equation}

\item The \emph{recoupling theorem}, which is perhaps the most important 
of these results:
\begin{equation}
\eqngraph[0.75]{Recoupling1} = \sum_{i} \left\{ \begin{array}{ccc}
a & b & i \\
c & d & i'
\end{array} \right\} \eqngraph[0.75]{Recoupling2}
\end{equation}
where the symbol on the right-hand side is the \emph{Kauffman-Lins 6-j 
symbol}.  Its value is given by
\begin{equation}
\left\{ \begin{array}{ccc}
a & b & e \\
c & d & f
\end{array} \right\} = \frac{ \mbox{Tet} \left[ \begin{array}{ccc}
a & b & e \\
c & d & f 
\end{array} \right] \Delta_e}{\theta(a,d,e) \theta(b,c,f)}
\end{equation}

The Kauffman-Lins 6-$j$ symbols are related to the more commonly known 
Wigner 6-$j$ symbols;  for the details of this relation, see Appendix
\ref{KLvsW}.

\item  Given a 2-edge grasping one edge of a trivalent vertex, it is 
possible to ``slide'' it onto the other two edges, using the relation:
\begin{equation} \label{Slide}
    r \eqngraph[0.75]{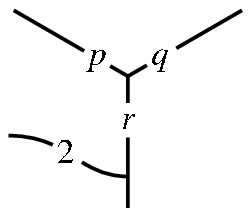} = p \eqngraph[0.75]{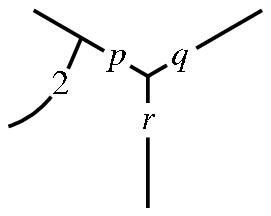} + q \eqngraph[0.75]{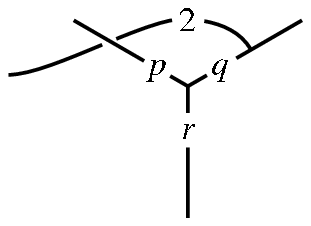}
\end{equation}

\item  A network with three edges branching from the vertices of a 
triangle can be written as a multiple of a simple trivalent vertex:
\begin{equation} \label{Triangle}
    \eqngraph[0.75]{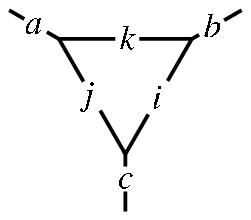} = \frac{\mathrm{Tet} \left[ \begin{array}{ccc} 
    a & b & c \\ i & j & k \end{array} \right]}{\theta(a,b,c)} 
    \eqngraph[0.75]{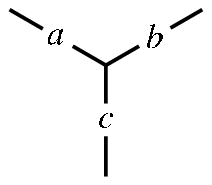}
\end{equation}

\end{itemize}

\setcounter{equation}{0}

\section{Relations Between Wigner 6-$j$ Symbols and Kauffman-Lins 6-$j$ 
Symbols}
\label{KLvsW}

The Wigner 6-$j$ symbols, used in transformations between different 
couplings of three angular momenta, are very closely related to the 
Kauffman-Lins 6-$j$ symbols used in the recoupling theorem.  Since the 
former of these symbols have been studied more thoroughly than the latter, 
it would be to our advantage to find a way to translate between them.  
However, there seems to be some confusion in the literature as to the 
proper translation.  It can be shown, using the explicit formulas for the 
Wigner 6-$j$ symbol in Varshalovich \cite{Varsh} and the Kauffman-Lins 
6-$j$ symbol in Kauffman and Lins \cite{KaufLins}, that
\begin{eqnarray}
\lefteqn{\left\{ \begin{array}{ccc} A/2 & B/2 & F/2 \\ C/2 & D/2 & E/2 
\end{array} \right\}_W = } \nonumber \\
& & \quad \left\{ \begin{array}{ccc} A & B & E \\ C & D & F \end{array} 
\right\}_{KL} \cdot \frac{\sqrt{\theta(B,C,E) \theta(A,D,E)}}{\Delta_E 
\sqrt{\theta(A,B,F) \theta(C,D,F)}} \cdot (-1)^{(E+F)/2}
\end{eqnarray}
where the subscript $KL$ indicates the Kauffman-Lins symbol and the $W$ 
subscript indicates the Wigner symbol.  Note the reversal of the arguments 
in the last column.

We will also have occasion to use the following identity, relating the Tet 
and the Wigner 6-$j$ symbol:
\begin{equation}
\left\{ \begin{array}{ccc} A/2 & B/2 & F/2 \\ C/2 & D/2 & E/2 \end{array} 
\right\}_W = \frac{\mbox{Tet} \left[ \begin{array}{ccc} A & B & E \\ C & D 
& F \end{array} \right]}{\sqrt{\theta (A,D,E) \theta (B,C,E) \theta 
(A,B,F) \theta (C,D,F)}}
\end{equation}

In general, we will omit the $KL$ subscript to indicate a Kauffman-Lins 
symbol;  however, a Wigner 6-$j$ symbol will, in this work, always have 
the $W$ subscript present.
\setcounter{equation}{0}

\section{Diffeomorphism Invariance \& the Angle Spectrum}
\label{Diffeo}

In our discussion of the definition and properties of spin networks, we 
scrupulously avoided any mention of what the exact nature of these networks 
might be.  Specifically, we did not mention whether these networks are 
merely bookkeeping tools, embedded in an underlying spatial manifold and 
drawing on the properties of that manifold to give values to the operators;
or whether this network has a deeper, more intrinsic structure.  While our
previous discussion proceeded without making this distinction, it is a
natural question one might have about the theory;  one might also wonder
whether the properties of the angle operator presented here might shed some
light on this subject.

One of the fundamental properties of general relativity is \emph{diffeomorphism 
invariance};  in other words, the equations of general relativity
are invariant under a certain class of smooth transformations known
as diffeomorphisms.  To get a better idea of what we mean by
``smooth'', suppose we have an infinitely stretchable, unpuncturable
balloon.  If we were to squash it or stretch it, we would have a smooth 
transformation of the sheet. If, however, we were to pinch a point on 
the balloon and pull it up, it would \emph{not} be a smooth transformation 
at this point of the sheet.  (It would, however, be a smooth 
transformation at every other point of the balloon.)

An important property of these diffeomorphisms is that they are 
locally linear;  in other words, if we look at a neighbourhood
around any point of the aforementioned balloon, we can 
approximate the transformation as a linear transformation.  Moreover,
this approximation can be made to arbitrary accuracy by selecting a 
sufficiently small neighbourhood.

Suppose, then, that our balloon has a network drawn on it.  If we 
examine a vertex of this network as we smoothly squash and stretch the 
balloon, we will see that the angles between the edges meeting at 
this vertex will change.
It is evident, then, that diffeomorphisms can change the angle
between two strands if they are embedded in an underlying manifold.
However, our angle operator does not depend on any properties of the 
geometry of the strands, merely on their labels.  This would seem to 
suggest, then, that the angle operator is diffeomorphism invariant.

There is another, more important conclusion
to be drawn from the nature of the angle operator.  The angle spectrum
for any given vertex only includes a finite number of angles;  even
if we consider the set of all vertices with a given valence (a
countable set), we would still only have a countable number of
possible angles in this larger set.  This is at odds with the
classical continuum model of spacetime, which predicts that within 
the set of all $n$-vertices (where we consider a set of $n$ rays
originating at a point in space to be an $n$-vertex), there are an
uncountable number of angles.   

We could try to explain this
uncountability away through diffeomorphism invariance;  it is 
certainly conceivable that through adequate stretching \& distorting
of the surrounding space, we could turn any vertex into any other
vertex;  at the very least, we might be able to define a countable
set of equivalence classes of vertices, where two vertices are defined
as equivalent if some diffeomorphism changes one vertex into the other.

Unfortunately, this is not the case.
Grot and Rovelli \cite{GrotRov} have shown that while
any intersection of $n$ lines in three-dimensional space can be deformed
into any other intersection through arbitrary transformations of the
underlying space, \emph{linear} transformations cannot do so for an
arbitrary intersection with $n \geq 5$.  Instead, for $n \geq 5$,
they show that if we define as equivalent two intersections which can 
be obtained from each other by a linear transformation, then there 
exist an uncountable number of equivalence classes of intersections.
These equivalence classes can therefore be parameterized by at least
one \emph{continuous} parameter;  Grot and Rovelli show that the 
dimension of this moduli space (i.e. the number of continuous parameters
required) for vertices of valence $n$ is given by
\begin{equation}
d(n) \geq (2n - 5) m - \frac{5}{2} m^2 - \frac{1}{2} m^3 \mbox{,}
\end{equation}
where
\begin{equation}
m(n) = \left\lfloor \frac{\sqrt{48n - 23} - 7}{6} \right\rfloor 
\mbox{.}
\end{equation}
This formula yields $d(n) \geq 0$ for $n = 2, 3, 4$, but $d(5) \geq 2$, 
$d(6) \geq 4$, and increases without bound as $n \to 
\infty$.  Grot and Rovelli also explicitly show (as an example)
that $d(5)$ is exactly equal to 2, and that the equivalence classes 
can be parameterized by the numbers
\begin{equation} \label{AngParam}
\lambda_1 = \frac{\angle^1_4 \angle^2_5}{\angle^2_4 \angle^1_5} \qquad 
\mbox{and} \qquad \lambda_2 = \frac{\angle^1_4 \angle^3_5}{\angle^3_4 
\angle^1_5}
\end{equation}
where we have labelled the edges $\{1,2,3,4,5\}$, and $\angle^i_j$ is
the angle between edges $i$ and $j$.

This result implies that even when we take diffeomorphism invariance
into account, there is still an uncountable spectrum of possible
``angles'' for vertices with valence greater than or equal to 5.  Of
course, these parameters are only functions of what one might 
normally consider to be the angles of the vertex, as in 
(\ref{AngParam});  nevertheless, classical vertices under 
diffeomorphism still possess more degrees of freedom in possible
angle values than the quantized angle operator would seem to allow.

What can we conclude from these facts?  For vertices with $n \geq 5$, 
it should not surprise us that we have been reduced from an 
uncountably infinite number of possible measurements to a countably 
infinite number.  After all, such a reduction occurs in many simple 
quantum mechanical systems, most notably in the measurements of 
angular momentum --- in the classical picture, the angular momentum 
of a particle can take on any real value, while in the quantum 
picture, it must be an integer multiple of $\hbar$.  The existence of 
these classical parameters may also imply, however, that the angle 
operator is ``not all there is'';  if these parameters do persist in 
the quantum theory, then some (if not all) angle eigenvalues are 
highly degenerate.

The $n = 3$ case, however, is much more surprising.  In this case, the angle 
operator can still produce a countably infinite number of 
measurements (from the set of all trivalent vertices);  
these measurements are, as we showed above, diffeomorphism-invariant.
However, under classical diffeomorphisms, a set of three vectors which span 
an $m$-dimensional space can be mapped to any other three vectors 
which also span an $m$-dimensional subspace.  If $m=3$, we have what 
Grot and Rovelli call the ``non-degenerate'' case;  in this case, any two
vertices are diffeomorphism equivalent.  Grot and Rovelli also cite 
the result of Arnold \cite{Arnold} that this holds for the $m=2$ case 
as well.  Finally, the $m=1$ case is trivial:  the only 
possible classical angle measurements are $0^{\mathrm{o}}$ and 
$180^{\mathrm{o}}$.  Hence, there is only a \emph{finite amount} of 
diffeomorphism-invariant information that can be extracted from a 
classical trivalent vertex!  This is a drastic difference 
from the normal state of affairs in quantum mechanics;  while 
quantizing a given theory normally \emph{restricts} the amount of 
knowledge one can have about a system, the angle operator actually 
seems to be \emph{increasing} it.  This suggests that there is 
something fundamentally different going on at these low-valence 
vertices;  the theory predicts that our naive continuum model of space 
breaks down at these tiny length scales.

\newpage

\addcontentsline{toc}{section}{Acknowledgements}
\section*{Acknowledgements}
I would like to thank my advisor, Seth Major, for his invaluable help 
in the preparation of this thesis;  without him, this work would 
certainly not have been possible.  I would also like to thank my 
professors and fellow students at Swarthmore College, for four years 
of superb academic stimulation.  Finally, I would like to thank my 
parents, for all of their support and encouragement over the past 21 years.

\end{document}